\pdfoutput=1
\documentclass[bibyear, longauth]{aa}
\usepackage{graphicx}
\usepackage{color}
\usepackage{amssymb}
\usepackage[varg]{txfonts}
\begin{document} 

   \title{Molecular line emission in NGC~1068 imaged with ALMA \thanks{Based on observations carried out with ALMA in Cycle 0.}}
   \subtitle{I. An AGN-driven outflow in the dense molecular gas}
            
             \author{S.~Garc\'{\i}a-Burillo \inst{1}
			\and
	   F.~Combes\inst{2}		
			\and
	   A.~Usero\inst{1} 		   
			\and
           S.~Aalto\inst{3}
           		\and
           M.~Krips\inst{4}
	   	        \and
	  S.~Viti\inst{5}
	   		\and    
	  A.~Alonso-Herrero\inst{6}		
			\and			
          L.~K.~Hunt\inst{7}
            		\and	
          E.~Schinnerer\inst{8}		
		        \and
%%%%%%%%%%%%%%%%%%%%%	
	  A.~J.~Baker\inst{9}
	  	        \and
	  F.~Boone\inst{10}
	  		\and	
	  V.~Casasola\inst{11}      
         		\and
          L.~Colina\inst{12}	 
          		\and
	 F.~Costagliola\inst{13}	 
	  		\and
          A.~Eckart\inst{14}
          	        \and
	  A.~Fuente\inst{1}
		\and		
	  C.~Henkel\inst{15, 16}
	  	\and	  
	  A.~Labiano\inst{1, 17}
	  	\and		
	  S.~Mart\'{\i}n\inst{4}	
	  	\and	
	  I.~M\'arquez\inst{13}
	  	\and
	  S.~Muller\inst{3}
	  	\and
	  P.~Planesas\inst{1}	
	                \and
	  C.~Ramos Almeida\inst{18, 19}
	    		\and
	  M.~Spaans\inst{20}
	  		\and			                
	  L.~J.~Tacconi\inst{21}
	  		\and
	  P.~P.~van der Werf\inst{22}}
   \institute{
          Observatorio Astron\'omico Nacional (OAN)-Observatorio de Madrid, Alfonso XII, 3, 28014-Madrid, Spain 
			  \email{s.gburillo@oan.es} 		  
			  	 \and
      Observatoire de Paris, LERMA, CNRS, 61 Av. de l'Observatoire, 75014-Paris, France 
	 \and
	 Department of Earth and Space Sciences, Chalmers University of Technology, Onsala Observatory, 439 94-Onsala, Sweden  	 
	 \and
	 Institut de Radio Astronomie Millim\'etrique (IRAM), 300 rue de la Piscine, Domaine Universitaire de Grenoble, 38406-St.Martin d'H\`eres, France 
	 \and
	 Department of Physics and Astronomy, UCL, Gower Place, London WC1E 6BT, UK
	 \and 	 
	 Instituto de F\'{\i}sica de Cantabria, CSIC-UC, E-39005 Santander, Spain. Augusto G.~Linares Senior Research Fellow.	 
	 \and
	 INAF-Osservatorio Astrofisico di Arcetri, Largo Enrico Fermi 5, 50125-Firenze, Italy 	 
	 \and
	 Max-Planck-Institut f\"ur Astronomie, K\"onigstuhl, 17, 69117-Heidelberg, Germany 	 	 
        \and        
         Department of Physics and Astronomy, Rutgers, The State University of New Jersey, Piscataway, NJ 08854, USA 
        \and  
       Universit\'e de Toulouse, UPS-OMP, IRAP, 31028, Toulouse, France   
        \and
       INAF - Istituto di Radioastronomia, via Gobetti 101, 40129, Bologna, Italy 
	\and
	Centro de Astrobiolog\'{\i}a (CSIC-INTA), Ctra de Torrej\'on a Ajalvir, km 4, 28850 Torrej\'on de Ardoz, Madrid, Spain         
      	\and
	 Instituto de Astrof\'{\i}sica de Andaluc\'{\i}a (CSIC), Apdo 3004, 18080-Granada, Spain 
	 \and
	I. Physikalisches Institut, Universit\"at zu K\"oln, Z\"ulpicher Str. 77, 50937, K\"oln, Germany	
	\and
	Max-Planck-Institut f\"ur Radioastronomie, Auf dem H\"ugel 69, 53121, Bonn, Germany	
	\and
	Astronomy Department, King Abdulazizi University, P.~O. Box 80203, Jeddah 21589, Saudi Arabia
	\and
	Institute for Astronomy, Department of Physics, ETH Zurich, CH-8093 Zurich, Switzerland	
	\and
	Instituto de Astrof\'{\i}sica de Canarias, Calle V\'{\i}a L\'actea, s/n, E-38205 La Laguna, Tenerife, Spain
	\and
	Departamento de Astrof\'{\i}sica, Universidad de La Laguna, E-38205, La Laguna, Tenerife, Spain
	\and
	Kapteyn Astronomical Institute, University of Groningen, PO Box 800, NL-9700 AV Groningen
	\and
         Max-Planck-Institut f\"ur extraterrestrische Physik, Postfach 1312, 85741-Garching, Germany 
	\and
	Leiden Observatory, Leiden University, PO Box 9513, 2300 RA Leiden, Netherlands 
	}
         
	\date{Received ---; accepted ----}
             
% \abstract{}{}{}{}{} 
% 5 {} token are mandatory
 
  \abstract
  {}
   {We investigate the fueling and the feedback of star formation and nuclear activity in NGC~1068, a nearby ($D=14$~Mpc) Seyfert 2 barred galaxy, by analyzing the distribution and kinematics of the molecular gas in the disk.  We aim to understand if and how gas accretion can self-regulate.}
   {We have used the Atacama Large Millimeter Array (ALMA) to map the emission of a set of dense molecular gas ($n$(H$_2$)~$\simeq10^{5-6}$~cm$^{-3}$) tracers (CO(3--2), CO(6--5), HCN(4--3), HCO$^+$(4--3), and CS(7--6)) and their underlying continuum emission in the central $r\sim2$~kpc of NGC~1068 with spatial resolutions $\sim0.3\arcsec-0.5\arcsec$ ($\sim20-35$~pc for the assumed distance of $D=14$~Mpc). }
   {The sensitivity and spatial resolution of ALMA give an unprecedented detailed view of the distribution and kinematics of the dense molecular gas 
 ($n$(H$_2$)~$\geq10^{5-6}$cm$^{-3}$) in NGC~1068. Molecular line and dust continuum emissions are detected from a $r\sim200$~pc off-centered circumnuclear disk (CND), from the 2.6~kpc-diameter bar region, and from the $r\sim1.3$~kpc starburst (SB) ring. Most of the emission in  HCO$^+$, HCN, and CS stems from the CND.   Molecular line ratios show dramatic order-of-magnitude changes inside the CND that are correlated with the  UV/X-ray illumination by the AGN, betraying ongoing feedback. We used the dust continuum fluxes measured by ALMA together with NIR/MIR data to constrain the properties of the putative torus using CLUMPY models and found a torus radius of $20^{+6}_{-10}\,$pc. 
 The Fourier decomposition of the gas velocity field indicates that rotation is perturbed by  an inward radial flow in the SB ring and the bar region.  However, the gas kinematics from $r\sim50$~pc out to $r\sim400$~pc reveal a massive ($M_{\rm mol} \sim 2.7^{+0.9}_{-1.2} \times 10^7$~M$_{\sun}$) outflow in all molecular tracers. The tight correlation between the ionized gas outflow, the radio jet, and the occurrence of outward motions in the disk suggests that the outflow is AGN driven.}
   {The molecular outflow is likely launched when the ionization cone of the narrow line region sweeps the nuclear disk. The outflow rate estimated in the CND, $\mathrm{d}M/\mathrm{d}t\sim63^{+21}_{-37}~M_{\odot}$~yr$^{-1}$, is an order of magnitude higher than the star formation rate at these radii, confirming that the outflow is AGN driven. The power of the AGN is able to account for the estimated momentum and kinetic luminosity of the outflow. The CND mass load rate of the CND outflow implies a very short gas depletion timescale of $\leq1$~Myr. The CND gas reservoir is likely replenished on longer timescales by efficient gas inflow from the outer disk.}

   \maketitle
%
%________________________________________________________________

\section{Introduction}

%\subsection{Feeding and feedback in Active Galactic Nuclei (AGNs)}

The study of the content, distribution, and kinematics of interstellar gas is a key to understanding the origin and maintenance 
of nuclear activity in galaxies. The processes involved in the fueling of active galactic nuclei (AGNs) encompass a wide range of scales, both spatial and temporal (Combes~\cite{Com03,Com06}; Jogee~\cite{Jog06}). Current mm-interferometers are instrumental 
in providing a sharp view of the distribution and kinematics of molecular gas, the dominant gas phase in galaxy nuclei, through 
extensive CO line mapping. Combined with high-resolution near infrared images, interferometric CO maps are used to derive 
the angular momentum transfer budget in the circumnuclear disks of AGNs (e.g., Garc\'{\i}a-Burillo et al.~\cite{Gar05}; 
Haan et al.~\cite{Haa09}; Meidt et al.~\cite{Meid13}). Maps of CO in nearby AGNs unveil a wide range of large-scale and
embedded $m=	1, 2$ instabilities in their central 1~kpc circumnuclear disks (CNDs). Results from the NUclei of Galaxies (NUGA) project, a CO 
interferometric survey of a sample of 25 nearby low-luminosity AGNs (Garc\'{\i}a-Burillo et al.~\cite{Gar03}), indicate that 
molecular gas is frequently stalled in rings, which are the signposts of gravity torque barriers, and that only $\sim 1/3$ of 
galaxies show smoking-gun evidence of AGN fueling (Garc\'{\i}a-Burillo \& Combes~\cite{Gar12}). In agreement with the picture 
drawn from observations, the most recent state-of-the-art numerical simulations show that several mechanisms,  namely large-scale and nuclear stellar bars as well as slowly precessing lopsided $m=1$ instabilities, are expected to 
cooperate to drain the gas angular momentum at the different spatial scales of galaxy disks (Hopkins et al.~\cite{Hop10a, Hop11, Hop12}).

Furthermore, the use of molecular tracers specific to the dense gas phase can probe the feedback of activity on the 
chemistry and energy balance/redistribution of the interstellar medium of galaxies. Observations suggest that the excitation 
and chemistry of the main molecular  species in AGNs are different with respect to those found in purely star-forming 
galaxies (Tacconi et al.~\cite{Tac94}; Kohno et al.~\cite{Koh01}; Usero et al.~\cite{Use04}; Graci\'a-Carpio et al.~\cite{Gra08}; Krips et al.~\cite{Kri08, Kri11}; Garc\'{\i}a-Burillo et al.~\cite{Gar10};  Imanishi et al.~\cite{Ima07, Ima13}; Aladro et al.~\cite{Ala13}). Although theoretical models have tried to explain these 
differences, the underlying reasons for the 
{\em apparent} AGN specificity are still debated (Lepp \& Dalgarno~\cite{Lep96}; Maloney et al.~\cite{Mal96}; Meijerink \& Spaans~\cite{Mei05}; Meijerink et al.~\cite{Mei07}; Yamada et al.~\cite{Yam07}; Harada et al.~\cite{Har13}). Molecular outflows, which are considered as a footprint of the mechanical feedback of activity, are being discovered in a 
growing number of nearby active galaxies including ultra luminous infrared galaxies (ULIRGs), radio galaxies, and Seyferts (Feruglio et al.~\cite{Fer10}; Sturm  et al.~\cite{Stu11}; Alatalo et al.~\cite{Ala11}; Chung et al.~\cite{Chu11}; Aalto et al.~\cite{Aal12}; Dasyra \& Combes~\cite{Das12}; Combes et al.~\cite{Com13}; Morganti et al.~\cite{Mor13}; Cicone et al.~\cite{Cic12, Cic14}). Radiative and mechanical feedback is often invoked as a mechanism of self-regulation in galaxy evolution (Di Matteo et al.~\cite{DiM05, DiM08}). Observations of nearby AGNs, where the distribution and kinematics of molecular gas can be spatially resolved, are thus instrumental if we are to understand  if and how gas accretion can self-regulate in galaxies.

\subsection{The prototypical Seyfert 2 galaxy NGC~1068}

 NGC\,1068 is a prototypical nearby ($D=14$~Mpc; Bland-Hawthorn et al.~\cite{Bla97}) Seyfert~2 galaxy. It has a large-scale oval and a nuclear bar with a pseudo-bulge which is overly massive with respect to its central black hole (e.g., Kormendy \& Ho~\cite{Kor13}). NGC~1068 has been the subject of numerous campaigns using molecular line observations to study the fueling and the feedback of activity. Schinnerer et al.~(\cite{Sch00}) used the Plateau de Bure Interferometer (PdBI) to map  the emission of molecular gas in the central $r\sim$1.5-2~kpc disk using the $J=1-0$ and $J=2-1$ lines of CO. The CO maps spatially resolved the distribution of molecular gas in the disk, showing a prominent starburst (SB) ring of $\sim1-1.5$\,kpc--radius, which contributes significantly to the total CO luminosity of NGC\,1068 (see also Planesas et al.~\cite{Pla91}, Helfer et al.~\cite{Hel95} and Baker~\cite{Bak00}). Furthermore CO emission is detected in a central $r\sim200$\,pc circumnuclear disk (CND) that surrounds the AGN. The CO($J=3-2$) map recently obtained by Tsai et al.~(\cite{Tsa12}) with the Submillimeter Array (SMA) gives a similar picture of the large-scale  distribution of molecular gas.

  The kinematics of molecular gas in NGC~1068 were first interpreted as being due to the action of two embedded bars. The gas 
response seen in the CND, characterized by strong non-circular motions observed in CO, SiO, and CN maps, indicates that molecular clouds would be trapped between the Inner Lindblad Resonances (ILRs) of the inner nuclear bar and thus cannot fuel the AGN at present (Schinnerer et al.~\cite{Sch00}; Baker~\cite{Bak00}; Garc\'{\i}a-Burillo et al.~\cite{Gar10}). Alternative scenarios invoke the existence of large-scale outflow motions in the 
CND (Galliano \& Alloin~\cite{Gal02}; Davies et al.~\cite{Dav08}; Garc\'{\i}a-Burillo et al.~\cite{Gar10}; Krips et al.~\cite{Kri11}); these models would also suggest that the AGN feeding is  
presently thwarted.

However, closer to the nucleus, at $r\leq50$~pc, the kinematics of the molecular gas  revealed by 
the 2.12~$\mu$m H$_2$ 1--0 S(1) map of M\"uller-S{\'a}nchez et al.~(\cite{Mue09}) give a completely different picture. These data, sensitive to hot
($T_{\rm K}\simeq10^3$~K) and moderately dense ($n$(H$_2$)~$\simeq10^3$~cm$^{-3}$) molecular gas, show elliptical streamers that bridge the CND and the central engine. M\"uller-S{\'a}nchez et al.~(\cite{Mue09}) suggest that these structures correspond to gas feeding the AGN.  Although on different spatial scales, inflowing and outflowing gas may therefore coexist in the CND. However, the H$_2$ map only traces a small fraction of the total molecular gas reservoir of the CND, which is known to be much denser ($n$(H$_2$)~$\simeq10^{5-6}$~cm$^{-3}$; Sternberg et al.~\cite{Ste94}; Usero et al.~\cite{Use04}; Krips et al.~\cite{Kri08, Kri11}; P\'erez-Beaupuits et al.~\cite{Per07, Per09}). The use of molecular tracers which are more representative of the total H$_2$ content is thus essential  to derive the mass and energetics associated with the different inflowing/outflowing gas components at the CND.

Interferometric images of NGC~1068, obtained in tracers specific to the dense molecular gas, have also revealed the existence of a strong  chemical differentiation in the disk of this Seyfert. Essential diagnostic line ratios are different in the SB ring compared to the CND. Tacconi et al.~(\cite{Tac94}) derived a high HCN/CO intensity ratio ($\sim$1) in the CND, about a factor of 5--10 higher than the ratio measured in the SB ring (Usero et al. in prep.). 
Radiative transfer calculations showed that the abundance of HCN relative to CO is globally enhanced in the CND: HCN/CO~$\sim$10$^{-3}$ (Sternberg et al.~\cite{Ste94}; Usero et al.~\cite{Use04}; Krips et al.~\cite{Kri11}; Kamenetzky et al.~\cite{Kam11}). The detection of molecular ions like HOC$^+$ and H$_3$O$^+$ has been interpreted as the signature of X-ray processing (Usero  et al.~\cite{Use04}; Aalto et al.~\cite{Aal11}). Garc\'{\i}a-Burillo et al.~(\cite{Gar10}) analyzed the likely drivers of chemical differentiation inside the CND with high-resolution observations of CN and SiO. The abundances of SiO and CN are enhanced at the extreme velocities of gas associated with non-circular motions/shocks close to the AGN ($r<70$~pc). On the other hand, the correlation of CN/CO and SiO/CO ratios with hard X-ray irradiation suggests that the CND is a giant X-ray-dominated region (XDR). Although these results imply a strong radiative and mechanical feedback in the CND, the mechanism that drives the excitation and chemistry of molecular gas is yet to be elucidated.

\subsection{This project}
 We use here the Atacama large millimeter array (ALMA) to map the emission of a set of molecular gas tracers 
 (the $J=3-2$ and $J=6-5$ lines of CO, the $J=4-3$ lines of HCN and HCO$^+$, and the $J=7-6$ line of CS) and their
 underlying continuum emission in the central $r\sim2$~kpc of NGC~1068 with spatial resolutions 
 $\sim0.3\arcsec-0.5\arcsec$ (20--35~pc).  These line transitions span a range of critical densities ($n_{\rm crit}[{\rm CO(3-2)}]\sim$~a few 10$^4$~cm$^{-3}$, $n_{\rm crit}[{\rm CO(6-5)}]\sim$~a few 10$^5$~cm$^{-3}$,  $n_{\rm crit}[{\rm HCO^+(4-3)}]\sim$~10$^7$~cm$^{-3}$, 
   $n_{\rm crit}[{\rm CS(7-6)}]\sim$~a few 10$^7$~cm$^{-3}$, and  $n_{\rm crit}[{\rm HCN(4-3)}]\sim$~10$^8$~cm$^{-3}$). The actual {\em average} densities of molecular gas probed by these lines in NGC~1068 are seen to be $\sim$10$^{5-6}$~cm$^{-3}$ due to subthermal excitation of the higher J-lines and radiative trapping (Krips et al.~\cite{Kri11}; Viti et al.~\cite{Vit14}, hereafter, paper~II).   
 The line and continuum maps of the dense molecular gas 
 and dust emission greatly improve the sensitivity and spatial  resolution of any 
 previous interferometric study of NGC~1068 in the (sub)millimeter range. We analyze  the distribution, 
 kinematics, and excitation of the molecular gas and study the processes associated with the fueling and the feedback of activity in this Seyfert. In paper~II we use the line ratio maps derived in this work to model the excitation and chemistry of molecular gas in the different environments of the CND and SB ring. Star formation laws will be analyzed in a future paper (Garc\'{\i}a-Burillo et al. in prep.; paper III).
 
 We describe in Sect.~\ref{Obs} the ALMA observations and the ancillary data used in this work.  Section~\ref{obs-cont} presents the continuum maps obtained at 349~GHz and 689~GHz. Section~\ref{dustmass} discusses dust masses derived from continuum observations and illustrates the use of the CLUMPY torus models  (Nenkova et al.~\cite{Nen08a, Nen08b}) to constrain the parameters of the torus. The distribution of the molecular gas derived from the CO, HCN, HCO$^+$, and CS line maps is discussed in Sect.~\ref{line}. We describe the kinematics of the molecular gas and derive the main properties of the outflow component in Sect.~\ref{outflow}. A first description of line ratio maps is presented in Sect.~\ref{lineratios}. The main conclusions of this work are summarized in Sect.~\ref{summary}. 
%The new X-ray images of NGC~1068 of Wang+2012 illustrate the role of mechanical and radiative feedback in this source.

 \section{Data}\label{Obs}

 \subsection{ALMA data}\label{ALMA-obs}
 
 We observed the CO($J=3-2$) emission and the CO($J=6-5$) emission in 
NGC~1068 with ALMA during Cycle~0 using Band~7 and Band~9 receivers (project-ID: $\#$2011.0.00083.S). 
The data in the two bands were calibrated 
using the ALMA reduction package {\tt CASA\footnote{http//casa.nrao.edu/}} while the calibrated uv-tables 
were subsequently exported to {\tt GILDAS\footnote{http://www.iram.fr/IRAMFR/GILDAS}} where the mapping and cleaning 
were performed as detailed below. Hereafter we adopt a distance to NGC~1068 of
$D\sim14$~Mpc (Bland-Hawthorn et al.~\cite{Bla97}); this implies a spatial scale of $\sim$70~pc/$\arcsec$.

\subsubsection{Band~7 maps}\label{B7-obs}

In order to cover the CND and the SB ring of NGC~1068 we used an eleven-field mosaic in Band~7 with 
a field-of-view of 17$\arcsec$ per mosaic pointing. In total four tracks were 
observed between June and August 2012 resulting in initial on-source times 
(i.e., before flagging) of $\sim24-39$~minutes per track or a total of 
138~minutes. Between 18 and 27 antennas were available during the 
observations with projected baselines ranging from 17~m to 400~m. Weather conditions 
were good with median system temperatures of T$_{\rm sys}=120-220$~K and peak values not exceeding 
300~K. Four spectral windows with a bandwidth of 1.875~GHz each and 
a spectral resolution of 488~kHz were placed in Band~7, two in the 
lower side band (LSB) and two in the upper sideband (USB); the centers of the two 
sidebands are separated by 12~GHz. The four windows were centered on 
the following sky frequencies: 341.955~GHz and 343.830~GHz in the LSB and 
353.830~GHz and 354.705~GHz in the USB. This setup allowed us to simultaneously observe 
CO($J=3-2$) (345.796~GHz at rest) and HCO$^+$($J=4-3$) (356.734~GHz at rest) in the 
LSB bands, and  HCN($J=4-3$) (354.505~GHz at rest) and CS($J=7-6$)(342.883~GHz at rest) 
in the USB bands. J0522-362 and 3C454.3 
were used as bandpass calibrators and J0217+017 was used to calibrate 
the amplitudes and phases in time. To set the absolute flux scale, 
Uranus and Callisto were observed using the Butler-JPL-Horizons 2012 
catalogue to model their visibilities as both were resolved at the 
given resolution of the observations. We estimate that the absolute flux accuracy is
 about 10--15$\%$.

The angular resolution obtained using natural weighting was 
$\sim0\farcs6\times0\farcs5$ at a position angle of 
$\sim60^\circ$ in all the line and continuum data cubes.  The conversion factor between 
Jy~beam$^{-1}$ and K is 37 K~Jy$^{-1}$~beam.  The line data cubes were binned to a common frequency 
resolution of 14.65~MHz (equivalent to $\sim$12-13~km~s$^{-1}$ in Band~7). The point source sensitivities in the line data cubes 
were derived selecting areas free from emission in all channels. They range from  2~mJy~beam$^{-1}$ (in the CS line)  
up to 2.8~mJy~beam$^{-1}$ (in the CO line) in channels of 12.4-12.8~km~s$^{-1}$ width. 
Images of the continuum emission were obtained by averaging in each of the four sub-bands those
channels free of line emission. The resulting maps were averaged to produce an image  of the
continuum emission of the galaxy at 349~GHz. The corresponding point source sensitivity for the
continuum is 0.14~mJy~beam$^{-1}$.

 As our observations do not contain short-spacing correction, we expect that a significant amount of flux will be filtered out on scales 
 $\geq6\arcsec$ for continuum 
 emission in Band~7.  A comparison 
 with the single-dish fluxes measured by Papadopoulos \& Seaquist~(\cite{Pap99}) with the James Clerk Maxwell Telescope (JCMT) in apertures of $\sim15\arcsec$ indicates that we are filtering up to $\simeq2/3$ of the total flux on these 
 scales. This will  affect mostly the low level emission that may extend on large-scales in the interarm and spiral arm region. However, fluxes estimated in small 
 ($\sim 1- 4\arcsec$) apertures centered on the brightest spots of emission of the CND and the SB ring are not expected to be significantly underestimated. The percentage of missing flux in the continuum maps in Band~7 represents the worst case scenario: the same estimate points to significantly lower missing fluxes in the molecular line ALMA  maps.  In order to estimate the missing flux due to  the lack of short-spacing correction in the CO(3--2) ALMA map we used as references the fluxes measured in two studies of the galaxy done with single-dish telescopes: 
the single-pointed observations done at 22$\arcsec$-angular resolution with the Heinrich Hertz Telescope (HHT) by Mao et al.~(\cite{Mao10}), and the fully-sampled map of the central 1$\arcmin$ region of NGC~1068
done at 14$\arcsec$-angular resolution with the JCMT by Israel~(\cite{Isr09a}). Compared on these apertures (14--22$\arcsec$), the ALMA map recovers 80--90$\%$ of the CO(3--2) flux at the CND and over the
SB ring. This percentage is lowered to 60--70$\%$  at the edges of the area mapped by the ALMA mosaic and over the interarm region. As expected, the clumpy distribution of gas aided by the velocity structure of the emission helps us
recover the bulk of the flux over the map on these spatial scales. Fluxes estimated on smaller
 ($\sim 1-4\arcsec$) apertures centered on the brightest emission spots of the CND and the SB ring are thus not expected to be significantly affected.

% (see Sect.~\ref{B9} and Sect.~\ref{COmissing}).

We set the phase tracking center of the central field of the mosaic to $\alpha_{2000}=02^{h}42^{m}40.771^{s}$, 
  $\delta_{2000}=-00^{\circ}00^{\prime}47.84\arcsec$, which is the galaxy's center according to SIMBAD (taken from the Two Micron All Sky Survey--2MASS survey; Strutskie et al.~\cite{Str06}).  By default,  ($\Delta\alpha$, $\Delta\delta$)-offsets 
  are relative to this position. Rest frequencies for all the lines were corrected for the 
  recession velocity initially assumed to be $v_{\rm o}$(LSR)$~=1135$~km~s$^{-1}=v_{\rm o}$(HEL)$=1146$~km~s$^{-1}$. 
  Systemic velocity is nevertheless re-determined in this work (Sect.~\ref{kinemetry}) as $v_{\rm sys}$(LSR)$~=1116$~km~s$^{-1}=v_{\rm sys}$(HEL)$=1127$~km~s$^{-1}$.  Relative velocities throughout 
  the paper refer to  $v_{\rm sys}$.  
 %______________________________________________ Fig1: dust-maps_________________________
 \begin{figure*}[tbh!]
   \centering  
 \includegraphics[width=18cm]{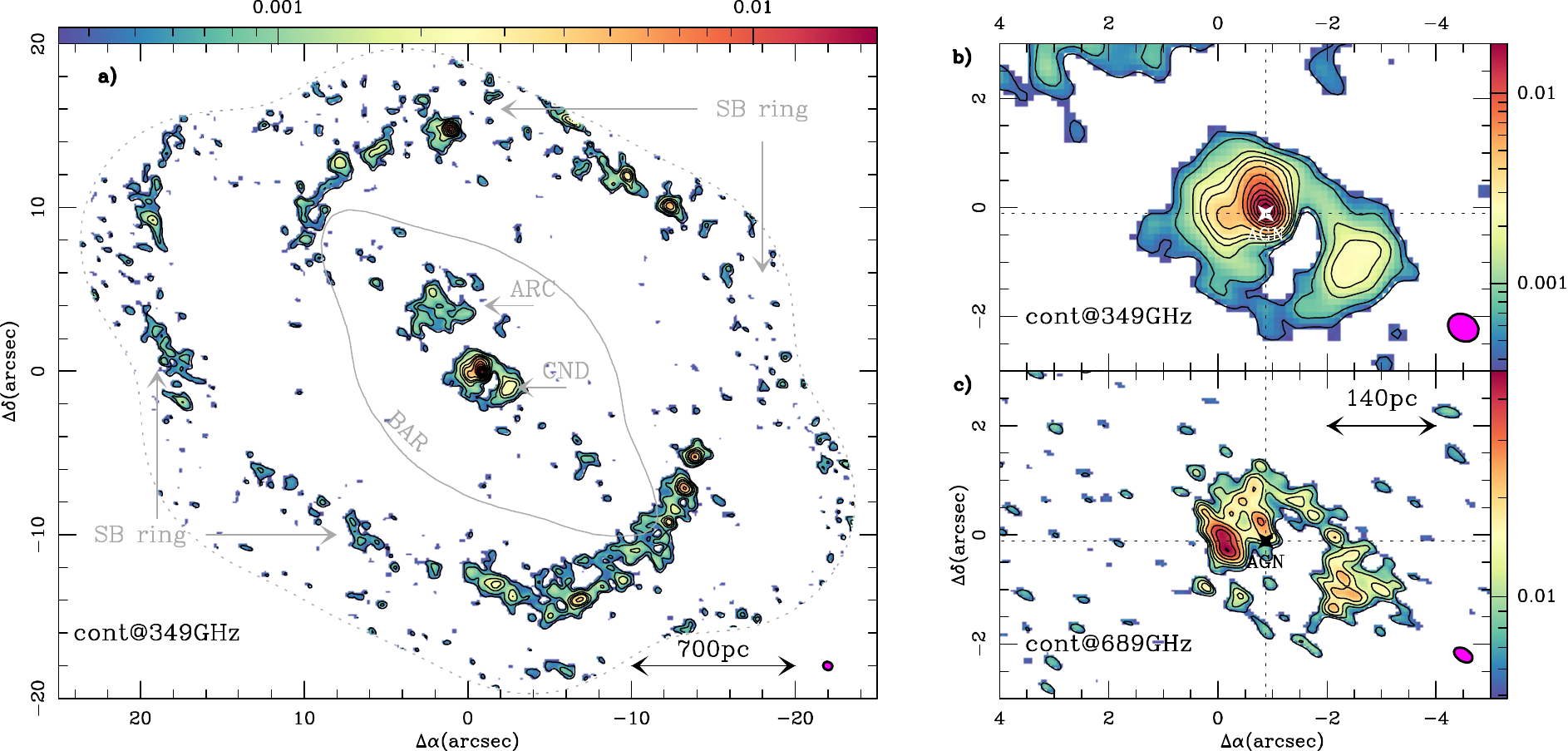}
       \caption{{\bf a)}~({\it Left panel})~The continuum emission map of NGC~1068 obtained with ALMA at 349~GHz. The map is shown in color scale (in Jy~beam$^{-1}$ units as indicated) with contour 
levels  3$\sigma$, 5$\sigma$, 10$\sigma$, 15$\sigma$, 20$\sigma$, 30$\sigma$ to  120$\sigma$ in steps of  15$\sigma$, where 1$\sigma=0.14$~mJy~beam$^{-1}$. 
($\Delta\alpha$, $\Delta\delta$)-offsets are relative to the location of the phase tracking center adopted in this work: 
$\alpha_{2000}=02^{h}42^{m}40.771^{s}$, $\delta_{2000}=-00^{\circ}00^{\prime}47.84\arcsec$.  We highlight the location of the regions and components of the emission described in Sect.~\ref{obs-cont}: the CND, the bar (identified by a representative isophote of the NIR K-band image of 2MASS), the bow-shock arc, and the SB ring. We also plot the edge of the eleven-field mosaic (gray dashed contour). The filled
ellipse at the bottom right corner represents the beam size at 349~GHz ($0\farcs6\times0\farcs5$ at  $PA=60^{\circ}$). {\bf b)}~({\it Upper right panel})~Same as {\bf a)} but zooming in on the CND region. The position of the AGN ([$\Delta\alpha$, $\Delta\delta$]~$\approx$~[--0.9$\arcsec$,~--0.1$\arcsec$]~=~[$\alpha_{2000}=02^{h}42^{m}40.71^{s}$, $\delta_{2000}=-00^{\circ}00^{\prime}47.94\arcsec$]) is highlighted by the star marker. {\bf c)}~({\it 
Lower right panel})~Continuum emission in the CND region at 689~GHz.  Color scale is given in Jy~beam$^{-1}$ units. Contour levels are 3$\sigma$, 5$\sigma$, 7$\sigma$, and 10$\sigma$ to 25$\sigma$ in 
steps of  5$\sigma$, where 1$\sigma=1.95$~mJy~beam$^{-1}$.  The filled
ellipse at the bottom right corner represents the beam size at 689~GHz ($0\farcs4\times0\farcs2$ at  $PA=50^{\circ}$). }
              \label{dust-maps}
\end{figure*}
   
   %___________________________________________________________________________________   

\subsubsection{Band~9 maps}\label{B9-obs}

Given the small field-of-view ($\sim9\arcsec$) and the difficulty of Band~9 
observations, we used only one field to cover the circumnuclear disk 
in NGC~1068. In total four different tracks were observed 
during July and August 2012 of which one had to be discarded because 
of poor quality. Weather 
conditions during the remaining three tracks were excellent with median system temperatures of T$_{\rm 
sys}=600$~K for a range of 400--1200~K. This results in an initial (i.e., prior to flagging) 
on-source time of 7--29~minutes per track or a total of 
$\sim 52$~minutes. Between 21 and 27 antennas were used during each 
track with projected baselines ranging between 17~m and 400~m. Three spectral windows with a 
spectral bandwidth of 1.875~GHz and a spectral resolution of 488~kHz 
were placed in the USB at sky frequencies of 686.899~GHz, 688.865~GHz, and 
690.884~GHz, while only one window with a spectral bandwidth of 
1.875~GHz and a spectral resolution of 488~kHz were placed in the LSB at 
a sky frequency of 674.944~GHz; the centers of the two sidebands are separated by 
$\sim 16$~GHz in Band~9. This setup allowed us to cover the CO($J=6-5$) 
line entirely and having enough line-free channels to determine the 
continuum emission. Bandpass calibration was performed on either 
Uranus or Ganymede while phase and amplitude calibration in time was 
done on J0339-017. The absolute flux scale was set by using models 
from the Butler-JPL-Horizons 2012 catalogue for Uranus, Ganymede, and/or Callisto. For most tracks, we had at least two solar bodies 
available so that we could crosscheck the flux calibration. However, 
we assume a conservative accuracy of $\sim30\%$ for the absolute 
flux scale. We performed a self-calibration on the final maps using the strongest 
peak emission, located in the eastern knot of the CND, which purposely coincides 
within 0.1$\arcsec$ with the adopted phase tracking center. Self-calibration helped to 
improve the fidelity and more importantly the dynamic range of the final images.

The synthesized beam obtained using natural weighting is $\sim0\farcs4\times0\farcs2$ at a position 
angle of $\sim50^\circ$.  The conversion factor between 
Jy~beam$^{-1}$ and K is 27 K~Jy$^{-1}$~beam.  The line data cube was binned to a frequency 
resolution of 29.30~MHz (equivalent to $\sim12.8$~km~s$^{-1}$). The point source sensitivity in the line data cube, 
derived by selecting areas free of emission is 23~mJy~beam$^{-1}$  in channels of 12.8~km~s$^{-1}$ width. 
Images of the continuum emission were obtained by averaging in each of the two nearby sub-bands at 
686.899~GHz and 690.884~GHz those channels free of line emission. The resulting maps were combined
 to produce an image  of the continuum emission of the galaxy at 689~GHz. The corresponding point source sensitivity for the
continuum is 1.95~mJy~beam$^{-1}$.

 As our observations do not include zero-spacings, we expect to filter a non-negligible amount of flux on scales $\geq3\arcsec$ for continuum emission at 689~GHz. The total integrated flux measured inside the 9$\arcsec$ field-of-view amounts to $\sim 815$~mJy. A comparison 
 with the fluxes measured with the JCMT by Papadopoulos \& Seaquist~(\cite{Pap99}), and with Herschel by Hailey-Dunsheath et al.~(\cite{Hai12}), on apertures of $\sim10\arcsec$ indicates that we are filtering at most $\simeq30-45\%$ of the total flux on these scales.  However, fluxes estimated on small 
 ($\sim 1- 3\arcsec$) apertures centered on the brightest spots of emission of the CND are not expected to be significantly affected. As it happens in Band~7 observations,  a similar estimate would also suggest that significantly lower missing fluxes are likely to be filtered out % in the 689~GHz-continuum of the CND and 
 in the CO(6--5)  maps, due to the high clumpiness and velocity structure of the emission.

The phase tracking center of the central field and the reference velocity
are the same as the ones assumed in Band~7 (see Sect.~\ref{B7-obs} and Fig.~\ref{dust-maps}).

%\subsection{Ancillary data}
 \subsection{Ancillary data}
 
We retrieved the {\it HST} NICMOS (NIC3) narrow-band (F187N, F190N) images of NGC~1068 from
the Hubble Legacy Archive (HLA). These images were completely reprocessed by
the HLA with a fully revamped calibration pipeline which includes
temperature-dependent dark frames, improved temperature measurements from bias
levels, and other routines to reduce the impact of image artifacts and improve
the calibrated data.  The pixel size of the HLA images is 0\farcs10 square.
The redshift of NGC\,1068 is sufficiently small that Pa$\alpha$ falls within
the F187N filter.  We checked the alignment and background subtraction, then
scaled the F190N and subtracted it from the F187N image.  Although the two
images are very close together in wavelength, their ratio is not exactly unity.
Following Thompson et al.(\cite{Tho01}), we determined the scaling factor by
examining the ratio of the two images at the nucleus, where the F190N emission
peaks.  Because the Pa$\alpha$ line is not centered at the filter central
wavelength, we corrected the flux and converted the data numbers in the image
to flux densities using the IRAF task {\tt synphot}.

 We also use the hard X-ray image obtained in the 6--8 keV band by Chandra.
  The X-ray map was obtained by applying an adaptive smoothing
on the Chandra archive data available in this
wave-band for NGC~1068, following the same procedure described by Usero et al.~(\cite{Use04}). This image is identical to the hard X-ray
map published by Ogle et al.~(\cite{Ogl03}) (see also Young et al.~\cite{You01}).

 \section{Continuum emission}\label{obs-cont}
 \subsection{Observations at 349~GHz} \label{B7}
   Figure~\ref{dust-maps}a shows the continuum emission map obtained in Band~7 (at 349~GHz). The emission detected inside the inner $r\sim25\arcsec$(1.7~kpc) region imaged by ALMA, with a total integrated flux $\sim355$~mJy, is distributed in three distinct regions:

  {\it 1.~The CND:} The brightest emission stems from this central component, which appears as an asymmetric elongated ring of  $4\arcsec\times2\farcs8$--(non-deprojected) size as shown in Fig.~\ref{dust-maps}b. The intrinsic (deprojected) shape of the ring ($5\farcs3\times2\farcs8$~$\sim$~$370$~pc$\times~200$~pc), derived assuming the disk geometry fit of Sect.~\ref{kinemetry}   ($PA=289\pm5^\circ$ and $i=41\pm2^\circ$), 
  reinforces the idea that the ring is of elliptical shape. The ring, which shows a rich substructure, with two knots of emission located east and west of the nucleus, is remarkably off-centered relative to the 
  location of the AGN. The latter is identified as the strongest emission peak in the continuum  map at [$\Delta\alpha$, $\Delta\delta$]~$\approx$~[--0.9$\arcsec$,~--0.1$\arcsec$]~=~[$\alpha_{2000}=02^{h}42^{m}40.71^{s}$, $\delta_{2000}=-00^{\circ}00^{\prime}47.94\arcsec$]. This maximum coincides within the errors with the position of the AGN core ($S1$) determined from VLBI radiocontinuum maps of the galaxy 
  (Gallimore et al.~\cite{Gal96, Gal04}). The striking off-centered ring morphology of the CND  is echoed by the molecular line emission maps discussed in Sect.~\ref{line}. This particular geometry questions  the 
  simplest scenario that interprets the gas ring as the signature of the ILR region located at $r\leq5\arcsec$ (Schinnerer et al.~\cite{Sch00}; Garc\'{\i}a-Burillo et al.~\cite{Gar10}).
  
 {\it 2.~The bar:}  Inside the domain occupied by the stellar bar, which extends along $PA\sim46\pm2^{\circ}$ up to a corotation radius 
 $R_{\rm COR}\sim18 \pm 2\arcsec$ ($\sim1.3$~kpc) 
  (Scoville et al.~\cite{Sco88}; Bland-Hawthorn et al.~\cite{Bla97}; Schinnerer et al.~\cite{Sch00}; Emsellem et al.~\cite{Ems06}), we identified an arc of dust emission on the northeast side of the disk at distances $\sim4-7\arcsec$ from the AGN (deprojected radii $r\sim350-650$~pc). The source, hereafter referred to as the {\em bow-shock arc}, has a V-shaped morphology on its far side. This feature coincides in 
  position with the northern edge of the AGN nebulosity  identified in ionized gas emission, and with the northeast radio lobe. Wilson \& Ulvestad~(\cite{Wil87}) interpreted the limb-brightened shape and the high polarization of the northeast radio lobe as the signature of a bow-shock in the galaxy disk. Leaving aside some isolated clumps, no 
  significant emission is detected anywhere else inside the bar region.  
  
 {\it 3.~The SB ring:}  Most of the emission in the disk ($\geq72\%$ of the flux integrated within the ALMA mosaic) is detected from a two-arm spiral structure that starts from the ends of the stellar bar and unfolds in 
  the disk over $\sim180^{\circ}$ in azimuth forming a pseudo-ring at $r\sim18\arcsec$($\sim1.3$~kpc), i.e., the position of the ILR resonance of the outer oval (Schinnerer et al.~\cite{Sch00}).  Continuum 
  emission at 349~GHz over the SB ring is highly clumpy.

 %                                     Two column figure (place early!)
%______________________________________________ Fig2: continuum-ratio
 \begin{figure}[t!]
  \centering  
     \includegraphics[width=8cm]{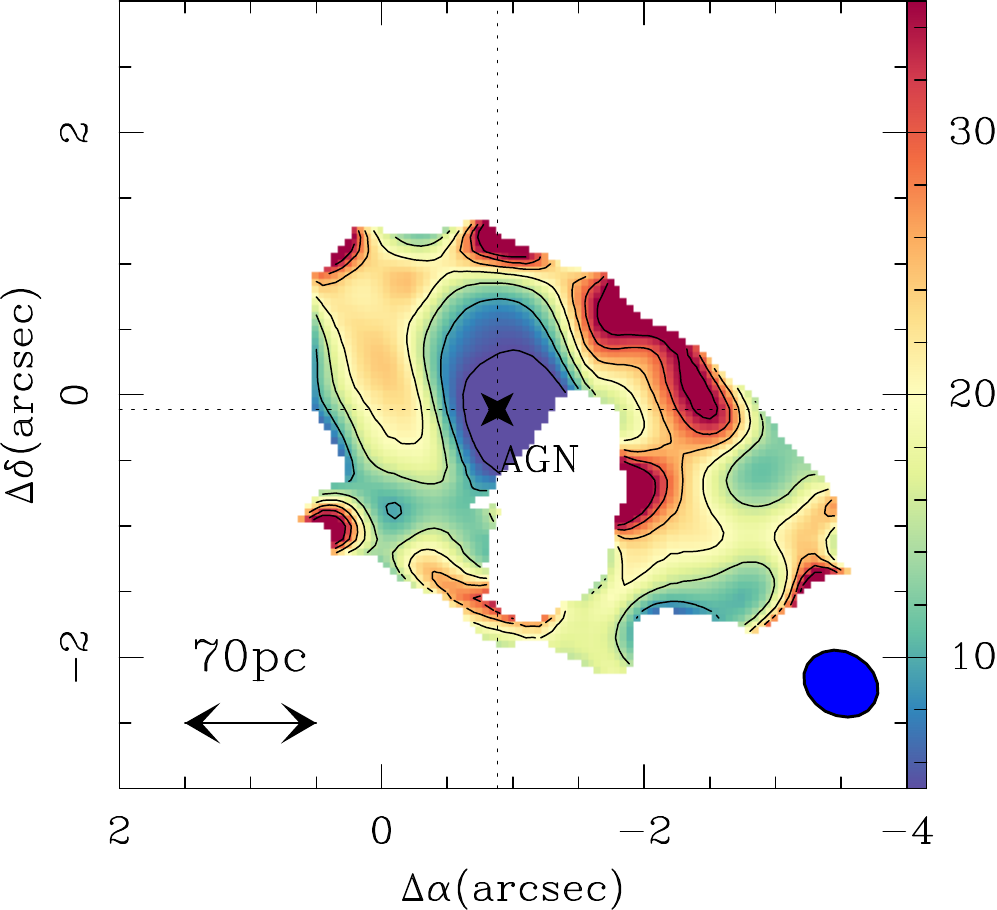}    
     \caption{The 689~GHz--to--349~GHz continuum flux ratio in the CND at the spatial resolution of observations in Band~7 ($0\farcs6\times0\farcs5$ at  $PA=60^{\circ}$, shown by the filled ellipse in the lower right corner). Levels go from 5 to 35 in steps of 5.}
              \label{continuum-ratio}
\end{figure}
 
 %______________________________________________ 

    %                                     Two column figure (place early!)
   %______________________________________________ Fig3: torus-SED
 \begin{figure}[h!]
 \centering
 \hskip 0.5cm
  \includegraphics[width=5cm, angle=-90]{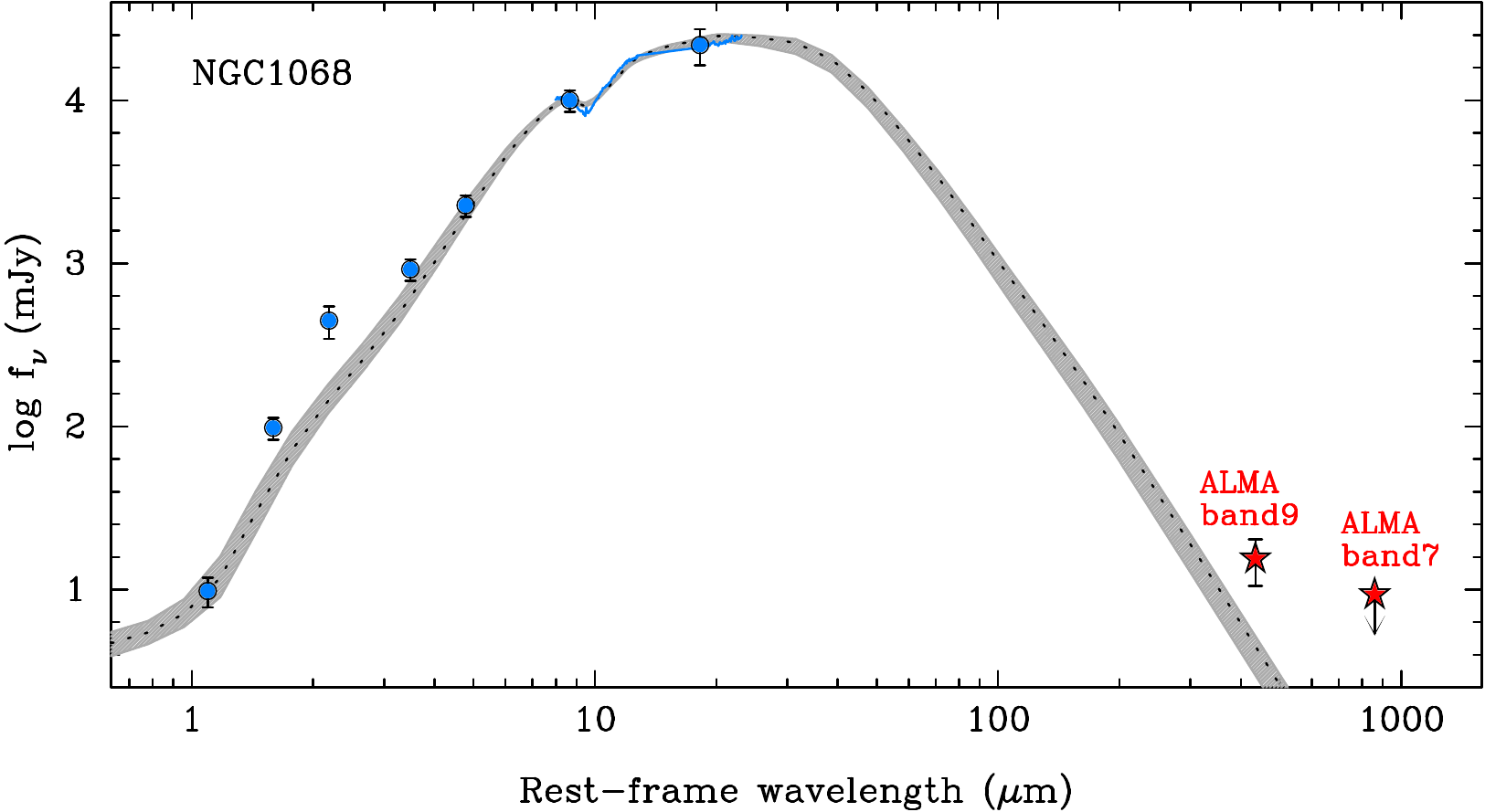}
  \vskip 0.3cm  
     \includegraphics[width=8cm]{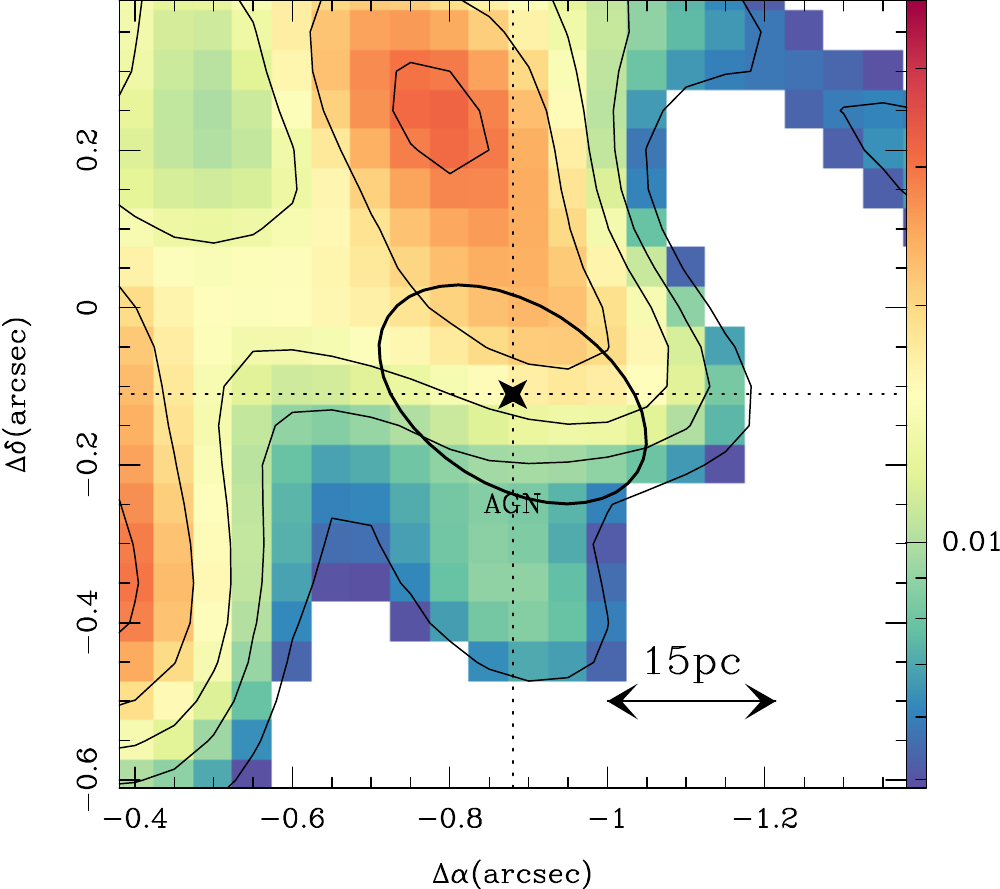}
                          \caption{{\bf a)}~({\it Upper panel})~The nuclear SED of the dust continuum emission in NGC~1068 derived  using NIR and MIR continuum and spectroscopy data (blue squares) from Alonso-Herrero et al.~(\cite{Alo11}), and the ALMA data points in Bands 7 and 9 (red stars). The SED was derived in apertures centered at the AGN that range from $0\farcs2$~($\sim$14~pc) for NIR data to $0\farcs5$~($\sim$35~pc) for MIR and Band-7 ALMA data.  The best CLUMPY model fit to the observations (curve) and the 1$\sigma$ uncertainty range of the fit (gray shaded region) are  superposed onto the data points. {\bf b)}~({\it Lower panel})~A close-up view of the dust continuum emission in Band~9. Levels and markers are as in Fig.~\ref{dust-maps}. The  ($0\farcs4\times0\farcs2$)--aperture
        used to extract the flux at the position of the AGN is plotted as an ellipse.}
              \label{torus}
\end{figure}
 
 %______________________________________________ 

\subsection{Observations at 689~GHz} \label{B9}
   Figure~\ref{dust-maps}c shows the continuum emission map obtained in Band~9 (at 689~GHz) in the inner $\sim9\arcsec$ field-of-view imaged by ALMA, which covers completely the CND region\footnote{The {\em bow-shock arc} region, not shown in Fig.~\ref{dust-maps}c, lies beyond the half power of the primary beam of ALMA at 689~GHz.}. Similar to the picture drawn from observations in Band~7, the CND emission in Band~9 stems from a clumpy ring that is off-centered relative to the AGN. The prominent east and west knots identified in Band~7 are spatially resolved into several clumps, due to the higher spatial resolution at 689~GHz. Furthermore, while
emission is  also detected at the AGN locus, like in Band~7 observations, the strongest peak coincides with the position of the east knot.  There is also a secondary peak of emission in Band~9 which is shifted to the northeast forming a spatially resolved  $\sim35$~pc--size elongated structure that connects the AGN with the CND along $PA\sim25^{\circ}$. This feature is reminiscent of the $\sim$north-south elongated source present in the mid-infrared (MIR) maps of the central $r\sim1-2\arcsec$ of the galaxy. This component has been attributed to emission of hot dust from narrow-line-region (NLR) clouds (Bock et al.~\cite{Boc98, Boc00}; Alloin et al.~\cite{All00}; Tomono et al.~\cite{Tom01,Tom06}; Galliano et al.~\cite{Gal05}; L\'opez-Gonzaga et al.~\cite{Lop14}). Gallimore et al.~(\cite{Gal01}) mapped the subarcsecond radio jet and found evidence of a jet-cloud  interaction around $0\farcs30$ (20~pc) north  and $PA\sim20^{\circ}$ relative to the AGN locus, based on the detection of strong  H$_2$O and OH maser emission and the bending of the radio jet at this location. This is very close to the position of a dust cloud identified in our maps, as shown in Fig.~\ref{torus}.

\section{Dust masses} \label{dustmass}

We combined observations of continuum emission in Bands 7 and 9 of ALMA with observations done in two PACS bands at 70~$\mu$m and 160~$\mu$m by Hailey-Dunsheath et al.~(\cite{Hai12}), and with high-resolution observations done at near- and mid-infrared  (NIR and MIR) wavelengths by Alonso-Herrero et al.~(\cite{Alo11}) to construct the spectral energy distributions (SED) and derive the mass of the dust in two different environments of NGC~1068: the $r\leq20$~pc region centered at the AGN (Sect.~\ref{dust-torus}), and the central $r\leq400$~pc of the galaxy, a region that includes the CND and the {\em bow-shock arc} (Sect.~\ref{dust-CND}). 
%We also used  the PdBI CO(1--0) map of NGC~1068 of Schinnerer et al.~(\cite{Sch00}) to calibrate the $N(H_2)$--to--CO conversion factor in these regions.
 
\subsection{The central $r\leq20$~pc} \label{dust-torus}

\subsubsection{Thermal versus non-thermal emission}\label{cont-ratio}
  
  Figure~\ref{continuum-ratio} shows the map of the 689~GHz--to--349~GHz continuum flux ratio (hereafter, $B9/B7$) in the CND derived at the (lower) spatial resolution of observations in 
Band~7, after convolution of the 689~GHz continuum image with a Gaussian beam with FWHM~$\sim0.4\arcsec$. The $B9/B7$ ratio is seen to change from $\simeq5$ close to the AGN ($r\leq0\farcs5$) up to a maximum of $\simeq20-25$ in a ring-like region of 
radius $r\simeq1.5\arcsec$. The highest ratios can be explained if we assume a steep dependence of the dust emissivity  $\kappa_{\rm \nu}$, which scales as $\sim\rm \nu^{\rm \beta}$, on frequency ($\beta\geq2.5$), 
 and attribute high temperatures to dust  $T_{\rm dust}\geq100$~K in the ring. On the other hand, the low value of $B9/B7$ close
to the AGN is indicative of a non-negligible contribution of non-thermal emission, which can be particularly relevant at 349~GHz. 
H{\"o}nig et al.~(\cite{Hoe08}) and Krips et al.~(\cite{Kri11}) discussed the relevance of three different mechanisms of non-thermal emission at the vicinity of the AGN: electron-scattered synchrotron emission, pure synchrotron emission, and thermal free-free emission. The inclusion of the two ALMA band  fluxes (integrated in 1$\arcsec$-apertures) into 
the AGN SED confirms that pure synchrotron emission is a poor representation of the SED in the submillimeter range, as also argued by Krips et al.~(\cite{Kri11}): this scenario 
over-predicts by 50$\%$ the total flux measured in Band~7, i.e., well beyond the associated uncertainties ($<10\%$).  
On the contrary the two alternative models discussed by Krips et al.~(\cite{Kri11}) account equally well for the AGN SED. 
In either case, the non-thermal contamination can at best be estimated to range  between 30 and 65$\%$ in Band~7. However, this percentage is much lower at the higher frequencies of 
Band~9 ($\leq5-18\%$).

\subsubsection{CLUMPY torus models: NIR/MIR and ALMA observations}\label{CLUMPY-models}

We used the ALMA Band 9 and Band 7 continuum thermal fluxes at
$435\,\mu$m  and $860\,\mu$m, respectively,  to investigate whether we
can set further constraints on the properties of the putative torus of NGC~1068 using {\sc CLUMPY}
torus models (Nenkova et al.~\cite{Nen08a, Nen08b}). To do
so we took into account the NGC~1068 SED 
and MIR spectroscopy presented in Alonso-Herrero et
al.~(\cite{Alo11}). We added the ALMA Band 9 measurement at the full
resolution of $0\farcs3$, which is similar to
that of the MIR data. 
The ALMA Band 7
continuum measurement  has a slightly worse resolution ($0\farcs5$)
and a higher contribution from non-thermal emission as discussed in Sect.~\ref{cont-ratio},
and therefore we used the estimated thermal component as an upper
limit. For the ALMA Band 9 flux uncertainties we included both the
error in the photometric calibration ($\sim 30\%$) and the
uncertainties associated with the modeling of the non-thermal
component at this wavelength estimated above ($\sim 10\%$). 
The details of the fitting procedure are explained in Appendix~A.

%As in Alonso-Herrero et
%al.~(\cite{Alo11}) we used the  detection of a maser in the nuclear region of
%NGC~1068, which implies 
%a close to edge-on view of the AGN accretion disk, to set the following prior
%for the viewing angle $i=60-90^{\circ}$. 

%Because we are including the ALMA
%far-infrared photometry we allowed the full range for the torus size
%$Y=5-100$. We used the new version of the BayesClumpy tool (Asensio
%Ramos \& Ramos Almeida~\cite{Ase09}) which now interpolates linearly between
%the {\sc CLUMPY} models. 

 %                                     Two column figure (place early!)
%______________________________________________ Fig4: CO-maps
 \begin{figure*}[tbh!]
  \centering  
   \includegraphics[width=18cm]{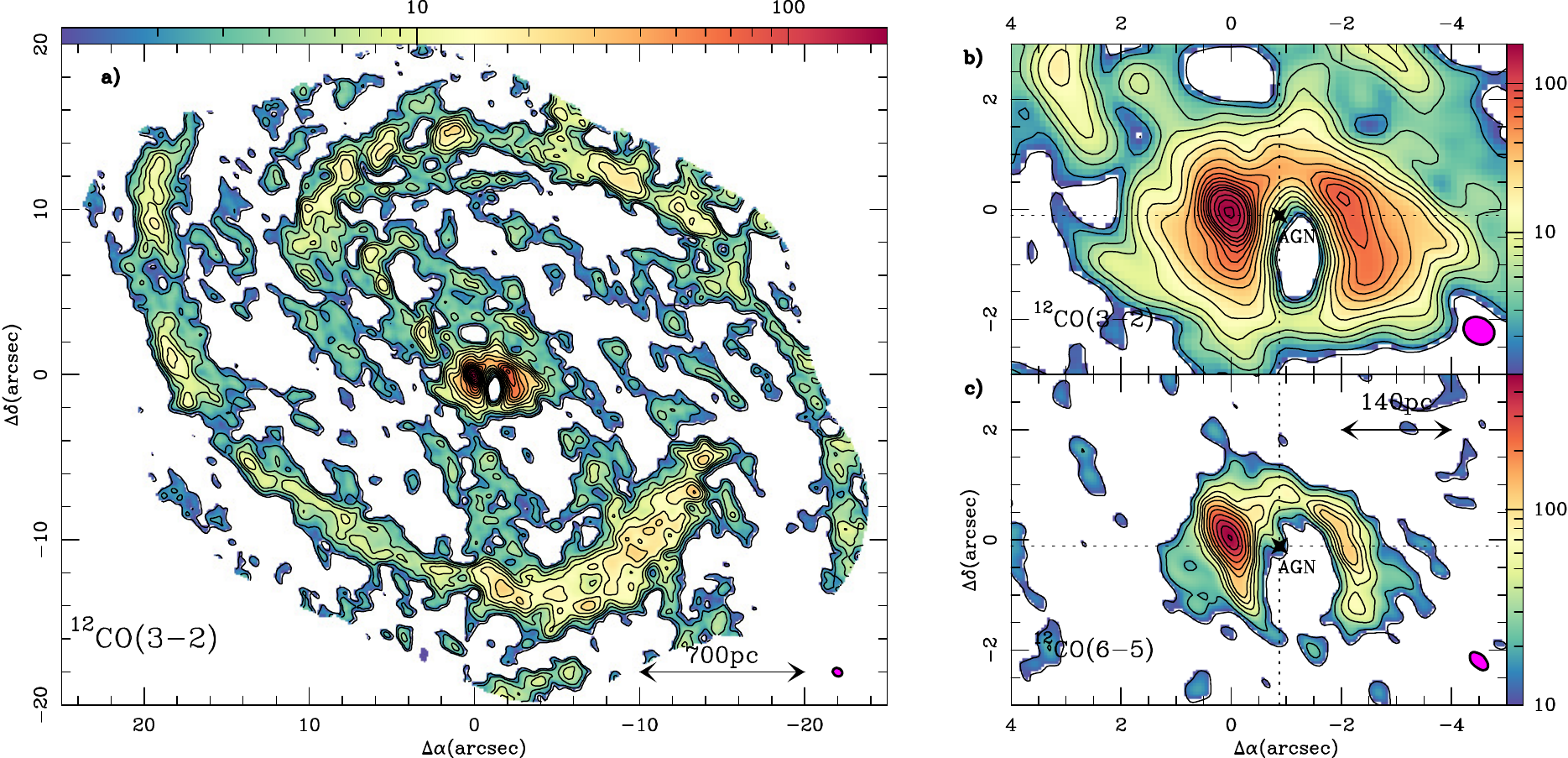}
                 \caption{{\bf a)}~({\it Left panel})~The CO(3--2) integrated intensity map obtained with ALMA using an eleven-field mosaic
in the disk of NGC~1068. The map is shown in color scale with contour
levels  5$\sigma$, 10$\sigma$, 20$\sigma$, 30$\sigma$, 45$\sigma$, 70$\sigma$, 100$\sigma$  to  500$\sigma$ in steps of  50$\sigma$,
and  600$\sigma$ to 800$\sigma$ in steps of 100$\sigma$, where 1$\sigma=0.22$~Jy~km~s$^{-1}$beam$^{-1}$. The filled
ellipse at the bottom right corner represents the CO(3-2) beam size ($0\farcs6\times0\farcs5$ at  $PA=60^{\circ}$).  {\bf b)}~({\it Upper right panel})~Same as {\bf a)} but zooming in on the circumnuclear disk (CND) region.  {\bf c)}~({\it Lower right panel})~Same as {\bf b)} but for the CO(6--5) line, obtained with a single field mosaic.  Contour
levels are: 5$\sigma$, 10$\sigma$, 20$\sigma$, 30$\sigma$ , 40$\sigma$,  70$\sigma$,  90$\sigma$, 120$\sigma$  to  240$\sigma$ in steps of  40$\sigma$
, where 1$\sigma=2$~Jy~km~s$^{-1}$beam$^{-1}$. The filled
ellipse at the bottom right corner represents the CO(6-5) beam size ($0\farcs4\times0\farcs2$ at  $PA=50^{\circ}$). }
              \label{CO-maps}
\end{figure*}
 
 %______________________________________________ 

The fitted torus parameters are all similar to those of
Alonso-Herrero et al.~(\cite{Alo11}) except for the torus size now fitted to
a factor $\sim10$ larger value, which corresponds to a torus outer 
radius of $20^{+6}_{-10}\,$pc, derived using the AGN bolometric luminosity of $\sim4.2^{+1.4}_{-1.1}\times10^{44}$~erg~s$^{-1}$ estimated
from scaling the best fit model to the data. The torus size is not well
constrained because the large value of the fitted index of the radial
distribution of the clouds, which implies that most of
the clouds are 
located close to the inner radius of the torus. 
%%%%%%%%(see Ramos Almeida et
%%%%%%%%al. 2013 for a discussion on this issue). 
For comparison the
modeled $12\,\mu$m interferometric half radius of the resolved and
unresolved components of NGC~1068 is 1.6\,pc (Burtscher et
al.~\cite{Bur13}). These differences might be explained because the NIR
and MIR emission is probing warm dust that is on average closer to the
AGN, whereas in the submillimeter range we are more sensitive to cold dust
distributed over larger distances from the AGN. 

The best CLUMPY model fit and the 1$\sigma$ uncertainty to the 
nuclear emission of NGC~1068 is presented in Fig.~\ref{torus}{\it a}. It is clear
from this figure that the measured continuum ALMA fluxes are above the
{\sc CLUMPY} torus fit. This could be explained if cold dust not
associated with the torus was included in the ALMA Band 7 and 9 flux
measurements. Figure~\ref{torus}{\it b} shows a close-up of the
nuclear region of NGC~1068 at $435\,\mu$m at full resolution. Clearly
there is much cold dust in this region including the ionization
cone, as discussed in Sect.~\ref{B9}, but more importantly there is no
point source associated with the position of the AGN. This seems to
confirm our suspicion that even at the $0\farcs4 \times
0\farcs2$ angular resolution of the ALMA Band 9 observations, there can be a contribution of cold dust not associated with
the clumpy torus. We note that the fitted outer torus radius would correspond
to an angular diameter of $0.57^{+0.17}_{-0.43}$$\arcsec$. The lower limit
is consistent with not having resolved the torus in  Band 9 with the
current angular resolution. Incidentally, the NIR data points are
also above the {\sc CLUMPY} torus model fit. The extra
flux at these wavelengths might come from the dust at the
base of the ionization cone, as can be the case with other Seyfert 
galaxies (see, e.g., Hoenig et al.~\cite{Hoe13}).

We can finally estimate the gas mass in the torus of NGC~1068 based on the fit, according to Eq.\ref{Mtor} of Appendix~A. 
%In the models of Nenkova et al.~(\cite{Nen08b}) the total mass in torus clouds can
%be written as $M_{\rm torus}=4\pi m_{\rm H} (\sin \sigma)N_{\rm
%  torus}^{\rm eq}R_{\rm sub}^2 Y I_q(Y)$,
%where the function $I_q(Y)=1$ for $q=2$ and $N_{\rm torus}^{\rm eq}$
%is the column density along the equatorial direction. The inner radius of 
%the torus $R_{\rm sub}$ was computed from the AGN bolometric luminosity
%inferred from the fit and we assumed a gas-to-$A_{\rm V}$ ratio of $N_{\rm H}/A_{\rm V} =
%1.79\times10^{21}\,{\rm cm}^{-2}\,{\rm mag}^{-1}$.  
We obtained: $M_{\rm 
torus}=2.1 (\pm 1.2)\times 10^5\, M_\odot$, where we considered that
the main uncertainties come from the relatively unconstrained torus
size and from the scatter  around the adopted  $N_{\rm H_2}/A_{\rm V}$  scaling ratio taken from Bohlin et al.~(\cite{Boh78}) (see Appendix~A). This mass estimate is comfortingly similar to the estimated molecular gas mass 
detected inside the central $r=20$~pc aperture derived from the CO(3--2) emission, as discussed in 
Sect.~\ref{XCO}.

\subsection{The central $r\leq400$~pc region: the CND and the bow-shock arc}\label{dust-CND}

 ALMA observations in Bands 7 and 9 were  combined with PACS observations obtained at 70~$\mu$m and 160~$\mu$m in comparable apertures by Hailey-Dunsheath et al.~(\cite{Hai12}) to constrain the overall SED of the dust emission in the central $r=400$~pc of the galaxy. The estimate of $M_{\rm gas}$ based on this fit and the implied conversion factor for CO ($X_{\rm CO}$), discussed in Sect.~\ref{XCO}, are used  in Sect.~\ref{outflow} to derive the mass load of the outflow identified in this region.

 The SED was fit using  a modified black-body  (MBB) model to derive the dust temperature 
 ($T_{\rm dust}$), dust mass ($M_{\rm dust}$) and emissivity index ($\beta$) in this region. In this approach, the measured fluxes, $S_{\rm \nu}$, can be expressed as $S_{\rm \nu}=M_{\rm dust}\times\kappa_{\rm \nu}\times B_{\rm \nu}$($T_{\rm dust}$)/$D^{2}$, where the emissivity of dust, 
 $\kappa_{\rm \nu}$, scales as $\sim\kappa_{{\rm 352~GHz}}\times(\rm \nu$[GHz]$/{\rm 352})^{\rm \beta}$, with $\kappa_{{\rm 352~GHz}}=0.09$~m$^2$~kg$^{-1}$ (a value rounded up from $\kappa_{{\rm 352~GHz}}=0.0865$~m$^2$~kg$^{-1}$ used by Klaas et al.~\cite{Kla01}), $B_{\rm \nu}$($T_{\rm dust}$) is the Planck function, and $D$ is the distance. Prior to the fit, the fluxes in the two ALMA bands were corrected for the contamination by non-thermal emission in the central $1\arcsec$ aperture at the AGN, as estimated in Sect.~\ref{cont-ratio}.  The best MBB fit is found for  $M_{\rm dust}=(8\pm2)\times10^{5}$~M$_{\sun}$, $T_{\rm dust}=46\pm3$~K and $\beta=1.7\pm0.2$. The errorbars on the parameters of the fit reflect the uncertainties due to the estimated range of missing flux in the ALMA bands, derived in Sects.~\ref{B7} and \ref{B9}.

 The value of $M_{\rm dust}$ can be used to predict the associated  neutral gas mass budget in this region.  Applying the linear dust/gas scaling ratio of Draine et al.~(\cite{Dra07}) (see also Sandstrom et al.~\cite{San13}) to the gas-phase oxygen abundances measured in the central 2~kpc of NGC~1068 ($\sim 12 + $log~(O/H)$\sim8.8$; Pilyugin et al.~\cite{Pil04, Pil07}), which yields a gas--to--dust mass ratio of $\sim60^{+30}_{-30}$, we estimate that the total neutral gas mass in the central 10$\arcsec$ aperture is 
$M_{\rm gas}=(5\pm3)\times10^{7}$\,M$_{\sun}$. This number is a good proxy for the molecular gas mass in this region because the HI distribution in the disk of NGC~1068, studied by Brinks et al.~(\cite{Bri97}), shows a bright ring between 30$\arcsec$ and 80$\arcsec$ with a true central hole.

%$M_{\rm gas}$=1.07$\times$10$^4$$\times$($D$/Mpc)$^2$($S_{\rm CO(1-0)}$/(Jy~km~s^${-1}$]) ($X_{CO}$ / [2x10$^{20}$cm$^{-2}$(K~km~s$^{-1}$pc$^{2}$)$^{-1}$])

%______________________________________________ Fig5: CO-bar
 \begin{figure}[th!]
  \centering  
   \includegraphics[width=8.5cm]{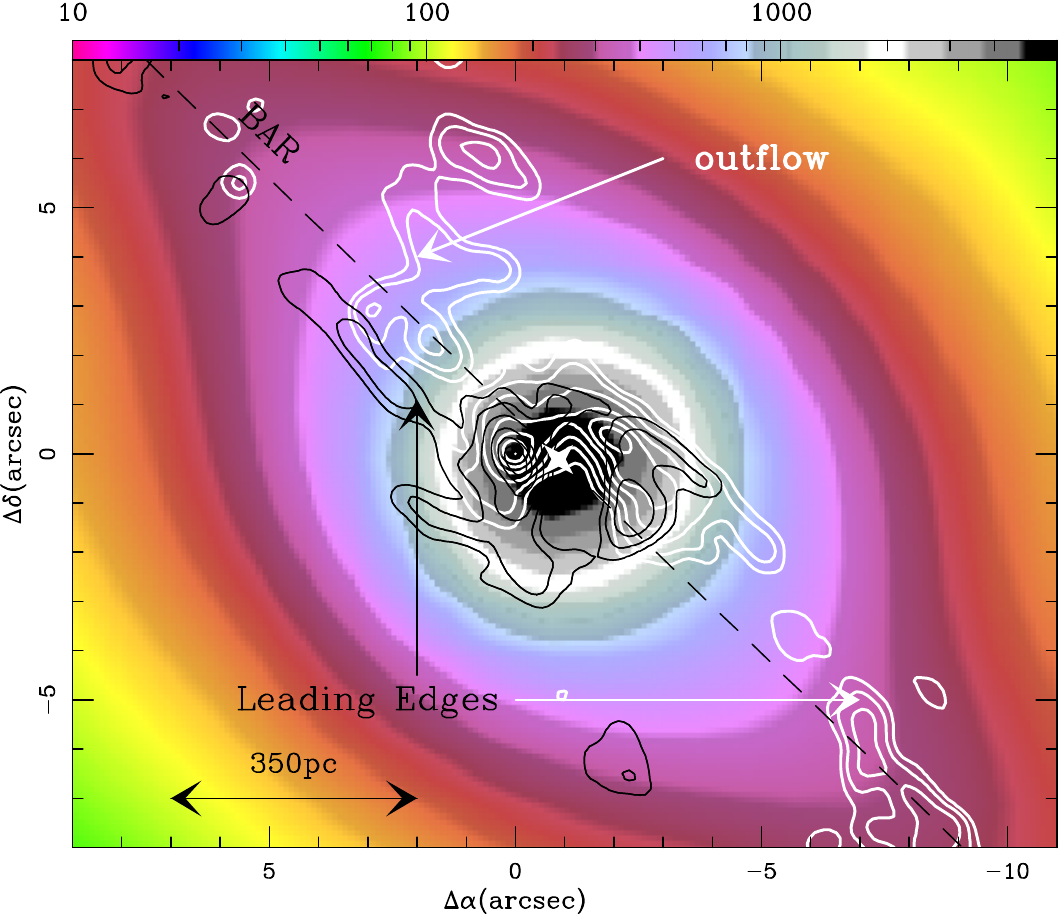}
            \caption{%{\bf a)}~({\it Left panel})~Overlay of the CO(3--2) intensity contours (from Fig.~\ref{CO-maps}{\it a)} on  the NIR K-band image (in color scale) of NGC~1068 obtained by the Two Micron All Sky Survey (2MASS).  {\bf b)}~({\it Right panel})~
       Zoom in of the inner disk of NGC~1068 to show the overlay of two CO(3--2) velocity channel maps  obtained for
  $v-v_{\rm sys}=-50$ (black contours: 5, 10, 20 to 100$\%$ in steps of 20$\%$ of the peak value$=1$~Jy) and 50~km~s$^{-1}$ (white contours: 5, 10, 20 to 100$\%$ in steps of 20$\%$ of the peak value$=0.5$~Jy), with $v_{\rm sys}$(HEL)$~=1127\pm$3~km~s$^{-1}$, on the NIR K-band image (in color scale and arbitrary units) of NGC~1068 obtained by the Two Micron All Sky Survey (2MASS). We highlight the location of gas emission along the bar's leading edges as well as the existence of an anomalous component associated with the outflow. The orientation of the stellar bar's major axis along $PA=46^{\circ}$ is marked by a dashed line.
 }
              \label{CO-bar}
\end{figure}
 
 %______________________________________________ 

\section{Molecular line emission}\label{line}
 
 In this section we analyze our newly acquired ALMA maps and compare the morphology of the dense-gas tracers.
 
  \subsection{CO maps} \label{co-maps}
 
Figure~\ref{CO-maps} shows the CO(3--2) and CO(6--5) velocity-integrated intensity maps of NGC~1068 obtained with ALMA.  The CO maps reveal the distribution of the dense molecular gas in NGC~1068 with unprecedented high-dynamic range capabilities: $\geq600$ and $\geq200$ for the CO(3--2) and CO(6--5) maps, respectively.
As for the dust emission, we identify three main regions in the disk of NGC~1068:

{\it 1.~The CND:}  The brightest CO(3--2) line emission comes from the CND, which appears as a closed asymmetric elliptical ring of  $6\arcsec \times4\arcsec$--size as shown in Fig.~\ref{CO-maps}{\it b}. The substructure of the CO(3--2) ring reveals two strong emission peaks located $\sim 1\arcsec$ east and $\sim 1.5\arcsec$ west of the AGN. Similar to the dust continuum ring, the CO(3--2) ring is noticeably off-centered relative to the location of the AGN: the two emission knots are bridged by lower-level emission that completes the ring north and south of the AGN and leaves a gas-emptied region to the southwest.  However, unlike for Band~7 continuum emission, CO(3--2) emission does not peak at the AGN. The morphology of the ALMA map of the CND is to a large extent similar to that of the SMA map of Krips et al.~(\cite{Kri11}). Nevertheless, the order of magnitude higher dynamic range of the ALMA image of the CND helps reveal that the ring closes south of the AGN, i.e., similarly to the molecular ring imaged with the Very Large Telescope (VLT) in the 2.12~$\mu$m H$_2$ line by M\"uller-S{\'a}nchez et al.~(\cite{Mue09}). The brightest CO(3-2) emission features are also detected and spatially resolved in the CO(6--5) map of  Fig.~\ref{CO-maps}{\it c}.

{\it 2.~The bar:}  The CND appears connected to lower-level emission which extends farther out in the disk into the stellar bar region. On these scales the CO(3--2) emission is detected along two lanes, offset by $2-4\arcsec$, which run mostly parallel to the bar's major axis %as shown in Fig.~\ref{CO-bar}{\it a} 
($PA=46\pm2^{\circ}$; Scoville et al.~\cite{Sco88}; Bland-Hawthorn et al.~\cite{Bla97}; Schinnerer et al.~\cite{Sch00}; Emsellem et al.~\cite{Ems06}). These are reminiscent of the gas leading edge morphology that is expected to prevail between the corotation of the bar and the ILR region. This is illustrated in Fig.~\ref{CO-bar}, which shows the emission of CO(3--2) at two velocity channels symmetrically located relative to $v_{\rm sys}$: $v-v_{\rm sys}=-50$~km~s$^{-1}$  and +50~km~s$^{-1}$, with $v_{\rm sys}$(HEL)~$=1127\pm3$~km~s$^{-1}$ as determined in Sect.~\ref{non-circ}.
Gas emission along the bar's leading edges appears at the velocity range expected for gas falling to the nucleus: at redshifted velocities on the southwestern (near) side and blueshifted velocities on the northeastern (far) side. We nevertheless detect a gas component with anomalously redshifted velocities on the northeastern side of the disk at distances $\sim 4-7\arcsec$ from the AGN. This CO feature, closely associated with the {\em bow-shock arc} identified in dust continuum emission, is interpreted in Sect.~\ref{outflow} as the signature of a molecular outflow. 

 %we identified an arc of emission on the southwest side of the disk at distances $\sim$4--7$\arcsec$ from the AGN (deprojected radii $r\sim$350--650~pc). The source, hereafter referred to as the {\em bow-shock arc}, has a V-shaped morphology on its farthest side. This feature coincides in position with the northern edge of the AGN nebulosity  (identified in ionized gas emission), and with the northern tip of the radio jet, as illustrated in Fig.~\ref{velres-zoom}. Leaving aside some isolated clumps, no significant emission is detected anywhere else on these intermediate scales inside the area of influence of the bar.  
 
\begin{figure*}[tbh!]
   \centering  
   \includegraphics[width=18cm]{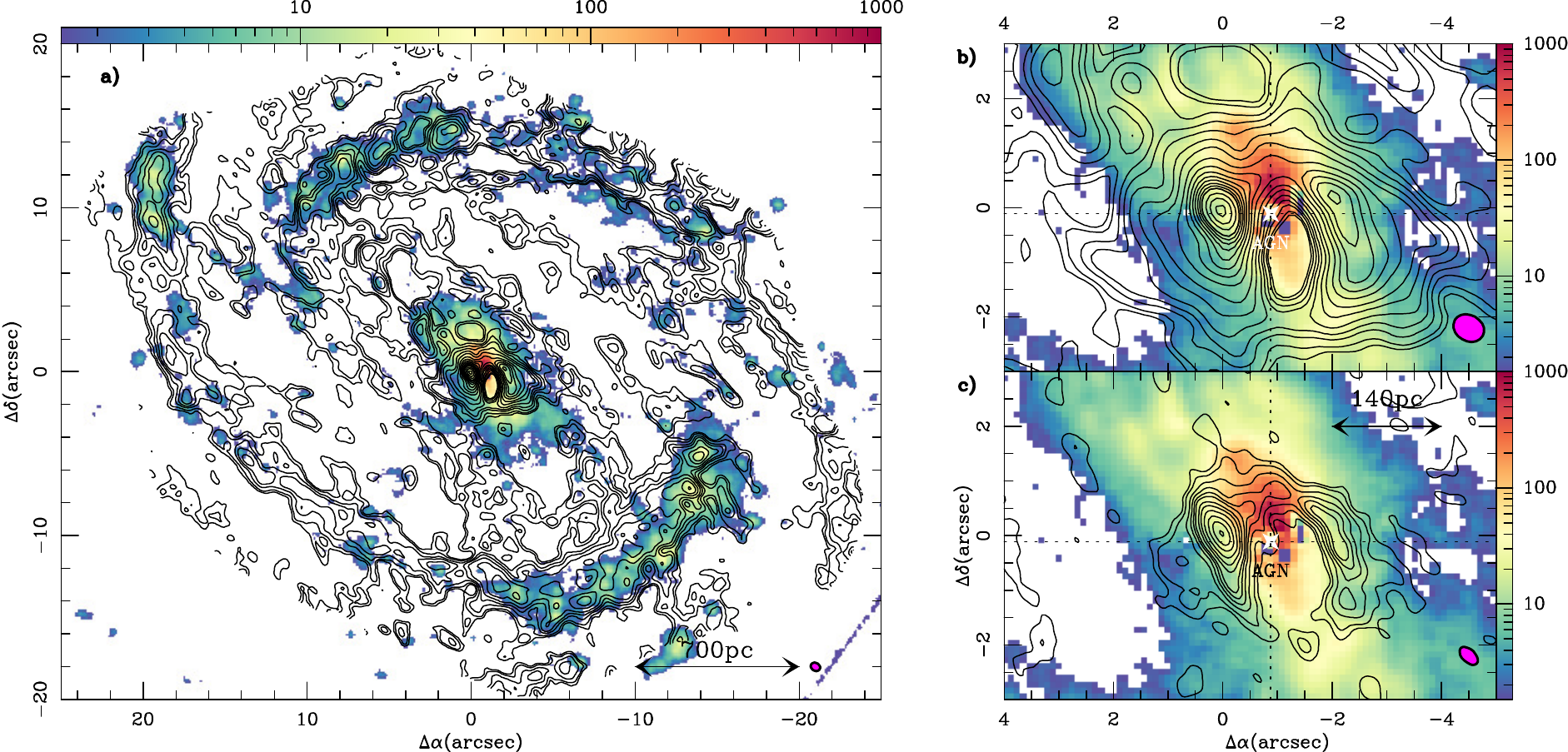}
               \caption{{\bf a)}~({\it Left panel})~Overlay of the CO(3--2) ALMA intensity contours (levels as in Fig.~\ref{CO-maps}{\it a}) on the Pa$\alpha$ emission HST map (color scale as shown in counts~s$^{-1}$pixel$^{-1}$). {\bf b)}~({\it Upper right panel})~Same as {\bf a)} but zooming in on the CND region. {\bf c)}~({\it Lower right panel})~Overlay of the CO(6--5) ALMA intensity contours (levels as in Fig.~\ref{CO-maps}{\it c}) on the HST Pa$\alpha$ emission  map (color scale as shown). The filled
ellipses at the bottom right corners represent the CO beam sizes.}
              \label{Paalpha-CO}
\end{figure*}

\begin{figure*}[tbh!]
   \centering  
   \includegraphics[width=18cm]{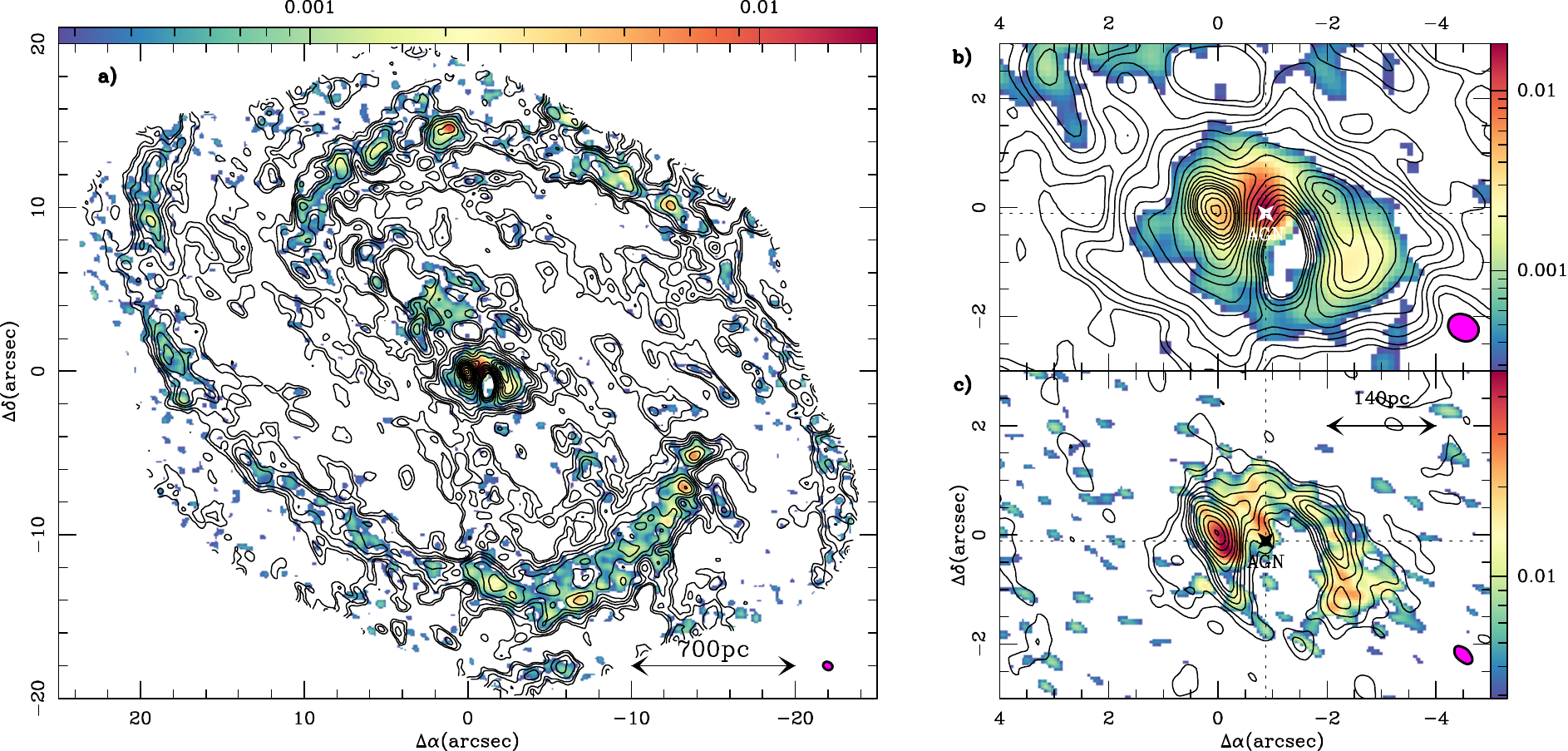}
                  \caption{{\bf a)}~({\it Left panel})~Overlay of the CO(3--2) intensity contours (levels as in Fig.~\ref{CO-maps}{\it a}) on the dust continuum emission at 349~GHz (color scale in Jy~beam$^{-1}$ units as indicated). {\bf b)}~({\it Upper right panel})~Same as {\bf a)} but zooming in on the CND region. {\bf c)}~({\it Lower right panel})~Overlay of the CO(6--5) intensity contours (levels as in Fig.~\ref{CO-maps}{\it c}) on the dust continuum emission at 689~GHz (color scale in Jy~beam$^{-1}$ units as indicated). The filled
ellipses at the bottom right corners represent the CO beam sizes.}
              \label{dust-CO}
\end{figure*}

 {\it 3.~The SB ring:}  Similar to the dust emission, most of the CO(3--2) flux in the disk mapped by ALMA ($\simeq 63\%$ of the grand total) is detected in the SB ring. The two spiral arms, which unfold over  $\sim 180^{\circ}$ in azimuth, are spatially resolved. The (transverse) widths of the arms go typically from 100~pc to 350~pc. Figure~\ref{Paalpha-CO} shows that the SB ring concentrates most of the ongoing massive star formation in the disk identified by strong Pa$\alpha$ emission. The emission of CO(3--2) over the SB ring is highly clumpy and it appears to be organized as coming from molecular cloud associations of $\geq50$~pc-size which are also identified in the dust continuum map, as shown in Fig.~\ref{dust-CO}. 
 
We also report the detection of CO(3--2) emission at different locations throughout the interarm region. These interarm complexes, which remained undetected in dust emission due to the lower dynamic range of the Band~7 continuum map, are organized into a network  of filaments that extends out to the edge of our mapped region.  

%______________________________________________ Fig6: pa-alpha-CO overlay
    
%
 %______________________________________________ Fig7: dust-CO overlay
    
%_______________

 %______________________________________________ Fig8: HCN-HCOp-CS-maps
 \begin{figure*}[tbh!]
%   \sidecaption
\centering
   \includegraphics[width=15cm]{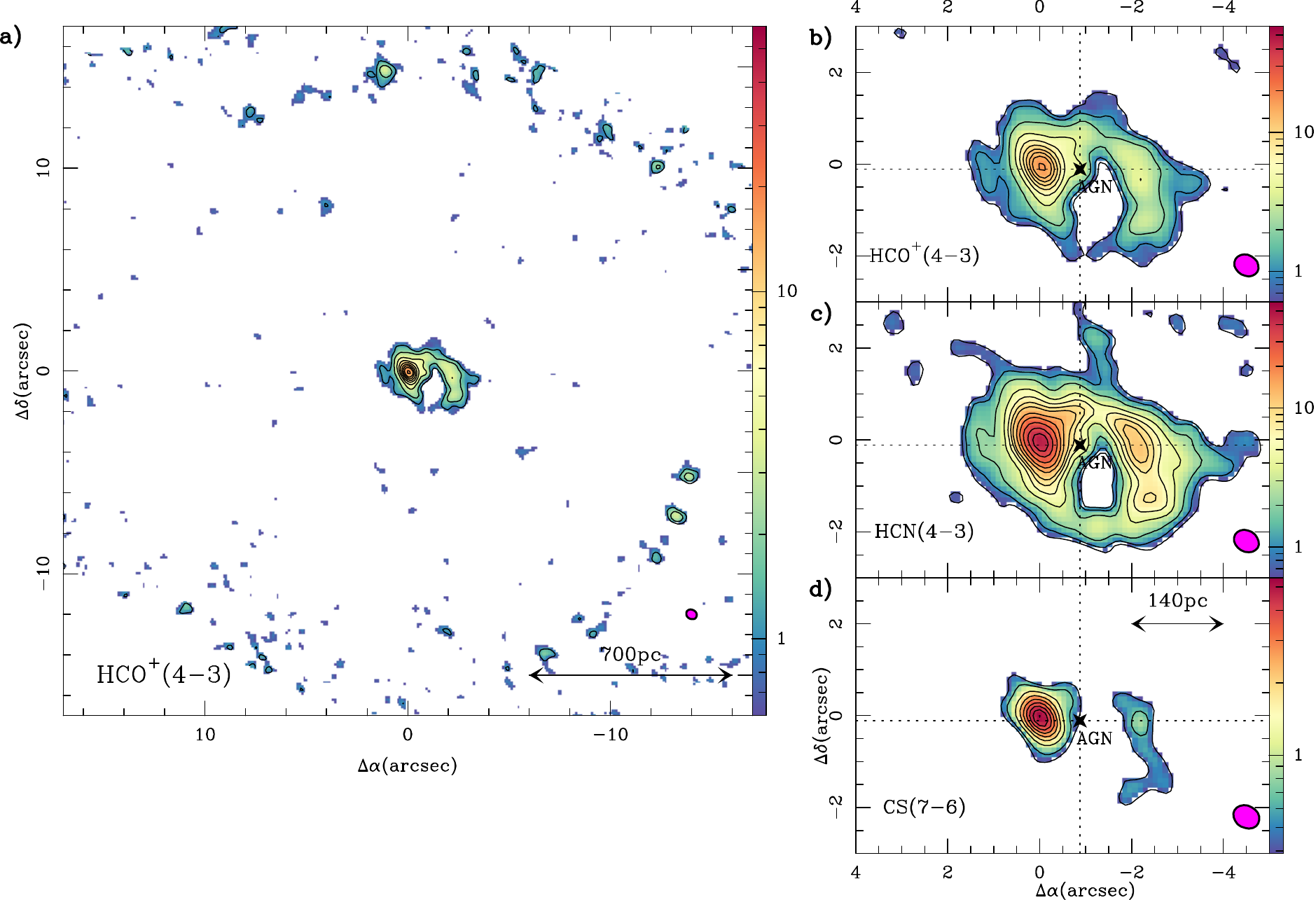}
               \caption{{\bf a)}~({\it Left panel})~The HCO$^+$(4--3) integrated intensity map obtained for NGC~1068. Contour levels are:   5$\sigma$, 10$\sigma$ to 70$\sigma$ in steps of 10$\sigma$, and 85$\sigma$. The filled ellipse shows the HCO$^+$ beam size,  
 similar to that of CO(3--2).  %{\bf b)}~({\it Lower left panel})~Same as {\bf a)} but for the HCN(4--3) line.  Contour levels
%are:   5$\sigma$, 10$\sigma$ to 40$\sigma$ in steps of 10$\sigma$, 50$\sigma$ to 90$\sigma$ in steps of 20$\sigma$, and 120$\sigma$ to 200$\sigma$ in steps of 40$\sigma$. 
Panels {\bf b)}~({\it upper right panel}) and {\bf c)}~({\it middle right 
panel}) show a zoomed-in view of the HCO$^+$ and HCN images, respectively, with 3$\sigma$ contours added to the list of displayed levels.   {\bf d)}~({\it Lower right panel})~ Same as {\bf b)} and {\bf c)} but for the CS(7--6) line. Contour 
levels are:  3$\sigma$, and 5$\sigma$ to 40$\sigma$ in steps of 5$\sigma$. The assumed value of 1$\sigma$ common for all lines is $\sim 0.20$~Jy~km~s$^{-1}$beam$^{-1}$. 
%Same centering as used in Fig.~\ref{CO-maps}~{\it a)}. The position of the AGN is highlighted.
}
              \label{dense-maps}
\end{figure*}

%      

 %______________________________________________ Fig9: AGN-spectra
 \begin{figure*}[tbh!]
 \centering
   \includegraphics[width=16cm]{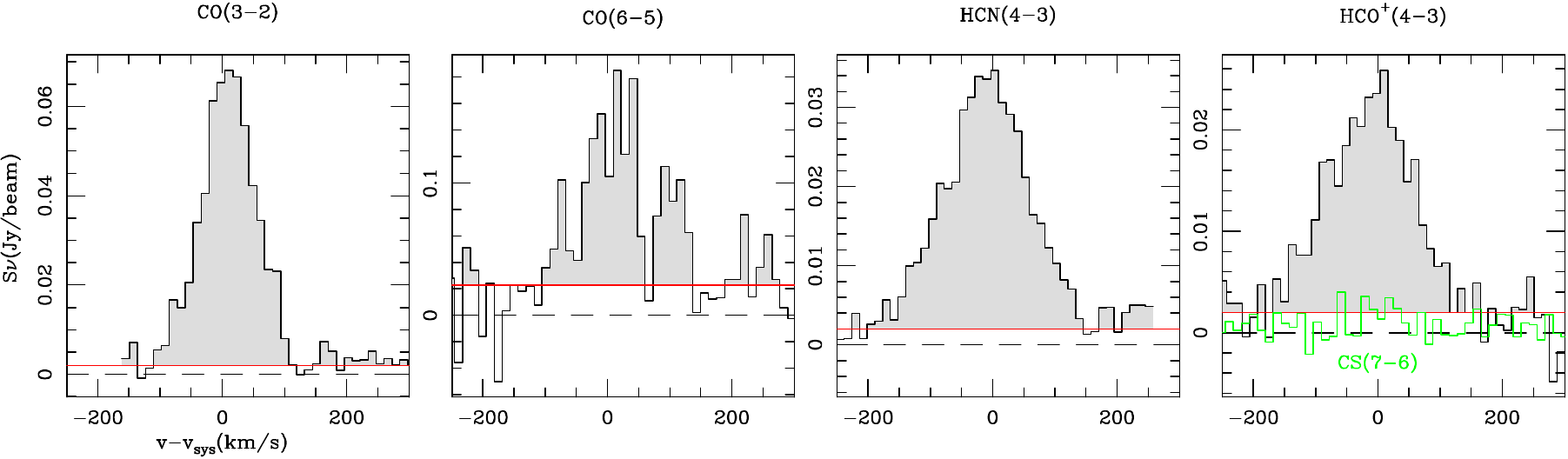}
                \caption{We show from left to right the CO(3--2), CO(6--5), HCN(4--3), and 
 HCO$^+$(4--3) emission line profiles towards the position of the AGN. The corresponding apertures are $0.6\arcsec \times 0.5\arcsec$ (40~pc) and $0.4\arcsec \times 0.2\arcsec$ (20~pc) for observations in Band~7 (CO(3--2), HCN(4--3) and 
 HCO$^+$(4--3)) and Band~9 (CO(6--5)), respectively. The red lines identify the 1$\sigma$ level for each transition in order to illustrate the reliability of detections. The undetected CS(7--6) line is included in the last panel (green histograms). Velocities refer to $v_{\rm sys}$(HEL)~$=1127\pm$3~km~s$^{-1}$.}
              \label{spectra}
\end{figure*}
 % _____________________

    \subsection{HCN, HCO$^+$, and CS maps}

Figure~\ref{dense-maps} shows the integrated intensity maps of NGC~1068 obtained with ALMA in the HCO$^+$(4--3), HCN(4--3), and CS(7--6) lines. In stark contrast with the CO(3--2) map, most of the emission in these {\em likely} denser gas tracers that are characterized by $\sim$a factor 100 comparatively higher critical densities, stems from the CND. Notwithstanding, a few ($\sim 12$) isolated clumps are detected in the  HCO$^+$(4--3) and HCN(4--3) lines at significant levels towards the SB ring (Figure~\ref{dense-maps}{\it a} shows the HCO$^+$ emission in the SB ring). The different distributions of the CO(3--2), HCN(4--3), and  HCO$^+$(4--3) line emission in the disk is reflected in the different line ratios measured in the CND and in the SB ring (see Sect.~\ref{lineratios} and paper~II). Other molecular tracers also show remarkably different distributions in the SB ring and the CND (e.g., Takano et al.~\cite{Tak14}).

The overall morphology of the CND in  HCO$^+$, and HCN is similar to that of the ALMA CO maps, i.e., an asymmetric closed molecular ring, which is noticeably off-centered relative to the AGN. However, the 
lower S/N ratio of the ALMA CS(7--6) map restricts the detected emission in this line mostly to the two prominent knots of the CND. The superior capabilities of ALMA dramatically improve the picture drawn for 
the distribution of the dense molecular gas traced by the HCN and HCO$^+$ lines in the CND compared to the previous SMA data of Krips et al.~(\cite{Kri11}): while most of the emission detected in the HCO$^+$(4--3) 
SMA map is restricted to the western knot, the ALMA map shows the emission to be widespread over the whole CND. The integrated flux of this line over the CND is a factor $\sim 1.7$ higher in the ALMA map 
($104\pm10$~Jy~km~s$^{-1}$) compared to the SMA map ($60\pm10$~Jy~km~s$^{-1}$).

\subsection{Molecular emission near the AGN}\label{line-AGN}

With the exception of the CS(7--6) line, emission is detected in all the molecular gas tracers probed by ALMA at the position of the AGN. Figure~\ref{spectra} shows  the CO(3--2), CO(6--5), HCN(4--3), and 
 HCO$^+$(4--3) emission line profiles towards the AGN. As expected, molecular emission is within the errors centered around  $v=v_{\rm sys}$(HEL)~$=1127\pm3$~km~s$^{-1}$ 
 (determined in this work: see Sect.~\ref{non-circ}) and  extends in contiguous channels across 200-300~km~s$^{-1}$ over significant levels ($>1\sigma$) in all tracers. Gaussian fits to the line profiles extracted from the same apertures ($\sim40$~pc) indicate that the higher density tracers 
 (HCN(4--3) and  HCO$^+$(4--3)) show significantly wider lines (FWHM~$\sim 180\pm10$~km~s$^{-1}$) compared to CO(3--2) 
 (FWHM~$\sim 106\pm3$~km~s$^{-1}$). This result indicates that the excitation of molecular gas, estimated from the HCN(4--3)/CO(3--2) and HCO$^+$(4--3)/CO(3--2) line ratios, is enhanced at the highest velocities, which in all likelihood correspond to gas lying closer to the central engine. This trend runs in parallel with the observed tendency to find higher velocity-integrated ratios at the smallest radii inside the CND (see discussion in Sect.~\ref{lineratios}). 
 
As shown in Fig.~\ref{spectra}  the half width of the CO(6--5) line derived using the velocity channels that show contiguous emission above a 1$\sigma$-level is $\sim125$~km~s$^{-1}$. Adopting an inclination angle $i=40-41^{\circ}$ (Bland-Hawthorn et al.~\cite{Bla97}; Sect.~\ref{kinemetry}), the implied spherical mass enclosed at $r\sim0\farcs15$ (10~pc) is $M_{\rm dyn}\sim8-9 \times 10^{7}~M_{\odot}$. This is a factor $\sim7-8$ higher than the black hole mass ($1-1.2\times10^7$~M$_{\odot}$) estimated from the H$_2$O maser kinematics in the inner $r\sim0.7$~pc of the galaxy (Greenhill et al.~\cite{Gre96}; Gallimore et al.~\cite{Gal01}), an indication that most of the dynamical mass at  $r\sim0\farcs15$ (10~pc) is contained in the central stellar cluster.

\subsection{The CO--to--H$_2$ conversion factor}\label{XCO} 
 
 Using the CO(1--0) map of  Schinnerer et al.~(\cite{Sch00}),  we find that the CO(1--0) flux ($S_{\rm CO(1-0)}$) integrated in the central 10$\arcsec$(700~pc)-aperture of the galaxy is $\sim100$~Jy~km~s$^{-1}$.  A comparison with the CO(1--0) flux derived using a similar aperture on the BIMA map of Helfer et al.~(\cite{Hel95}), which, unlike the Schinnerer et al. map, includes zero spacings, suggests that the PdBI map recovers $\simeq 80\%$ of the total flux in this region. The implied conversion factor 
$X_{\rm CO}=N({\rm H_2})/I_{\rm CO(1-0)}$ that is required to match the value of $M_{\rm gas}$ derived from the dust SED fitting discussed in Sect.~\ref{dust-CND} is $\sim1/(4^{+6}_{-1})\times X_{\rm CO}^{\rm MW}$, where 
$X_{\rm CO}^{\rm MW}=2\times10^{20}$cm$^{-2}$(K~km~s$^{-1}$)$^{-1}$ is the average CO conversion factor assumed to hold in the molecular clouds of the Milky Way disk (Strong et al.~\cite{Str88}; Strong \& Mattox~\cite{Str96}; Dame et al.~\cite{Dam01}).  A similar conclusion, pointing to an $X_{\rm CO}\sim$ a factor of 4--8 lower than $X_{\rm CO}^{\rm MW}$ in the CND of NGC~1068, was reached by Usero et al.~(\cite{Use04}) who used LVG models to fit the CO line ratios observed in this region. This result is also in line with the observational work of Israel~(\cite{Isr09a, Isr09b}) and Sandstrom  et al.~(\cite{San13}), who found that  $X_{\rm CO}$ can be up to a factor 4--10 lower than $X_{\rm CO}^{\rm MW}$ in the central $\sim 1$~kpc of a subset of galaxy disks of solar metallicities. Bell et al.~(\cite{Bel07}) also derived conversion factors which are typically $\sim$a factor of 10 lower than $X_{\rm CO}^{\rm MW}$  in their models of dense PDRs ($n$(H$_2$)~$\geq10^{4}$~cm$^{-3}$) of starburst nuclei. Lower $X_{\rm CO}$ factors have also been commonly  assumed for ULIRGs;  however,  it is still debated whether this is a common feature of extreme starburst systems (e.g., see discussion in Papadopoulos et al.~\cite{Pap12}).

 An estimate of the  molecular gas mass detected inside the central $r=20$~pc aperture  ($M_{\rm gas}$[AGN]) can be made based on the detected CO(3--2) emission, provided that we assume a conversion factor for this line.  Taking the  average value derived above for $X_{\rm CO}$, which is representative for the 1--0 line in the CND,  the conversion factor for the 3--2 line has to be scaled down by an additional factor 3,  based on the 3--2/1--0 line ratio $\sim 3$ (in T$_{\rm mb}$ units) derived in the neighborhood of the central engine of NGC~1068 (see Sect.~\ref{lineratios}). This implies that $M_{\rm gas}$[AGN]$~=1.6^{+0.5}_{-0.9}\times10^{5}$~M$_{\sun}$, i.e., the typical mass of a GMC and a value compatible within the errors with the gas mass derived from the dust emission in Sect.~\ref{dust-torus}. This is equivalent to an average H$_2$   column density $N_{\rm H_2}=5\times10^{21}$~mol~cm$^{-2}$.
%Based on PDR models, Bell et al.~(\cite{Bel07}) also derived conversion factors for the 3--2 line that are to be scaled-down by about a factor of 2--3 relative to  the  value of $X_{\rm CO}$ commonly applied to the 1--0 line in the nuclear environment of starburst galaxies. 
     
 Similarly, if we use the CO(6--5) integrated intensities in the ALMA aperture  for Band~9, i.e., $r=10$~pc, and the factor of $\sim 3-4$ scaled-down version of  $X_{\rm CO}$ for this line estimated from the 6--5/1--0 ratio $\sim3-4$ (in T$_{\rm mb}$ units) measured at the AGN, we derive $M_{\rm gas}$[AGN]$~=1.2^{+0.4}_{-0.7}\times10^{5}$~M$_{\sun}$. Considering the smaller aperture used in the 6--5 line, this value is, not surprisingly, slightly below the estimate derived from 3--2. This is equivalent to an average H$_2$ column density $N_{\rm H_2}=1.2\times10^{22}$~mol~cm$^{-2}$.
%This suggests that the CO(6--5) conversion factor derived by Bell et al.~(\cite{Bel07}) cannot be applied to the particular environment of the NGC~1068 central engine, as otherwise hinted at by the high CO(6--5)/CO(1--0) temperature ratio ($\sim 3-4$) measured close to the AGN (see Sect.~\ref{lineratios}). %Conversion factors for all molecular lines are re-evaluated in paper II, where we use radiative transfer models to fit the different line ratios observed.

%\section{Star formation laws}\label{KS} [{\bf TB written}]

%*****Main results include

%\begin{itemize}
%\item KS laws obtained with CO10, CO32, HCN10, HCO+10 and Palpha at different spatial resolutions 
%\item scatter in KS laws increases at higher spatial resolution for all tracers, as expected
%\item CO10 KS law is highly scattered. No KS law for scales<700pc. Dense(r) tracers (CO32, HCN10, HCOp10) show linear KS laws down to 100pc
%\end{itemize}

%*****
 
%______________________________________________ Fig10: co isovels
 \begin{figure*}[tbh!]
   \centering  
   \includegraphics[width=18cm]{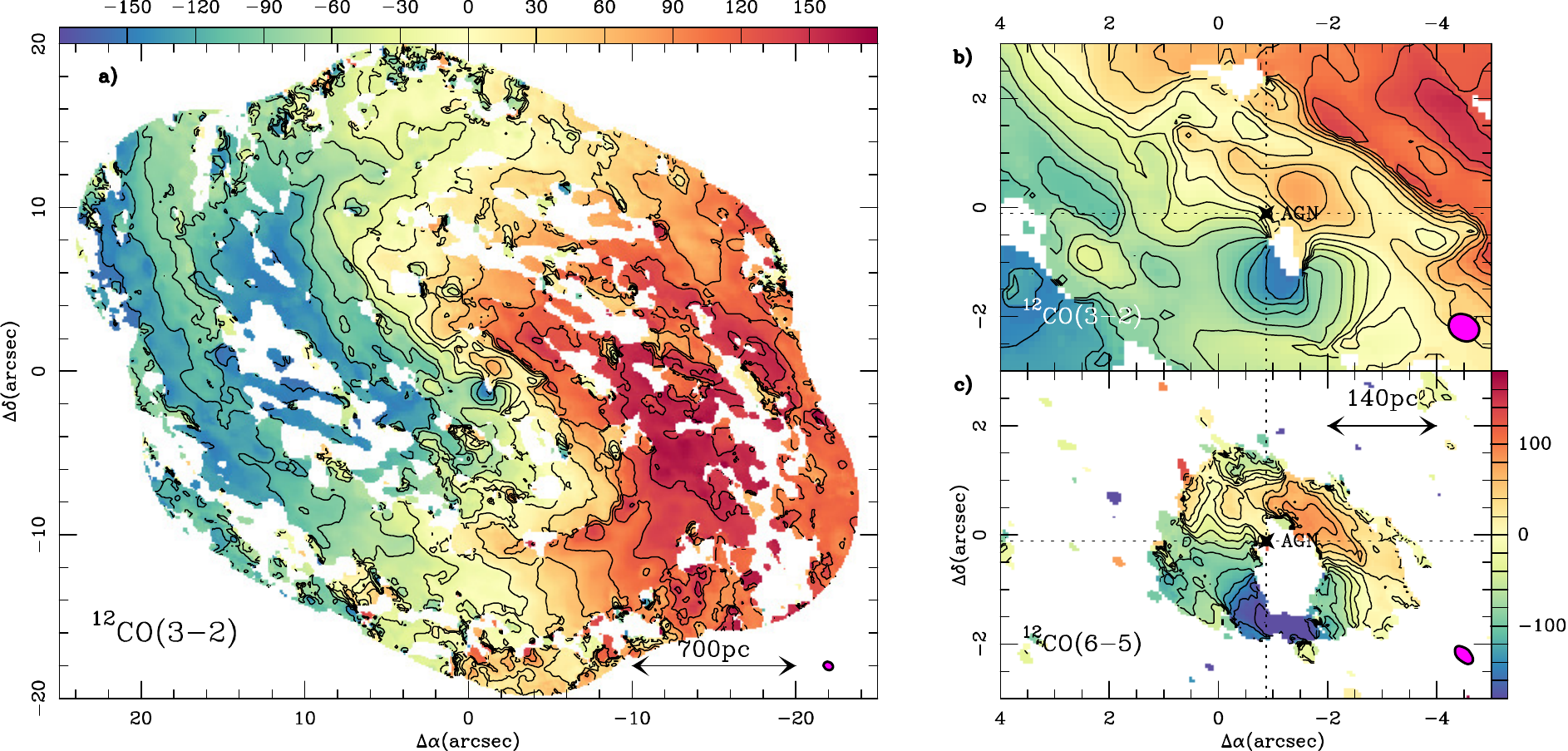}
   \includegraphics[width=18cm]{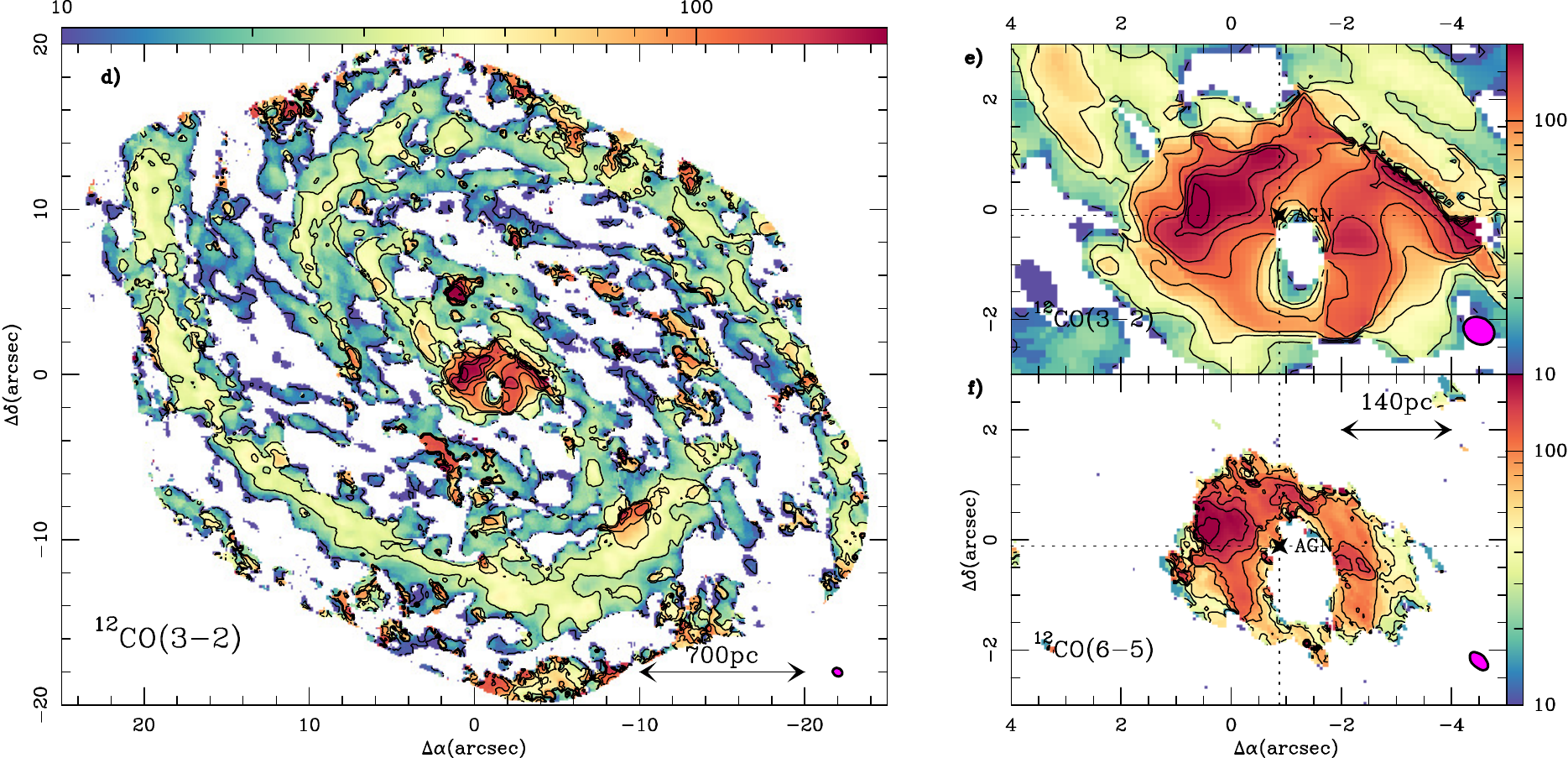}
               \caption{{\it Upper panels}:~{\bf a)}~Overlay of CO(3--2) isovelocity contours that span the range (--180~km~s$^{-1}$, 180km~s$^{-1}$) in steps of 30~km~s$^{-1}$ on a false-color velocity map (linear color scale as shown). Velocities refer to $v_{\rm sys}$(HEL)$~=1127$~km~s$^{-1}$. {\bf b)}~Same as {\bf a)} but zooming in on the CND region with a 20~km~s$^{-1}$-velocity spacing. {\bf c)}~Same as {\bf b)} but derived from the CO(6--5) data. 
  {\it Lower panels}:~{\bf d)}~Overlay of the CO(3--2) line widths (FWHM) shown in contours (10, 30, 50 to 200~km~s$^{-1}$ in steps of 30~km~s$^{-1}$)
on a false-color width map (logarithmic color scale as shown). {\bf e)}~Same as {\bf d)} but zooming in on the CND region. {\bf f)}~Same as {\bf e)} but 
derived from the CO(6--5) data.}
              \label{CO-vels}
\end{figure*}
    
%_____________________

%______________________________________________Fig11: HCN-HCOp-CS-isovels
 \begin{figure*}[tbh!]
 %  \sidecaption
 \centering
   \includegraphics[width=15cm]{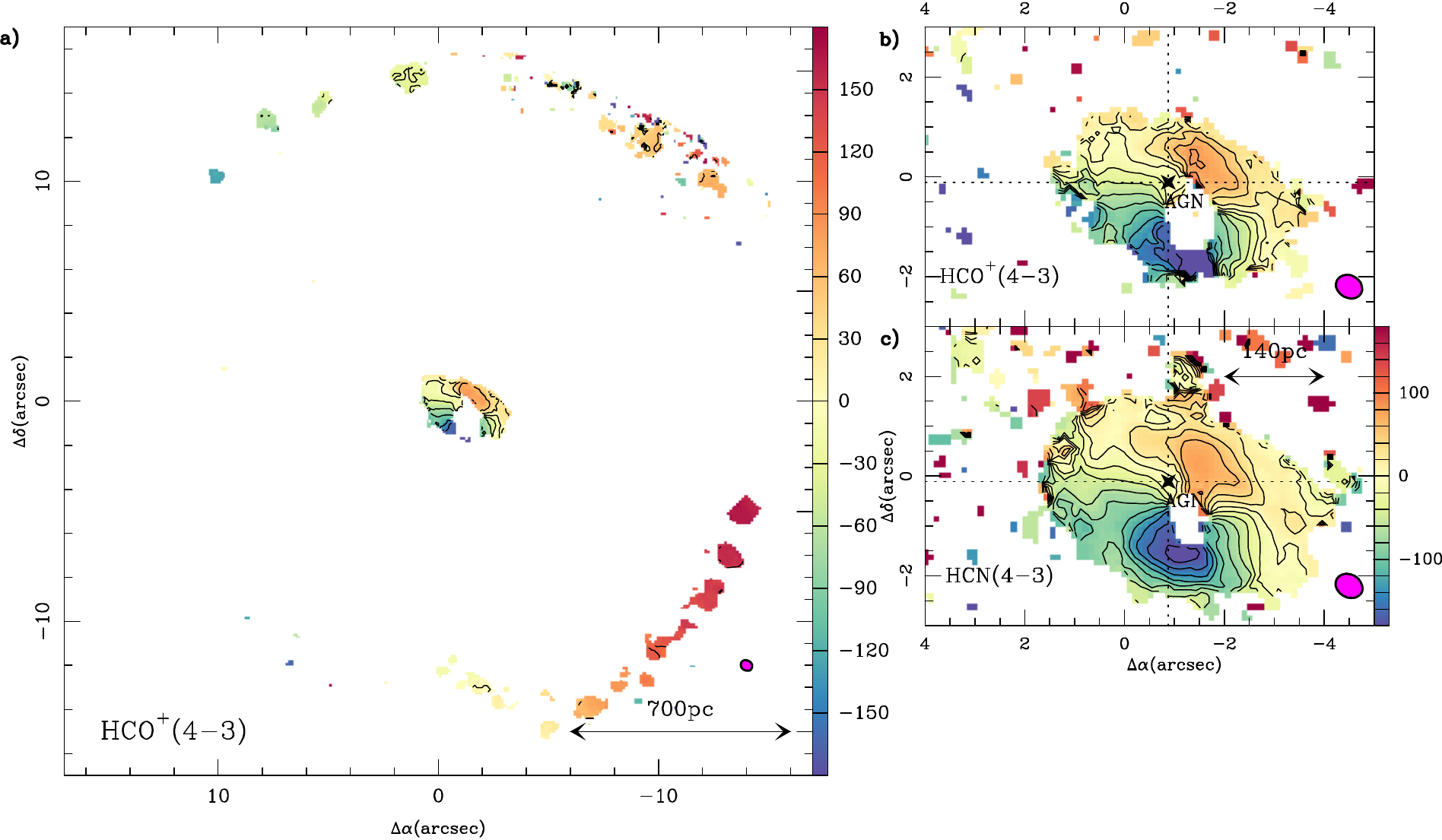}
                  \caption{Mean-velocity fields derived from the HCO$^+$(4--3) ({\bf a)}~{\it left panel}) data set. Panels  {\bf b)}~({\it upper right panel}) and {\bf c)}~({\it lower right panel}) show a close-up of the CND isovelocities derived from HCO$^+$(4--3) and  HCN(4--3), respectively. %{\bf e)}~({\it Lower right panel})~Same as {\bf c)} and {\bf d)} but derived from the CS(7--6) data. 
        Contour levels, color scales and velocity reference in all panels as in Fig.~\ref{CO-vels}. See Sect.~\ref{outflow} for details on the different clipping thresholds used.
        } %and masking methods used.}
              \label{dense-vels}
\end{figure*}

%______________________________________________  

 \section{Gas kinematics: the molecular outflow}\label{outflow}

 Figures~\ref{CO-vels} and  \ref{dense-vels} show the mean-velocity field of molecular gas in the disk of NGC~1068 derived from the CO(3--2), 
CO(6--5), HCN(4--3), and  HCO$^+$(4--3) lines.  By default, isovelocities are derived by integrating the emission above 5$\sigma$-levels throughout the disk in all tracers. The clipping is lowered to 3$\sigma$--levels when we zoom in on the CND region for HCN(4--3) and HCO$^+$(4--3) in Fig.~\ref{dense-vels}.

The CO(3--2) isovelocities of Fig.~\ref{CO-vels}{\it a}, which sample the kinematics of molecular gas 
throughout the central $\sim 40\arcsec$ (2.8~kpc)--region, show the expected pattern for a rotating disk characterized by an overall east-west 
orientation of its kinematic major axis ($PA=278^{\circ}\pm10^{\circ}$; Bland-Hawthorn et al.~\cite{Bla97}; Schinnerer et al.~\cite{Sch00}\footnote{The position angle of the kinematic major axis is measured east from north for the receding side.}). At close sight, the orientation 
of the major axis is seen to change from the CND region ($PA\sim330^{\circ}$) to the bar region  
($PA\sim290^{\circ}$) and farther out to the spiral arm region  ($PA\sim260^{\circ}$). The $PA$ of the CND is close to the orientation of the kinematic major axis derived for the much smaller $r\sim0.7$~pc rotating disk traced by H$_2$O maser emission published by Greenhill et al.~(\cite{Gre96}) ($PA=315^{\circ}$). The $PA$ values for the CO disk, 
derived from a qualitative fit of isovelocities, are confirmed by the kinematic modeling of Sect.~\ref{non-circ}. If we assume that gas orbits 
are roughly coplanar, the observed trend can be interpreted as  the footprint of non-circular motions on the gas flow. As discussed in 
Sect.~\ref{non-circ}, these distortions of the gas kinematics are caused by different mechanisms which are at work on different spatial scales: 
the spiral arm structure in the outer disk, the stellar bar on intermediate scales and the nuclear outflow in the CND.  All molecular tracers show a similar tilt of the CND kinematic major axis relative to the large-scale disk as shown in Fig.~\ref{dense-vels}.

The observed line widths (FWHM), displayed in Fig.~\ref{CO-vels}{\it d}, go from 15--20~km~s$^{-1}$ in the well detected ($ > 5\sigma$) molecular complexes of the interarm region, to an average value of 35--40~km~s$^{-1}$ in the SB ring. The corresponding velocity dispersion estimates would be $\sigma_{\rm v} \sim 6-8$~km~s$^{-1}$ in the interarm region and $\sigma_{\rm v} \sim 15-17$~km~s$^{-1}$ in the SB ring. The measured widths at the CND, where FWHM values range from 50 to 200~km~s$^{-1}$, likely overestimate $\sigma_{\rm v}$ for molecular gas in this region as they reflect the superposition of different velocity components inside the beam associated with rotation but also with an underlying strong outflow pattern (see Sect.~\ref{kinemetry}). Figures~\ref{CO-vels}{\it e} and \ref{CO-vels}{\it f} show that the region in the CND where FWHM values are above $\sim$100~km~s$^{-1}$ has a ring-like morphology. %with three maxima located at the two emission knots east and west of the AGN, and at the northern bridge connecting the latter. 
The widespread  distribution of high line width values in this region suggests that the outflow, which is found in Sect.~\ref{kinemetry} to be a mode superposed onto rotation, spatially extends over most of the CND. 

   We analyze below the evidence supporting  the existence of a massive molecular outflow in NGC~1068 based on the modeling of the velocity field derived from the CO(3--2) data (Sect.~\ref{kinemetry}). We examine in Sect.~\ref{outflow-dense} that the evidence for an outflow is also found in the molecular gas traced at higher densities by CO(6--5), HCN(4--3), and HCO$^{+}$(4--3). 
 The possible powering sources of the molecular outflow are discussed in Sect.~\ref{power}. We explore in Sect.~\ref{alternatives} if other alternative scenarios can explain the observed kinematics.  Section~\ref{outflow-others} compares the properties derived for the molecular outflow detected by ALMA with those derived from previous studies of NGC~1068.

  %______________________________________________ Fig12: harmonics-I
 \begin{figure*}[tbh!]
   \centering  
   \includegraphics[width=17cm]{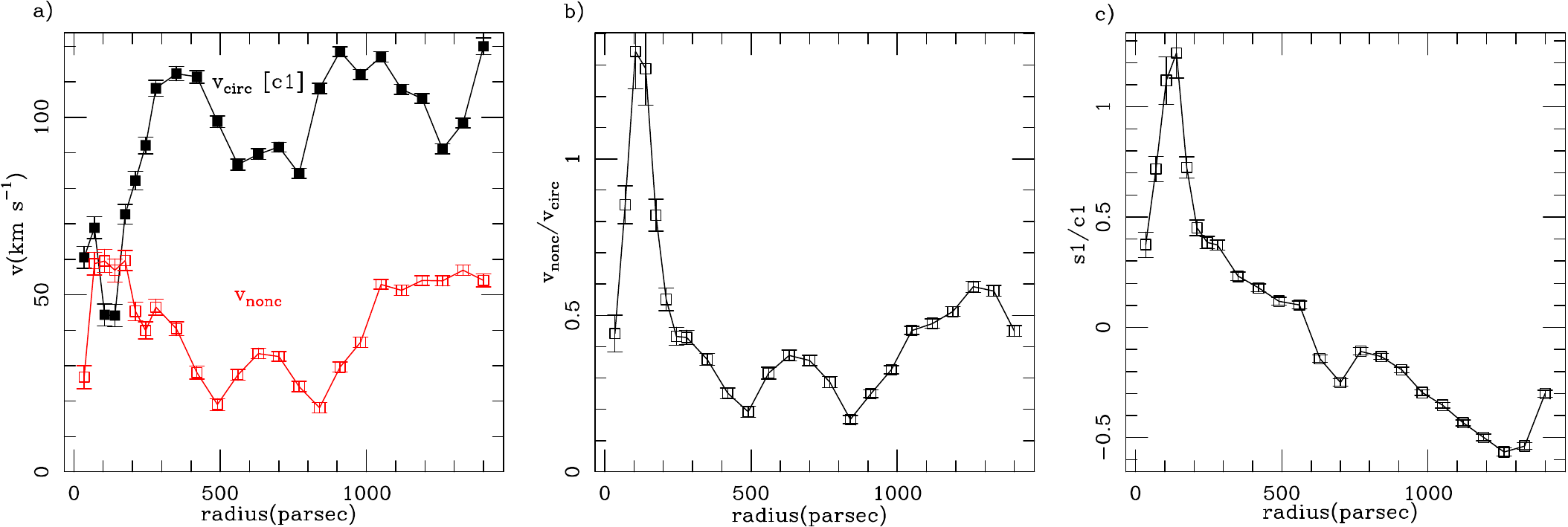}
              \caption{{\bf a)}~({\it Left panel})~The magnitude of the $c_1$ term of the Fourier decomposition of the velocity field of NGC~1068 
described in Sect.~\ref{kinemetry}  as a function of radius (black curve). The $c_1$ term represents the best fit of the (projected) axisymmetric circular component of the velocity field ($v_{\rm circ}$). 
We also plot the radial variation of the overall magnitude of the (projected) non-circular motions ($v_{\rm nonc}$) derived from the Fourier decomposition till third order (red curve). {\bf b)}~({\it  Middle panel}). The variation of the $v_{\rm nonc}/v_{\rm circ}$ ratio as function of radius. {\bf c)}~({\it Right panel})~The variation of the  $s_{1}/c_{1}$ ratio as function of radius. The $s_{1}$ term represents the best fit of the (projected) axisymmetric radial component of the velocity field.}
              \label{harm-I}
\end{figure*}

 \subsection{The CO(3--2) molecular outflow}\label{kinemetry}
      
 \subsubsection{Fourier decomposition of the velocity field}\label{non-circ}
 
The general description of the two-dimensional line-of-sight velocity field of a galaxy disk can be expressed as
\begin{equation}
v_{\rm los} (x,y) = v_{\rm sys} + v_\theta (x,y) \cos\psi\sin i + v_R (x,y) \sin\psi\sin i\;,
\label{eq-1}
\end{equation}
where $(v_R,v_\theta)$ is the velocity in polar
coordinates, $\psi$ is the phase angle measured in the galaxy plane from the receding side
of the line of nodes, and $i$ is the inclination angle restricted to the range $0<i<\pi/2$.   With this convention $v_\theta > 0$ always, while $v_R > 0$ ($v_R < 0$) indicates 
outflow (inflow) for counterclockwise rotation and inflow (outflow) for clockwise rotation, respectively. We note that in the case of NGC~1068, where the receding side of the major axis is to the west, rotation should be counterclockwise in the likely scenario where spiral arms are trailing  (e.g., Contopoulos~\cite{Con71}; Toomre~\cite{Too81}; Binney \& Tremaine~\cite{Bin87}).

 %______________________________________________ Fig13: harmonics-II
 \begin{figure}[tbh!]
 \centering
    \includegraphics[width=6cm]{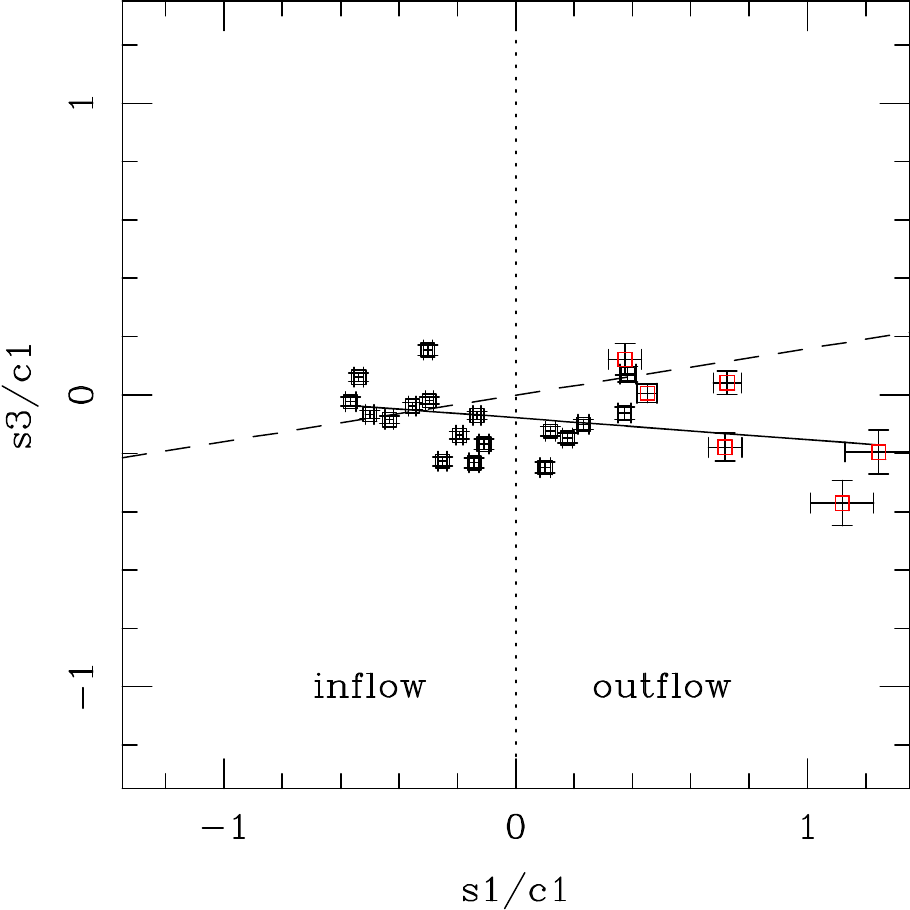}
                \caption{Comparison of $s_1$ and $s_3$ terms normalized by the circular rotation term $c_1$ derived from the adopted best fit of the CO(3-2)     
 velocity field.  The continuous line represents the least-squares fit to the data points, and the dashed line represents the expected warp line location 
 predicting a relation between the $s_1$ and $s_3$  terms for an error in the position angle adopted throughout the disk ($PA=289^\circ$) (see discussion in Wong et al.~\cite{Won04}). The sign of 
 $s_1$ is taken as a signature of  inflow or outflow as shown, for the assumed geometry of the galaxy. Black symbols correspond to points at radii~$>3\arcsec$,  
 while red symbols correspond to radii~$\leq3\arcsec$.}
              \label{harm-II}
\end{figure}
    
%_______________________________  

    %______________________________________________ Fig14: co-velres-big
 \begin{figure}[tbh!]
   \centering  
   \includegraphics[width=9cm]{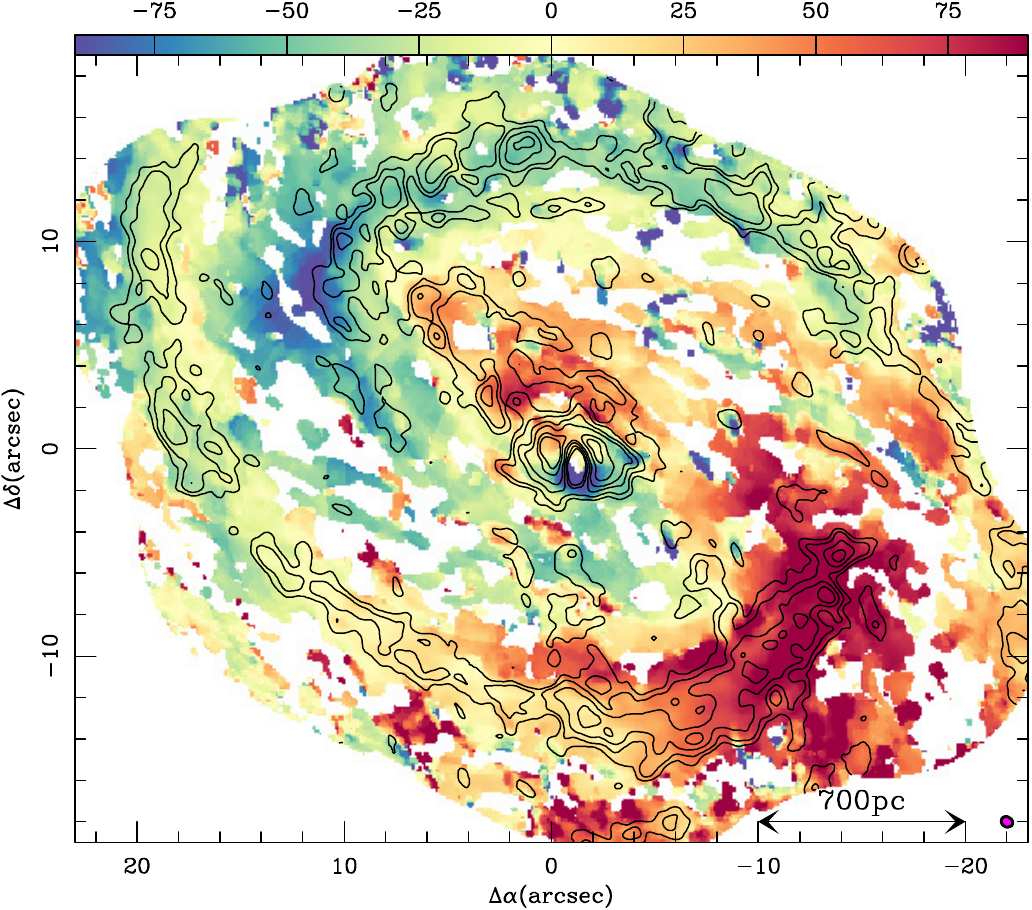}
                \caption{Overlay of the integrated intensity map of CO(3--2) (in contours: 15$\sigma$, 30$\sigma$, 60$\sigma$, 100$\sigma$, 200$\sigma$, and 300$\sigma$; with 1$\sigma=0.22$~Jy~km~s$^{-1}$beam$^{-1}$) on the residual mean-velocity field ($\langle  V_{\rm res} \rangle$, in false-color scale spanning the range: --90~km~s$^{-1}$ to 90~km~s$^{-1}$) obtained after subtraction of the best fit rotation component, as described in Sect.~\ref{non-circ}. Velocities refer to $v_{\rm sys}$(HEL)$~=1127$~km~s$^{-1}$.}
              \label{velres-big}
\end{figure}

Alternatively, the line-of-sight velocity field can be divided into a number of elliptical ring profiles defined by ($PA$, $i$)  for a given 
radius $r$, and $v_{\rm los}(r, \psi)$ is  decomposed as a Fourier series with harmonic coefficients $c_j(r)$ and $s_j(r)$, where 
\begin{equation}
v_{\rm los} (r, \psi)= c_0 + \sum_{j=1}^n [c_j(r) \cos (j\psi) + s_j(r) \sin (j\psi)].
\label{eq-2}
\end{equation}
In the most general case, $c_1$ reflects  the contribution from circular rotation while all remaining terms represent contributions from noncircular 
motions (Schoenmakers et al.~\cite{Sch97}; Schoenmakers~\cite{Sch99}). It is known that expanding the series of Eq.~\ref{eq-2} out to $n=3$ 
provides a fair description of $v_{\rm los}$ in most models (Trachternach et al.~\cite{Tra08}). In a disk with simple (axisymmetric) circular rotation 
($v_c$), $c_0=v_{\rm sys}$ and $c_1=v_c \sin i$ and all remaining terms can be neglected. In the case
of a pure (axisymmetric) radial flow ($v_R$), $c_0=v_{\rm sys}$ and $s_1=v_R \sin i$, with the rest of coefficients being 0.
 
 We derived the Fourier terms that best describe the $v_{\rm los}$ extracted from the CO(3--2) mean-velocity field presented in Sect.~\ref{line}. We adopted an iterative three-step process using the software package {\tt kinemetry} developed by Krajnovic et al.~(\cite{Kra06}). We used in the fit  28 ellipses with semi major axes covering the disk from $r=0\farcs5$ to $r=24\arcsec$, with 
 a spacing $\Delta r=0.5\arcsec$ in the inner 4$\arcsec$ and $\Delta r=1\arcsec$ farther out.  At each step, {\tt kinemetry} finds the best fitting ellipses by minimizing the contribution of noncircular motions ($v_{\rm nonc}$), evaluated as
  
 \begin{equation}
v_{\rm nonc} (r)= \sqrt{s_1^2(r)+s_2^2(r)+c_2(r)^2+s_3^2(r)+c_3^2(r)}. 
\label{eq-3}
\end{equation}
 
 In a first step we left the position angle $PA(r)$ and the inclination $i(r)$  as free parameters in the fit and assumed that the dynamical center coincides with the AGN. We then derived the average values of $PA(r)$ and $i(r)$:   
 $\langle PA \rangle~=290\pm5^\circ$ and $\langle i \rangle~=41\pm2^\circ$ (excluding outliers). In a second step, we fixed  $\langle i \rangle=41\pm2^\circ$ and re-determined the  
 $PA(r)$ profile.  This profile shows a changing $PA$ value from the CND region ($PA\sim330^{\circ}\pm10^{\circ}$ at 
 $r<5\arcsec$) to the bar ($PA\sim290^{\circ}\pm10^{\circ}$ at 5$\arcsec < r < 10\arcsec$) and the spiral arm region  
 ($PA\sim 260^{\circ}\pm10^{\circ}$ at $r>12\arcsec$). From this second iteration, we derived $\langle PA \rangle~=289\pm5^\circ$. These values of $\langle PA \rangle$ and $\langle i \rangle$ are compatible within the errors with previous determinations of the disk orientation 
 available in the  literature ($PA=286\pm5^\circ$ and $i=40\pm3^\circ$, e.g., see compilation by  Bland-Hawthorn et al.~\cite{Bla97}). Finally, we derived the Fourier decomposition of the 
 NGC~1068 velocity  field fixing  $PA=289\pm5^\circ$ and $i=41\pm2^\circ$ at all radii.  Implicit in this assumption is that the gas kinematics at all radii can be satisfactorily modeled by coplanar orbits (alternative scenarios are described in Sect.~\ref{alternatives}). As an output parameter of this final iteration we obtained  the best fit for the systemic velocity, 
 $v_{\rm sys}$(HEL)~$=1127\pm3$~km~s$^{-1}$. This coincides within the errors with the value of $v_{\rm sys}$ inferred from the kinematics of the water maser emission detected in the central $r\sim0.7$~pc, as derived from the Very Long Base Interferometry (VLBI)  observations of Greenhill \& Gwinn~(\cite{Gre97}).

 Figures~\ref{harm-I}{\it ab} compare the magnitude of  the (projected)  circular component of the CO(3--2) velocity field ($v_{\rm circ} = c_{1}$) obtained in the fit with 
that of the non-circular motions ($v_{\rm nonc}$) defined in Eq.~\ref{eq-3}. While the value of $v_{\rm nonc}/v_{\rm circ}$ stays within the range $\sim 0.2-0.6$ over a 
sizable fraction of the outer disk (3$\arcsec < r < 20\arcsec$), reflecting the expected order of magnitude of the contribution from the bar and the spiral structure
to $v_{\rm nonc}$ in this region, the same ratio reaches extremely high values ($\sim 0.8-1.30$) in the inner $r\leq3\arcsec$ of the galaxy. As further illustrated in  Fig.~\ref{harm-I}{\it c}, the 
main contribution to $v_{\rm nonc}$ comes from the $s_1$ term. In particular, the sign ($>0$) and normalized strength of $s_1$  ($s_1/c_1 > 0.3$) indicates that strong 
outflow motions prevail at $r\leq3\arcsec$.
%, a region  associated with the bow-shock feature and the CND. 
The $s_1$ term changes sign at $r\sim 8\arcsec$ and stays 
moderately strong and negative ($-0.5 < s_1/c_1 < -0.1$) farther out in the disk; this behavior fairly reflects the expected influence of the bar and the spiral structure on the 
gas flow provided that we are inside corotation of the relevant perturbation (bar or spiral) at these radii (Schinnerer et al.~\cite{Sch00}; Wong et al.~\cite{Won04}; Emsellem et al.~\cite{Ems06}).

Figure~\ref{harm-II}, where we compare the $s_1$ and $s_3$ terms of the Fourier decomposition of $v_{\rm los}$, also supports that bar-and-spiral induced 
streaming motions (inside corotation) are the simplest explanation for the $s_1$ profile in the outer disk: the dominance of the $s_1$ term over the  $s_3$ term indicates that we 
remain inside corotation in this region. This supports the conclusion that
the molecular gas pseudo-ring at $r\sim18\arcsec$ (1.3~kpc) is not part of the inner bar pattern, but that it constitutes an independent wave feature characterized by a lower pattern speed (e.g., see Rand \& Wallin~\cite{Ran04}). By contrast, this mostly excludes the streaming motion scenario to explain the strong outward radial motions identified in the inner disk, which we rather interpret as a molecular outflow: values of $s_1\gg0$ would be  prevalent {\em outside} corotation for the assumed geometry of NGC~1068.

    Figure~\ref{velres-big} shows the residual mean-velocity field ($\langle  V_{\rm res} \rangle$) obtained after subtraction of the rotation component derived in the analysis above. The residuals clearly show a pair of approaching-receding regions in the outer disk. This morphology follows the expected 
2D-pattern produced by the combined action of the bar and the spiral on the gas flow provided that we are inside corotation of the perturbations 
on these scales (Canzian~\cite{Can93}; Sempere et al.~\cite{Sem95}, Colombo et al.~\cite{Col14}). This is in agreement with the predominance of 
an inward radial flow that characterizes the fit to non-circular motions in the outer disk. 

Closer to the nucleus the comparison between the residual velocity field, shaped by outward radial motions, and the gas/dust distribution in the 
inner disk suggests that a significant fraction of the gas in this region is associated with the outflow, as shown in Figs.~\ref{velres-zoom}{\it ab}. Most remarkably, 
Figs.~\ref{velres-zoom}{\it cd} show a noticeable spatial correlation between the AGN ionized nebulosity, traced by Pa$\alpha$ emission, the radio jet 
plasma, traced by the radio continuum emission at 22~GHz, and the molecular outflow signature identified  in the CO velocity field. This close association, which applies to a large range of radii and to a wide angle in the disk,  
suggests that a sizable fraction of the dense molecular gas traced by CO(3--2) is being  
entrained due to AGN feedback in the CND, and farther out in the disk, in the bow-shock arc region (located at $r \sim 400$~pc).

\subsubsection{Mass outflow rate}\label{load}

To derive the mass load of the outflow we need to estimate the characteristic mass ($M_{\rm mol}$) as well as the projected values of  the radial size ($R_{\rm out}$) and velocity ($V_{\rm out}$) of the outflow, and also assume a certain geometry (i.e., a certain angle $\alpha$ between the outflow and the line of sight). If we assume a {\em multi-conical} outflow {\em uniformly filled} by the outflowing clouds (Maiolino et al.~\cite{Mai12}; Cicone et al.~\cite{Cic14}), the mass load rate ($\mathrm{d}M/\mathrm{d}t$) can be estimated from the expression

\begin{equation}
\frac{\mathrm{d}M}{\mathrm{d}t}=3 \times V_{\rm out} \times M_{\rm mol}/R_{\rm out} \times \tan(\alpha) \label{out}.
\end{equation}

 As discussed by Maiolino et al.~(\cite{Mai12}), if instead of a homogeneous multi-conical outflow, we assume a {\em single shell-like} geometry of thinness $dR_{\rm out}$ (where generally $dR_{\rm out} \ll R_{\rm out}$), the above estimate should be replaced by
 
\begin{equation}
\frac{\mathrm{d}M}{\mathrm{d}t}= V_{\rm out} \times M_{\rm mol}/dR_{\rm out} \times \tan(\alpha) \label{out-2}.
\end{equation}

In the following we use Eq.~\ref{out}, as it provides a more conservative lower limit to the outflow rate estimated globally over the CND.

The mass ($M_{\rm mol}$) has been calculated from the CO(3--2) data cube, after subtraction of the projected rotation 
curve (derived in Sect.~\ref{kinemetry}), by 
integrating the emission of the line outside a velocity range $\langle  V_{\rm res} \rangle=[-50,+50]$~km~s$^{-1}$, which encompasses most of the expected virial motions 
around rotation.  We determine that $\simeq 50\%$ of the total CO(3--2) flux in the CND stems from the outflow component. A similar percentage is derived for the bow-shock arc region, where we also find the signature of the outflow at larger radii. This translates into a total molecular mass $M_{\rm mol} \sim 1.8^{+0.6}_{-1.1} \times 10^7$~M$_{\sun}$ for the CND, including the mass of helium, if we assume a CO conversion factor $\sim1/(4^{+6}_{-1})$ of the MW value (see discussion in Sect.~\ref{XCO}). We note that this estimate is rather conservative as the $X_{\rm CO}$ conversion factor for the fraction of molecular gas that participates in the outflow could be higher if this component consists of an ensemble of optically thick dense clumps embedded in a diffuse medium.

\begin{figure*}[tbh!]
   \sidecaption 
   \includegraphics[width=11cm]{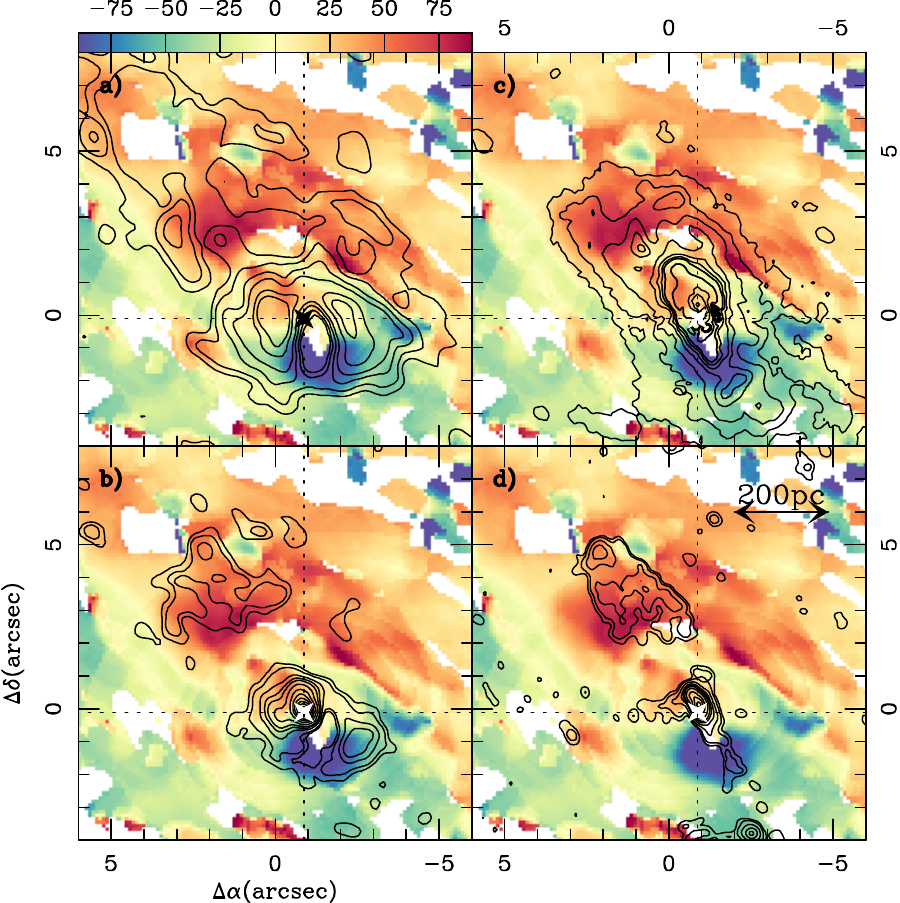}
           \caption {Overlay of the residual mean-velocity field ($\langle  V_{\rm res} \rangle$) of Fig.~\ref{velres-big}, in color scale, with the contours representing: the integrated intensity of CO(3--2) ({\bf a)}~{\it upper left panel}; contours as in Fig.~\ref{velres-big}),  the 349~GHz continuum emission  ({\bf b)}~{\it lower left panel}; contours as in Fig.~\ref{dust-maps}), the HST Pa$\alpha$ emission ({\bf c)}~{\it upper right panel}: with contours
 0.2$\%$, 0.5$\%$, 1$\%$ to 5$\%$ in steps of 1$\%$, 10$\%$ to 90$\%$ in steps of 20$\%$ of the peak value=1000~counts~s$^{-1}$pixel$^{-1}$), and the 22~GHz VLA map of Gallimore et al.~(\cite{Gal96}) 
  ({\bf d)}~{\it lower right panel}: with contours
 1$\%$, 2$\%$, 5$\%$,10$\%$, 15$\%$, 20$\%$, 30$\%$, 50$\%$, 70$\%$, and 90$\%$ of the peak value=36~mJy~beam$^{-1}$).}
 
              \label{velres-zoom}
\end{figure*}

\begin{figure*}[tbh!]
  \sidecaption
   \includegraphics[width=12cm]{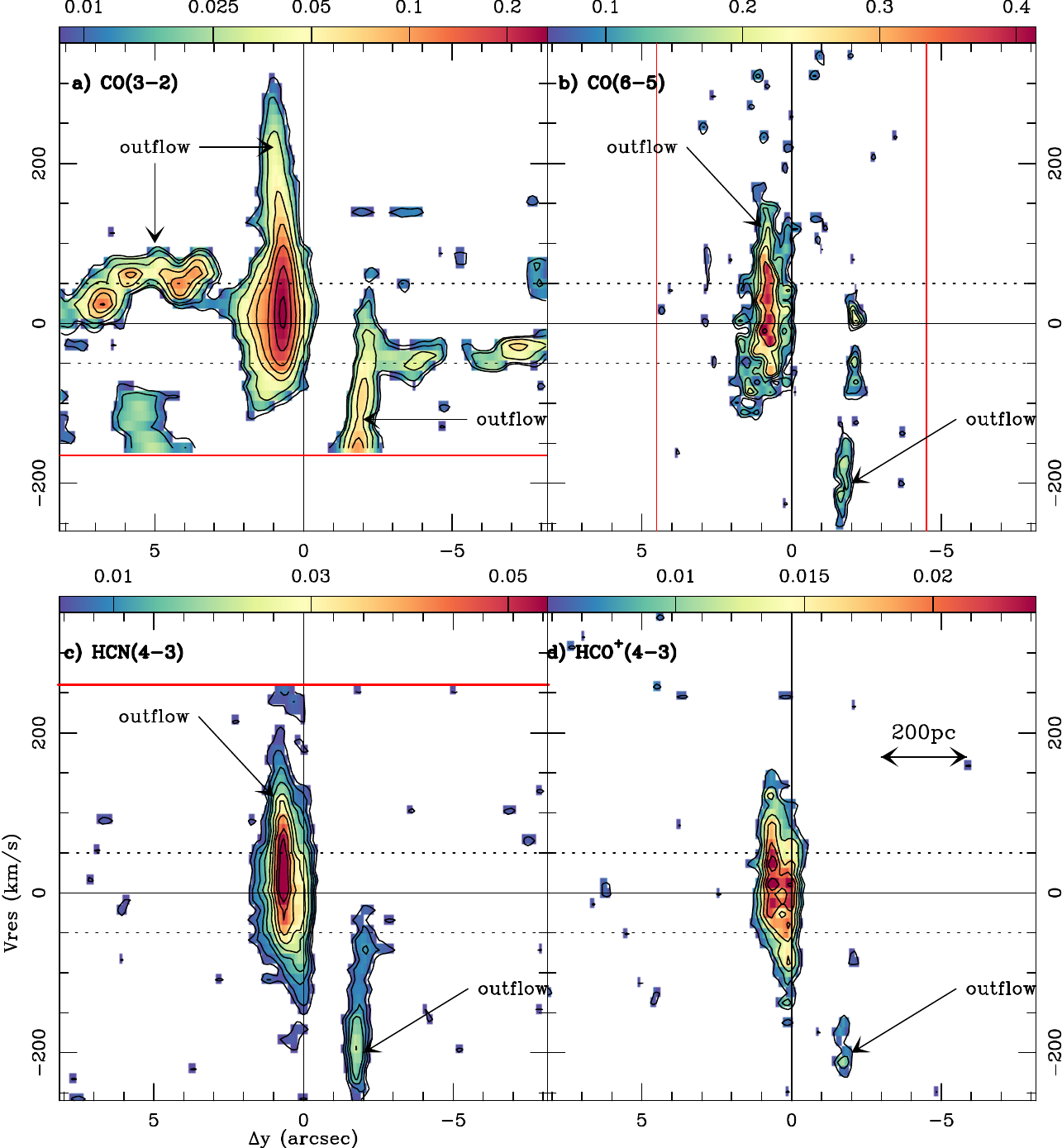}    
           \caption {Position-velocity (p-v) plots taken along the kinematic minor axis ($PA=19^{\circ}$) of NGC~1068 help identify outflow signatures in 
several molecular tracers: {\bf a)}~({\it Upper left panel})~CO(3--2) (color scale in Jy~beam$^{-1}$ and contours:  3$\sigma$,  5$\sigma$, 10$\sigma$, 20$\sigma$, 30$\sigma$, 45$\sigma$, 70$\sigma$, and 95$\sigma$; 1$\sigma=2.8$~mJy~beam$^{-1}$).    {\bf b)}~({\it Upper right panel})~CO(6--5) (color scale in Jy~beam$^{-1}$ and contours: 2.5$\sigma$, 4$\sigma$, 6$\sigma$, 8$\sigma$,10$\sigma$, 14$\sigma$, and 18$\sigma$; 1$\sigma=23$~mJy~beam$^{-1}$). {\bf c)}~({\it Lower left panel})~HCN(4--3) (color scale in Jy~beam$^{-1}$ and contours: 
2.5$\sigma$, 4$\sigma$, 6$\sigma$, 8$\sigma$, 12$\sigma$, 16$\sigma$, 20$\sigma$, 26$\sigma$, and 30$\sigma$; 1$\sigma=1.8$~mJy~beam$^{-1}$). {\bf d)}~({\it Lower right panel})~HCO$^{+}$(4--3) (color scale in Jy~beam$^{-1}$ and 
contours: 2.5$\sigma$, 4$\sigma$, 6$\sigma$, 8$\sigma$, 10$\sigma$, and 12$\sigma$; 1$\sigma=2.0$~mJy~beam$^{-1}$). The velocity scale ($\langle  V_{\rm res} \rangle$) is identical to that of Fig.~\ref{velres-big}. The spatial 
scale ($\Delta$y) along the minor axis refers to the AGN locus; positive offsets on the northern side. The black dashed lines at  $\langle V_{\rm res} \rangle=\pm50$~km~s$^{-1}$ identify the 
expected range of virial motions along the minor axis.         The red solid lines in panels {\bf a)} and {\bf c)} indicate the edges of the bands 
beyond which data have been flagged ($\langle  V_{\rm res} \rangle<-170$~km~s$^{-1}$ in CO(3--2) and $\langle  V_{\rm res} \rangle > 260$~km~s$^{-1}$ in HCN(4--3)). 
Similarly we also identify the 9$\arcsec$ field-of-view in the CO(6--5) p-v plot of panel ${\bf b)}$.}
              \label{pvs}
\end{figure*}

The average projected radial extent of the outflow in the CND is $R_{\rm out}\sim1.5\arcsec$ (100~pc) and the projected radial velocities $V_{\rm out}$ are close to $\sim100$~km~s$^{-1}$ according to the residual velocity field shown in Figs.~\ref{velres-big}, \ref{velres-zoom} and \ref{pvs}. 

The implied outflow rate given by Eq.~\ref{out} is $\mathrm{d}M/\mathrm{d}t\sim54^{+18}_{-32}\times \tan(\alpha)$~M$_{\sun}$~yr$^{-1}$ , where $\alpha$ reflects the unknown angle between the outflow and the line-of-sight. If we assume that the outflow is coplanar with the main disk, $\tan(\alpha)=1/\tan(i)$, where $i=41^{\circ}$, then
$\mathrm{d}M/\mathrm{d}t\sim63^{+21}_{-37}$~M$_{\sun}$~yr$^{-1}$. %(or $\sim63$~M$_{\sun}$~yr$^{-1}$ in the \emph{uniformly filled} model). 
This is significantly above the mass load rate estimated  on similar spatial scales for the ionized gas outflow: $\sim9$~M$_{\sun}$~yr$^{-1}$ (M\"uller-S\'anchez et al.~\cite{Mue11}). The molecular mass load rate implies a very short gas depletion timescale of $\leq1$~Myr in the CND.

%______________________________________________ Fig15: co-velres-zoom

%______________________________________________ 

Krips et al.~(\cite{Kri11}) used a simplified model to fit by eye the CO(3--2) spectra of the CND obtained by the SMA. While based on a different approach,  Krips et al.~(\cite{Kri11}) also concluded that a significant fraction ($\geq30\%$) of the molecular gas in the CND could be participating in an outflow, in qualitative agreement with our findings. Similarly, Davies et al.~(\cite{Dav08}) argued that the kinematics of molecular gas at the CND traced by the 2.12~$\mu$m H$_2$ line are suggestive of an outflow with typical projected velocities  $V_{\rm out}\sim100$~km~s$^{-1}$, i.e., in quantitative agreement with the values derived in this work. While the NIR line traces a minor fraction of the total H$_2$ content and a distinctly different molecular gas phase to that probed by the ALMA data, the similar picture drawn from these different data sets supports the outflow scenario in the CND.    

The same estimate applied to the outflow detected in this work in the bow-shock arc region, with $V_{\rm out} \sim 75$~km~s$^{-1}$, $M_{\rm mol} \sim 9^{+3}_{-5.4} \times 10^6$~M$_{\sun}$ and $R_{\rm out} \sim 5\arcsec$ ($\sim$400~pc), gives $\mathrm{d}M/\mathrm{d}t \sim 6^{+2}_{-3.6}$~M$_{\sun}$~yr$^{-1}$ for the \emph{uniformly filled} model. The bow-shock arc component was not detected in the SMA map of Krips et al.~(\cite{Kri11}), nor is there  a signature of the outflow on these scales in the VLT map of Davies et al.~(\cite{Dav08}).

            %______________________________________________ Fig16: pvs
    
%_________________

 \subsection{The molecular outflow in dense gas tracers ($n$(H$_2$)~$\geq$~10$^{5-6}$~cm$^{-3}$)}\label{outflow-dense}

In the following we will show that the analysis of the kinematics derived from CO(6--5), HCN(4--3), and HCO$^{+}$(4--3) confirms that the molecular outflow detected in the CO(3--2) line has a higher density ($n$(H$_2$)~$\geq$~10$^{5-6}$~cm$^{-3}$, according to paper~II) counterpart in NGC~1068. 

As discussed in Sect.~\ref{line},  the mean-velocity field of the CND shows in all tracers a similar pattern: the kinematic major axis of the 
CND is tilted by $\geq 40^{\circ}$ relative to that of the large-scale disk traced by CO(3--2), as shown in Fig.~\ref{dense-vels}. This distinct kinematic 
feature has been modeled as an outflow in CO(3--2)  (see Sect.~\ref{kinemetry}), and as such can be similarly interpreted as a 
signature of outflow in the dense gas tracers.
Figure~\ref{pvs} shows the position-velocity (p-v) plots taken along the minor axis, determined in Sect.~\ref{kinemetry} ($PA=19^{\circ}$), for  
various dense molecular tracers. In these diagrams any deviation of the emission from $v_{\rm sys}$ beyond the virial range, determined by the expected cloud-cloud velocity dispersion, is indicative of radial inflow/outflow motions. An inspection of this figure shows that the outflow signatures identified in CO(3--2) out to radii 
$r \sim 400$~pc are echoed in CO(6--5), HCN(4--3), and HCO$^+$(4--3) on the scales of the 
CND ($r \sim 100-200$~pc). 
A  sizable fraction of the emission of these molecular tracers lies outside the expected range of virial 
motions attributable to rotation and  dispersion: on average, emission is $\geq50$~km/s-redshifted on the northern side of the CND, while it is $\geq100$~km/s-blueshifted on the southern side. This reflects the sign and the right order of magnitude of the mean-velocity field deviations seen in the CO(3--2) map of Fig.~\ref{velres-zoom} at 
the CND.

The limited velocity coverage of the CO(3--2) data at the blue velocity end ($\langle  V_{\rm res} \rangle < -170$~km~s$^{-1}$) implies that we underestimate the most extreme outflow velocities on the southern side of the CND in this line. We note that outflow velocities up to $\langle  V_{\rm res} \rangle \sim+280$~km~s$^{-1}$ are detected in CO(3--2) on the northern side. The reality of these highly redshifted velocities in the outflow is confirmed by the HCN(4--3) data, which show emission from $\langle  V_{\rm res} \rangle=$--260 to +260~km~s$^{-1}$ in the CND, as shown in Fig.~\ref{pvs}. This implies that the outflow mass load rate estimated in Sect.~\ref{load} from CO(3--2) should be taken as a lower limit.

\subsection{The powering source of the molecular outflow: star formation or AGN jet driven?}\label{power}

Evidence of a young stellar population in the CND has been recently found by  Storchi-Bergmann et al.~(\cite{Sto12}), who 
located the young star formation (SF) episode inside the expanding molecular ring at $r\sim100$~pc. While the presence of young stars  
is well established, the total energy possibly injected by SF in the CND is much lower compared to the AGN. Davies et al.~(\cite{Dav07}) (see also discussion in Hailey-Dunsheath et al.~\cite{Hai12}) estimated that only a fraction
of the total FIR continuum luminosity ($L_{\rm FIR}$) from the CND can be attributed to ongoing/recent SF. Based on the luminosity measured in the K-band in the inner $r\sim 35$~pc, Davies et al.~(\cite{Dav07}) conclude that $L_{\rm FIR}$ due to SF is $\sim (1.7-3) \times 10^{9}$~L$_{\sun}$; this implies an integrated star formation rate of $SFR \sim 0.4-0.7$~M$_{\sun}$~yr$^{-1}$. This estimate is similar to the nuclear $SFR$ measured from the 11.3~$\mu$m PAH luminosity in the inner $r\sim12$~pc of NGC~1068 by Esquej et al.~(\cite{Esq14}), who estimated that  $SFR_{\rm nuclear} \sim 0.4$~M$_{\sun}$~yr$^{-1}$. The SFR estimated by Esquej et al.~(\cite{Esq14}) for the circumnuclear region out to a radius $r\sim2\arcsec$ (140~pc) is about 1~M$_{\sun}$~yr$^{-1}$.  The total SFR in the CND is therefore about an order of magnitude lower than the estimated mass load rate of the molecular outflow. Taken at face value this discrepancy suggests that SF is not able to drive the molecular outflow in the CND of NGC~1068 (see Murray et al.~\cite{Mur05} and 
Veilleux et al.~\cite{Vei05} for a general discussion).

The kinetic luminosity of the CND outflow can be derived from the expression

\begin{equation}
 L_{\rm kin}=1/2 \times \frac{\mathrm{d}M}{\mathrm{d}t} \times \left(\frac{V_{\rm out}}{\cos(\alpha)}\right)^2. \label{kin}
\end{equation}

Assuming that the gas in the outflow is coplanar,  $L_{\rm kin}\sim 5^{+1.7}_{-3}\times10^{41}$~erg~s$^{-1}$ .  AGN feedback models require that a significant fraction  of the radiated luminosity 
 ($\sim5\%L_{\rm bol}$;  di Matteo et al.~\cite{DiM05}) should be coupled to the ISM to produce an outflow. This fraction is lowered to 
$\sim0.5\%$ in the two-phase feedback model of Hopkins \& Elvis~(\cite{Hop10b}). The bolometric luminosity of the AGN in NGC~1068 
($L_{\rm bol}$), estimated from MIR and X-ray wavelengths, is at least three orders of magnitude larger than $L_{\rm kin}$: $L_{\rm bol}\geq10^{44-45}$erg~s$^{-1}$ (Bock et al.~\cite{Boc00}; Laurent et al.~\cite{Lau00}; Matt et al.~\cite{Mat00}; Raban et al.~\cite{Rab09}; Prieto et 
al.~\cite{Pri10}; Alonso-Herrero et al.~\cite{Alo11} and Sect.~\ref{CLUMPY-models}). This result  indicates that the AGN can power the outflow in the CND of NGC~1068.

The momentum flux of the CND outflow can be computed from the expression
\begin{equation}
 \frac{\mathrm{d}P_{\rm out}}{\mathrm{d}t}=\frac{\mathrm{d}M}{\mathrm{d}t}  \times \frac{V_{\rm out}}{\cos(\alpha)} . \label{mom}
\end{equation}

In the case of coplanar gas, Eq.~\ref{mom} yields $\mathrm{d}P_{\rm out}/\mathrm{d}t\sim6^{+2}_{-3.6}\times10^{34}$g~cm~s$^{-2}$.  Compared to the momentum provided by the AGN photons, derived as $L_{\rm bol}/c \sim (0.3-3)\times10^{34}$g~cm~s$^{-2}$,  $\mathrm{d}P_{\rm out}/\mathrm{d}t$ is a moderate factor 1--27 larger. The required {\em boost} is only $\sim$1.7--6 if we adopt $L_{\rm bol}=4.2\times10^{44}$erg~s$^{-1}$ from Sect.~\ref{CLUMPY-models}. The molecular outflow could thus be in the momentum-conserving regime (where cooling is {\em fast}), and radiation pressure could be the driving mechanism.  
This is also within the range of values of the momentum {\em boost} factors predicted by AGN feedback models under the assumption that molecular outflows are energy-conserving (adiabatic Sedov phase): $(\mathrm{d}P_{\rm out}/\mathrm{d}t)/(L_{\rm bol}/c)\sim10-50$ (Faucher-Gigu\`ere \& Quataert~\cite{Fau12}).

%______________________________________________ Fig17:scheme
 \begin{figure*}[th!]
  \centering
%  \hskip 2cm
  %\vskip -2cm
  %\includegraphics[width=14cm]{scheme-new-revised.ps}
   \includegraphics[width=14cm]{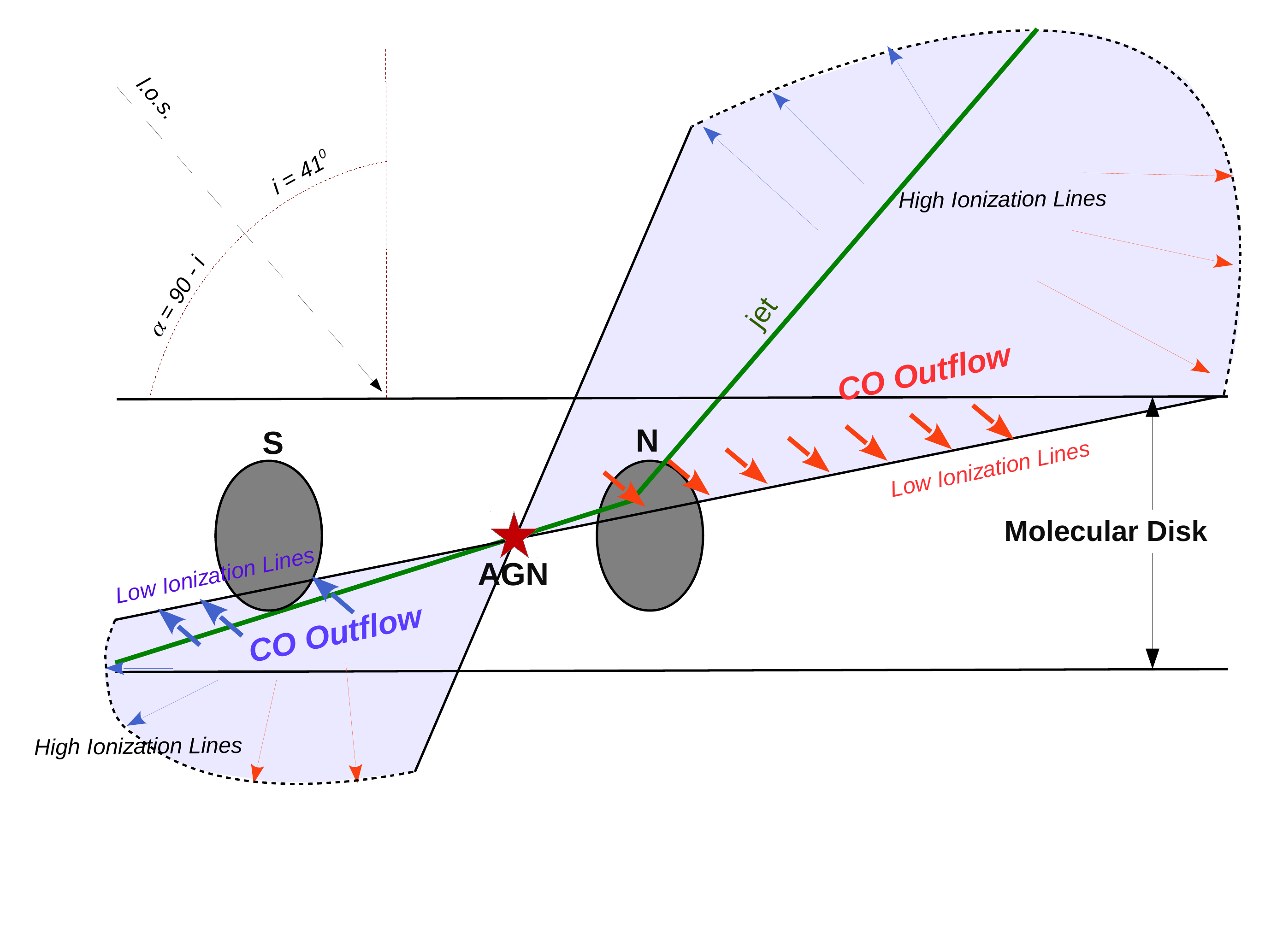}
     \vskip -1.35cm 
        \caption{A revised version of the kinematic model of the NLR, first proposed by Tecza et al.~(\cite{Tec01}) and Cecil et al.~(\cite{Cec02}), which now accounts for the molecular outflow (denoted in figure as {\em CO outflow}) detected by ALMA in the CND (seen in projection in N and S, for northern and southern knots) and farther north in the molecular disk. The figure shows a crosscut of the NLR as viewed from inside the galaxy disk along the projected direction of the radio jet ($PA\sim30^{\circ}$; shown by the green line). We highlight the extent of the ionized gas outflow (light purple shade) and the assumed geometry defined by angles $\alpha$ and $i$.}
              \label{scheme}
\end{figure*}
    
%__________________________________________________

Alternatively, the radio jet of NGC~1068 could also inject the required outflow power.  The jet power ($W_{\rm jet}$) can be estimated 
from the monochromatic luminosity at 1.4~GHz, according to B{\^i}rzan et al.~(\cite{Bir08}). In the case of NGC~1068, the spatially integrated flux of the jet at 1.4~GHz, including the main components (known in the literature as $NE$, $C$, and $S$), is $\sim840$~mJy from the radio continuum map of 
Gallimore et al.~(\cite{Gal96}). This implies $W_{\rm jet}=1.8\times10^{43}$erg~s$^{-1}$. Since 
$W_{\rm jet}\sim(30-100) \times L_{\rm kin}$, we conclude that the jet can drive the molecular outflow in the CND even assuming a low coupling efficiency. %The jet interaction with the
%interstellar medium has been simulated by Wagner et al.. (2012)
%who show that the jet is able to drive a flow efficiently as soon
%as the Eddington ratio of the jet Pjet/LEdd is larger than 10?4. In
%NGC 1433, this ratio is about 3.2 ? 10?3.

\subsection{Alternatives to the molecular outflow}\label{alternatives}

The velocity field of the outer disk from $r \sim 400$~pc out to $\sim1.8$~kpc shows a regularly rotating pattern with conspicuous distortions that reflect the expected 
perturbation on the gas flow due to the bar and the spiral structure inside corotation of the patterns (Sect.~\ref{non-circ}). 
While the outer disk and, also, the molecular gas component detected at the AGN (see Sect.~\ref{line-AGN}) seem to share the latter as the common dynamical center, the CND is a
strongly off-centered ring, which is {\em apparently} rotating but with a different kinematic axis, as shown in Fig.~\ref{CO-vels}. As an alternative to the outflow scenario described in Sects.~\ref{kinemetry} and \ref{outflow-dense},  the abrupt shift of $\geq 40^{\circ}$ degrees in the kinematic $PA$ of the
two disks can be interpreted as due to the CND being a non-coplanar disk which is far from being dynamically relaxed. Two types of non-coplanar instabilities can be invoked to account for the decoupled kinematics observed in the CND: a nuclear warp or a non-coplanar lopsided instability.

\subsubsection{A nuclear warp}

Schinnerer et al.~(\cite{Sch00}) explored a nuclear warp scenario in an attempt  to model the  kinematics of molecular gas in the nuclear region of NGC~1068, based on CO(2--1) observations obtained with the PdBI.  They concluded that gas motions could be equally fit with either a warp or a nuclear bar in the CND. One of the main predictions of their warp model was that the CO disk should become edge-on at a radius of $\sim70$~pc. However, we do not see the signature of an edge-on disk on these spatial scales in the higher-resolution ALMA maps presented in this work. Furthermore the internal kinematics of the CND do not show the typical {\em S-shaped distortion} attributable to a nuclear warp instability.

The diagram shown in Fig.~\ref{harm-II}, originally introduced by Wong et al.~(\cite{Won04}), can also be used as a diagnostic tool to identify nuclear warp signatures in the velocity field of galaxies in the particular case where the hypothesized tilted orbits share a common center. An inspection of Fig.~\ref{harm-II} indicates that the 
continuous line, which represents the least-squares fit to the NGC~1068 data points, shows no correspondence to the expected location of the {\em warp line}, which would 
relate the $s_1$ and $s_3$  terms in the case of a warp having the AGN as a common center of the tilted non-coplanar orbits. This disagreement, more evident in the inner disk, leads us to conclude  that the simplest nuclear warp scenario described above is not a satisfactory explanation for the velocity residuals observed in this region.

\subsubsection{A non-coplanar lopsided instability}

If we abandon the restriction of having the AGN as common orbiting center for both the CND and the outer disk, the scenario of non-coplanarity remains nevertheless viable. In this scenario, a {\em non-coplanar} CND would be orbiting around a {\em secondary} nucleus at $(\Delta$$\alpha$, $\Delta$$\delta) \sim (-1.5\arcsec, -1\arcsec)$, i.e., 1$^{\arcsec}=70$~pc offset to the southwest relative to the AGN.

Two types of mechanisms have been described as potential triggers of lopsided non-coplanar gas instabilities in galactic nuclei:
%\begin{itemize}
%\item 

{\it 1.~External trigger:}  The postulated {\em secondary} nucleus  could be the footprint of a recent minor merger with a nucleated satellite. In this scenario dynamical friction would make the satellite quickly sink toward the nucleus of the host dragging the accreted gas disk to the central region  (Taniguchi \& Wada~\cite{Tan96}; Taniguchi~\cite{Tan99, Tan13}). Minor mergers could thus explain a {\em random} orientation of circumnuclear gas disks 
in the NLR of Seyfert galaxies, as the orbital plane of the resulting nuclear disk would be determined  by the orbital   parameters of the accreted satellite (Taniguchi~\cite{Tan13}). As the CND is noticeably off-centered, this would imply that NGC~1068 would be at an early stage in the merging process where 
the secondary nucleus has not yet sunk toward the nucleus. However, the absence of any signature of a secondary nucleus in the high-resolution HST/NICMOS images of the CND, which should be at the position of the {\em apparent} guiding center of the non-coplanar disk,  makes the minor merger scenario implausible in NGC~1068.

{\it 2.~Internal trigger:}  Alternatively, a lopsided instability triggered on the gas could have an internal origin (Jog \& Combes~\cite{Jog09}). %This phenomenon happens frequently in galaxy nuclei, given the very different dynamical timescales involved.  The period of rotation of the outer disk in NGC~1068
%is of the order of  58~Myr, if we adopt a projected velocity of $\sim 100$~km~s$^{-1}$ at
%a 22" radius, which implies an intrinsic rotation velocity of $\sim 150$~km~s$^{-1}$ at 1.5~kpc, while
%in the CND, at 1$^{\arcsec}=70$~pc radius, the corresponding rotation period is $\sim 1.4$~Myr.  
The CND could be the result of a recent episode of gas infall produced by the feedback of a star formation burst in the disk.
This feedback action would be able to eject gas perpendicular to the plane, at
heights of 100-200~pc, as frequently seen in numerical simulations (e.g., Emsellem et al.~\cite{Ems14}). The gas fountain would fall
back while settling initially in an inclined disk, the perpendicular orientation
being preferred since differential precession is then canceled. This would explain the offset between the AGN and the guiding center of the instability. While settling back to the disk, the nuclear gas would  trigger a slow and long-lasting lopsided
perturbation, which  could remain during
more than 100 dynamical times at this radius, i.e., 400~Myr for the CND (e.g., Jog \& Combes~\cite{Jog09}). 
In this scenario it is difficult to predict what evolutionary stage during the settling of the disk best represents the case of NGC~1068.
However, at the end of the process, the $m=1$ instability and the main molecular disk would  become coplanar.

As will be discussed in Sect.~\ref{outflow-others}, the orientation of the jet and the geometry of the ionized gas outflow in NGC~1068 imply that these can efficiently interact with molecular gas in a {\em coplanar} CND. Overall, the outflow hypothesis is the simplest explanation for the distorted  kinematics of molecular gas at the CND, but also for the bow-shock arc region, the signature of the interaction at larger radii. In this scenario the remarkable off-centering of the molecular ring would be triggered before the jet or the ionized gas wind hits the disk and would launch the outflow rather than being the result of it. Furthermore, the lopsided morphology in a {\em coplanar} CND, which is characterized by a strong near side/far side asymmetry in NGC~1068, could also be enhanced due to opacity effects in molecular lines, as discussed by Boone et al.~(\cite{Boo11}).

  %______________________________________________ Fig 18:co-ratios
 \begin{figure*}[th!]
   \centering  
   \includegraphics[width=17cm]{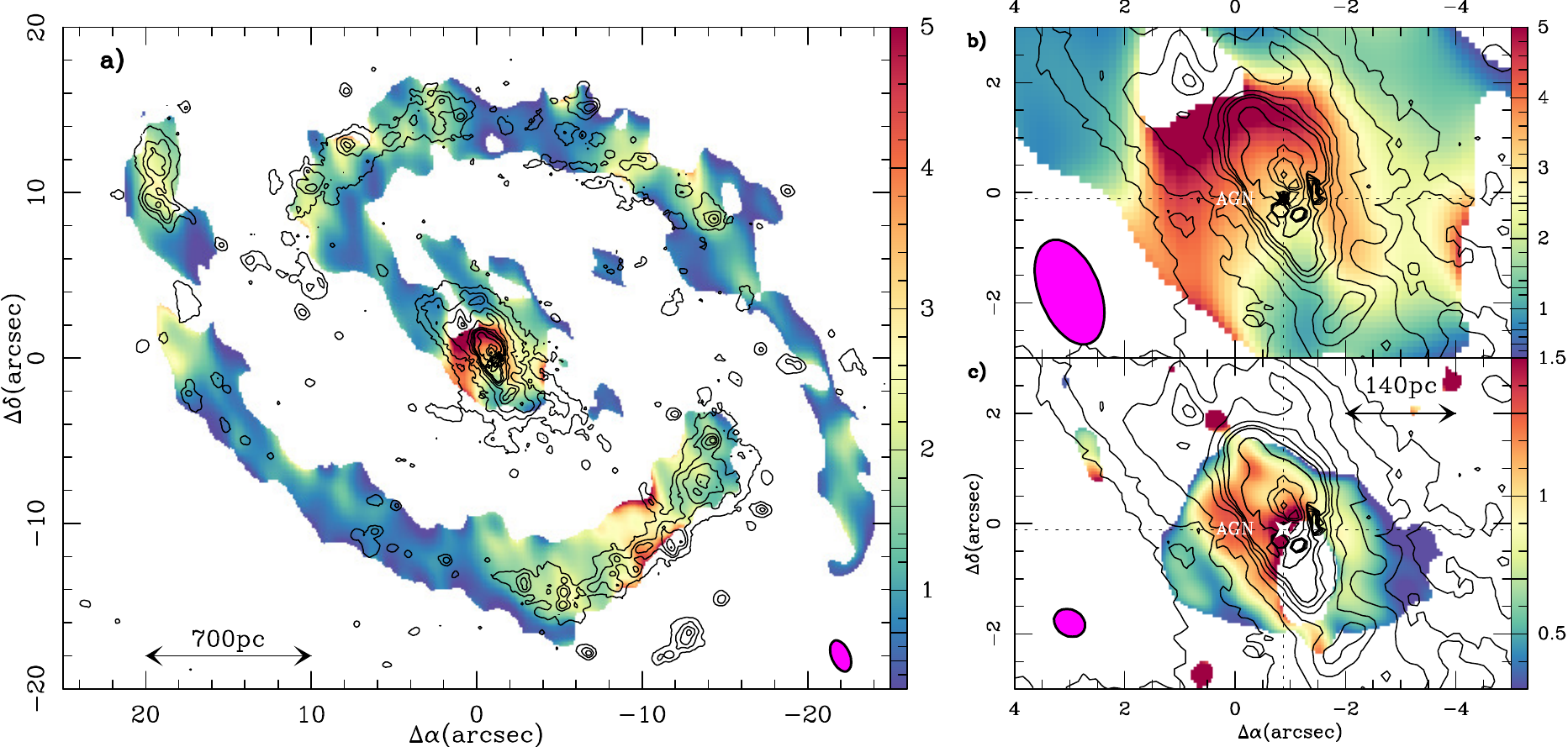}
              \caption{{\bf a)}~({\it Left panel})~Overlay of the  HST Pa$\alpha$ emission map (in contours as in Fig.~\ref{velres-zoom}{\it c}) on the CO(3--2)/CO(1--0) brightness temperature ratio map (in color scale and $T_{\rm mb}$ units) derived at the spatial resolution of the 1--0 observations of Schinnerer et al.~(\cite{Sch00}) ($2\arcsec \times1\farcs1$ at $PA=22^{\circ}$; ellipse in lower right corner). {\bf b)}~({\it  Upper right panel})~Same as {\bf a)} but showing a zoom in on the CND region. {\bf c)}~({\it Lower right panel})~Same as {\bf a)} but showing the CO(6--5)/CO(3--2) brightness temperature ratio map (in color scale and $T_{\rm mb}$ units) derived at the spatial resolution of the 3-2 observations ($0\farcs6\times0\farcs5$ at  $PA=60^{\circ}$; ellipse in lower left corner).}
              \label{co-ratios}
\end{figure*}
    
%__________________________________________________

\subsection{The molecular outflow in context: relation to other tracers}\label{outflow-others}

NGC~1068 harbors a wide-angle  (FWHM$\sim50-60^{\circ}$) biconical outflow of ionized gas (Macchetto et al.~\cite{Mac94}; Arribas et al.~\cite{Arr96}; Crenshaw et al.~\cite{Cre00}; Tecza et al.~\cite{Tec01}; Cecil et al.~\cite{Cec02}; Mueller-S\'anchez et al.~\cite{Mue11}). The orientation of the outflow is not perpendicular to the galaxy disk: $i_{outflow}\sim70-80^{\circ}$, while $i_{disk}\sim41^{\circ}$. This particular geometry favors interaction with the molecular disk out to a galactocentric radius of $r\sim400$~pc.  An interaction between the jet and the ISM was already identified close to the AGN ($r\sim20-40$~pc) by Gallimore et al.~(\cite{Gal96, Gal01}) (see also Bicknell et al.~\cite{Bic98}).

Figure~\ref{scheme} shows a simple kinematic model of the NLR, first proposed by Tecza et al.~(\cite{Tec01}) (see also: Cecil et al.~\cite{Cec02}; Das et al.~\cite{Das06, Das07}), here revised to account for the molecular outflow detected by ALMA. The figure shows a scaled crosscut of the NLR along the projected direction of the radio jet ($PA\sim30^{\circ}$). The observer's line-of-sight, which makes an angle $i_{disk}\sim41^{\circ}$ relative to the direction orthogonal to the molecular disk, as shown, is at 90$^{\circ}$ with respect to the viewer of this figure. The northeast side of the ionized gas cone is at the upper right of the figure. The CND, represented by two molecular knots located asymmetrically north and south of the AGN (N and S in figure), is embedded in the molecular disk.

 According to Tecza et al.~(\cite{Tec01}) (see also Cecil et al.~\cite{Cec02} and Mueller-S\'anchez et al.~\cite{Mue11}), high-ionization lines, like [Si{\small VI}] or [O{\small III}], are produced where the radio lobes encounter diffuse material located outside the galaxy plane. The interaction of the  radio lobes with this medium generate two bow-shock fronts that are responsible for the simultaneous detection of redshifted and blueshifted emission in both ionization cones, as shown in Fig.~\ref{scheme}. The low-ionization lines, like [Fe{\small II}], are produced when the bow-shock sweeps the denser molecular disk. This explains why low-ionization lines are exclusively  detected as redshifted (blueshifted) emission on the northern (southern) side of the disk. The amplitude of velocity shifts are larger for the high-ionization lines, reaching $\sim3000$~km~s$^{-1}$ in the northeastern cone, while they remain $\leq500$~km~s$^{-1}$ for the low-ionization lines. Furthermore, the sign of the velocity shifts for the strongest emission components of the high-ionization lines are noticeably reversed with respect to low-ionization lines.

 The molecular outflow detected by ALMA stands out as a redshifted (blueshifted) kinematic component on the northern (southern) side of the disk, as discussed in Sects.~\ref{kinemetry} and ~\ref{outflow-dense}, consistent with the low-ionization lines of Tecza et al.~(\cite{Tec01}). In this scenario, depicted in Fig.~\ref{scheme}, the molecular outflow is launched when the ionization cone of the NLR sweeps the disk in the CND and farther out north in the bow-shock arc region where Wilson \& Ulvestad~(\cite{Wil87}) found that the radio-lobe nebulosity becomes limb-brightened and highly polarized, the signature of a bow-shock in the disk. The amplitudes of the velocity shifts in the molecular outflow are significantly smaller than for the  high-ionization lines.  There is also evidence that the molecular outflow becomes decelerated at $r\sim400$~pc: the terminal velocities of the outflow in the bow-shock arc region are a factor 2 smaller than in the CND, as shown in Fig.~\ref{pvs}.

 The detection of emission from dense gas  tracers at the extreme velocities of the molecular outflow in NGC~1068 ($\geq150$~km~s$^{-1}$) raises the question of how molecular clouds can survive during the launching of the outflow, since shocks of velocities greater than $\sim50$~km~s$^{-1}$ are known to be fast enough to destroy molecules (Hollenbach \& McKee~\cite{Hol89}; Neufeld \& Dalgarno~\cite{Neu89a, Neu89b}). However, the chemistry in both dissociative J-type and non-dissociative C-type shocks, which involve dust-mantle disruption and a high-temperature environment, favors an efficient  fast  reformation of molecules in the gas. Molecules can reform in the post-shocked gas on timescales $\leq$~hundreds of yrs. 
There is ample observational evidence that the emission of dense gas tracers ($\geq 10^{5-6}$cm$^{-3}$) can coexist with fast shocks ($\sim 50-100$~km~s$^{-1}$) in galactic bipolar outflows of Young Stellar Objects (YSOs) (e.g., Arce et al.~\cite{Arc07} and references therein). These observations show that the abundance of some molecular species can undergo spectacular enhancements in the outflow gas, due to the onset of shock chemistry. Besides classical shock tracers such as SiO, whose abundance can be enhanced by factors of about 10$^4$,  molecular species which are less specific to shock environments, such as HCN or CS, can also undergo significant order-of-magnitude enhancements (e.g., Tafalla et al.~\cite{Taf10}; Tafalla~\cite{Taf13}). The detection of strong emission from some of these tracers, such as HCN or CS, at the velocities identified as {\em abnormal} in NGC~1068 suggests that molecular clouds survive in the outflow and that shock chemistry is likely at work in this component. 
  
    %______________________________________________ Fig 19: dense ratios
 \begin{figure}[tbh!]
   \centering 
   \includegraphics[width=8cm]{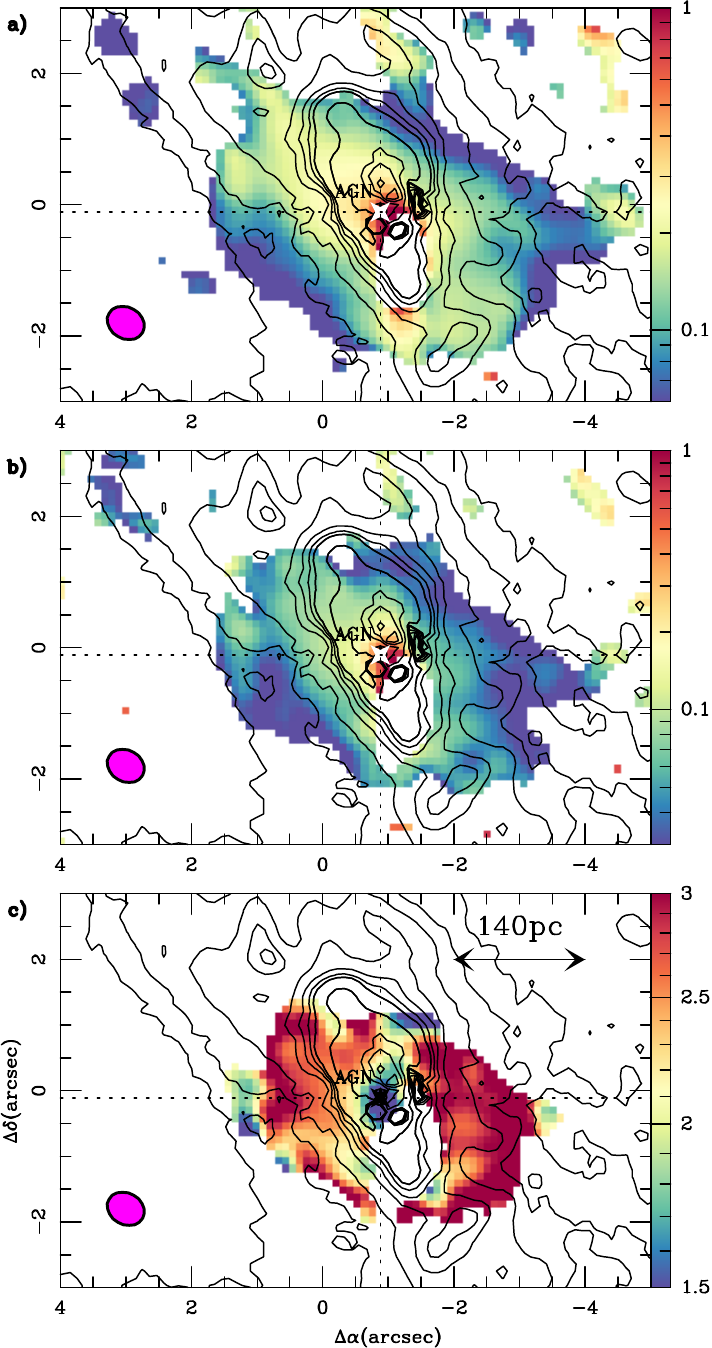}
               \caption{{\bf a)}~({\it Upper panel})~Overlay of the  HST Pa$\alpha$ emission map (in contours as in Fig.~\ref{velres-zoom}{\it c})  on the HCN(4--3)/CO(3--2) brightness temperature ratio map (in color scale and $T_{\rm mb}$ units). {\bf b)}~({\it Middle panel})~Same as {\bf a)} but showing the HCO$^+$(4--3)/CO(3--2) brightness temperature ratio map. {\bf c)}~({\it Lower panel})~Same as {\bf a)} but showing the HCN(4--3)/HCO$^+$(4--3) brightness temperature ratio map. All the ratio maps have been derived at the spatial resolution of the CO(3-2) observations ($0\farcs6\times0\farcs5$ at  $PA=60^{\circ}$; ellipses in lower left corners).}
              \label{dense-ratios}
\end{figure}
    
%________________________________________________

 \section{Molecular line ratios and environment}\label{lineratios}
 
  \subsection{The SB ring versus the CND}

   Figure~\ref{co-ratios} shows the overlay of the  HST Pa$\alpha$ emission map on the CO(3--2)/CO(1--0) (hereafter $R32/10$) and CO(6--5)/CO(3--2) (hereafter $R65/32$) line ratio maps. Line ratios were derived in $T_{\rm mb}$ units. To derive the 3--2/1--0 brightness temperature ratio, shown in panels ab, we degraded the 3--2 map to the spatial resolution of the 1--0 observations of Schinnerer et al.~(\cite{Sch00}) ($2\arcsec \times1\farcs1$). The CO(6--5)/CO(3--2) brightness temperature ratio map, shown in panel c, was derived at the spatial resolution 
 of the 3--2 ALMA observations ($0\farcs6\times0\farcs5$). Line ratios were obtained assuming a common 
 3$\sigma$ clipping on the integrated intensities to assure image reliability.

 The $R32/10$ ratio changes  significantly  depending on the particular environment of the disk. The average ratio is $\sim 1.2\pm0.02$ in the 
 SB ring yet $R32/10$ shows a large dispersion of values:  $R32/10\sim0.7-1$ in the less actively star-forming regions of the SB ring, which are 
 characterized by low or undetected  Pa$\alpha$ emission, while the strongest Pa$\alpha$ emitting regions show $\sim$a factor 3--4 higher ratios ($R32/10\sim2-3$). %The few interarm complexes simultaneously detected in the 3--2 and 1--0 lines show low values of $R32/10\sim0.5-0.7$. 

 Overall, the $R32/10$ ratio in the CND, $\sim2.7\pm0.1$, is higher 
 than in the SB ring, in agreement with the global values measured with the PdBI and the SMA by  Krips et al.~(\cite{Kri11}) and Tsai et al.~(\cite{Tsa12}). In summary, the $R32/10$ ratios
  measured in NGC~1068, most particularly those of the CND,  are at the high end of the typical values  observed in the centers of nearby normal 
  and starburst galaxies: $\sim0.5-1.4$ (Devereux et al.~\cite{Dev94}), $\sim0.2-0.7$ (Mauersberger et al.~\cite{Mau99}), $\sim0.2-1.9$ (Mao et al.~\cite{Mao10}), an indication that the excitation of molecular gas is extreme in the CND (Krips et al.~\cite{Kri11}; paper II).

 \subsection{Line ratio  changes in the CND: the footprint of AGN feedback?}

  At the spatial resolution used in Figs.~\ref{co-ratios}{\it ab}, the  $R32/10$ ratio shows a range of values in the CND:  $R32/10$ goes from  $\sim1$ at the southern end of the ring to $\sim5$ at the northern end, and it is $\sim3-4$ at the AGN and at the E and W CO knots.  The $R32/10$ ratio shows a north-south gradient which runs in parallel with 
 the orientation of the bipolar AGN/jet nebulosities. Overall, the $R65/32$ ratio in the CND is $\sim0.8$, with a significant range of values: R$65/32$ goes from  $\sim0.7$ at the southern end of the CND ring to a maximum value $\sim2.0$ at the AGN.  At the higher spatial resolution of Fig.~\ref{co-ratios}{\it c}, the excitation of CO in the CND probed by the  $R65/32$ ratio unveils the footprint of AGN feedback: the  $R65/32$ ratio is enhanced by the degree of {\em illumination} of the molecular gas by the photons of the ionized gas outflow traced by Pa$\alpha$ emission.     

 Other molecular line ratios show dramatic changes of up to an order of magnitude inside the CND on the spatial scales probed by ALMA ($\sim35$~pc). Figure~\ref{dense-ratios} shows the overlay of the  HST Pa$\alpha$ emission map on a set of four molecular line ratio maps (in $T_{\rm mb}$ units) obtained at the resolution of ALMA observations in  Band 7: 
 (a) HCN(4--3)/CO(3--2) ($R_{\rm HCN/CO}$), (b) HCO$^+$(4--3)/CO(3--2) ($R_{\rm HCO^+/CO}$), and
 (c) HCN(4--3)/HCO$^+$(4--3) ($R_{\rm HCN/HCO^+}$).

 A visual inspection of Fig.~\ref{dense-ratios} indicates that, as for CO, the $R_{\rm HCN/CO}$ and $R_{\rm HCO^+/CO}$ ratios also show the footprint of AGN feedback: these line ratios go from $\sim 0.02-0.03$ at the outer radii of the CND to $\sim$0.8 at the AGN, showing an enhancement which runs in parallel with the irradiation of molecular gas  by the photons of the bipolar ionized gas nebulosity.

 The  $R_{\rm HCN/HCO^+}$ ratio is globally very high in the CND with an average value of $\sim$2.5. These high values have been widely used as a diagnostic ratio to identify an AGN-like environment (Kohno et al.~\cite{Koh01}; Usero et al.~\cite{Use04}; Garc\'{\i}a-Burillo et al.~\cite{Gar06}; Graci\'a-Carpio et al.~\cite{Gra06, Gra08}; Imanishi et al.~\cite{Ima07, Ima13}; Krips et al.~\cite{Kri08, Kri11}). High HCN/HCO$^+$ ratios measured in galaxy nuclei have also been interpreted as the signature of significant mechanical heating in shock/mechanically dominated regions (MDRs) (Kazandjian et al.~\cite{Kaz12}).  Nevertheless, the ALMA maps show that the lowest value of this ratio is found precisely at the AGN locus, where $R_{\rm HCN/HCO^+}\sim1.3$, an indication that a simplistic interpretation of these ratios can be misleading (see paper II). The comparatively lower value of $R_{\rm HCN/HCO^+}$ at the AGN might reflect departures from chemical equilibrium: the HCN/HCO$^+$ mass ratio is strongly time-dependent with variations that can reach orders of magnitude for times $\geq10^4$~years (Meijerink et al.~\cite{Meij13}).

          %______________________________________________ Fig 20: scatter-plots
 \begin{figure*}[tbh!]
 \centering
   \includegraphics[width=7.5cm]{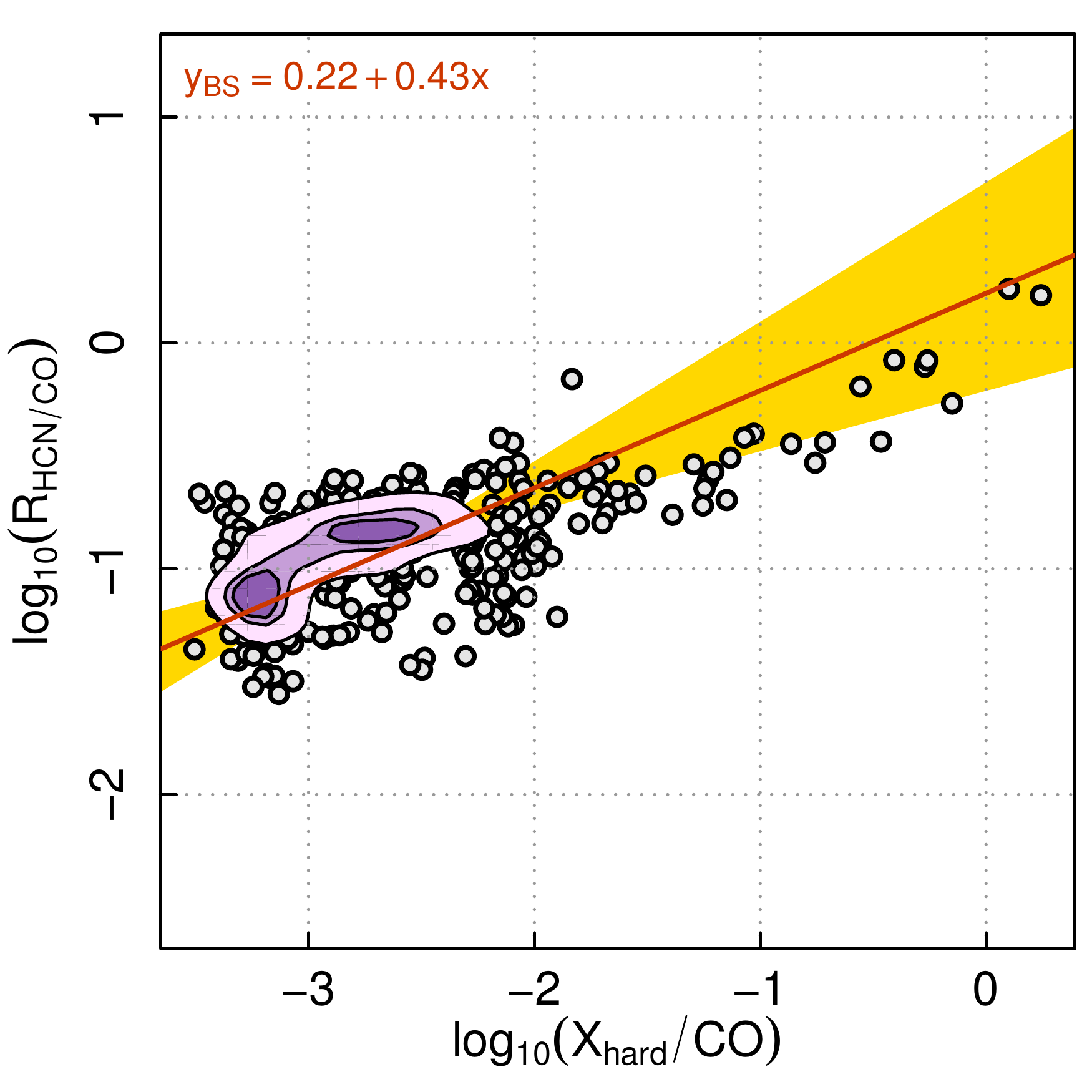}
   \includegraphics[width=7.5cm]{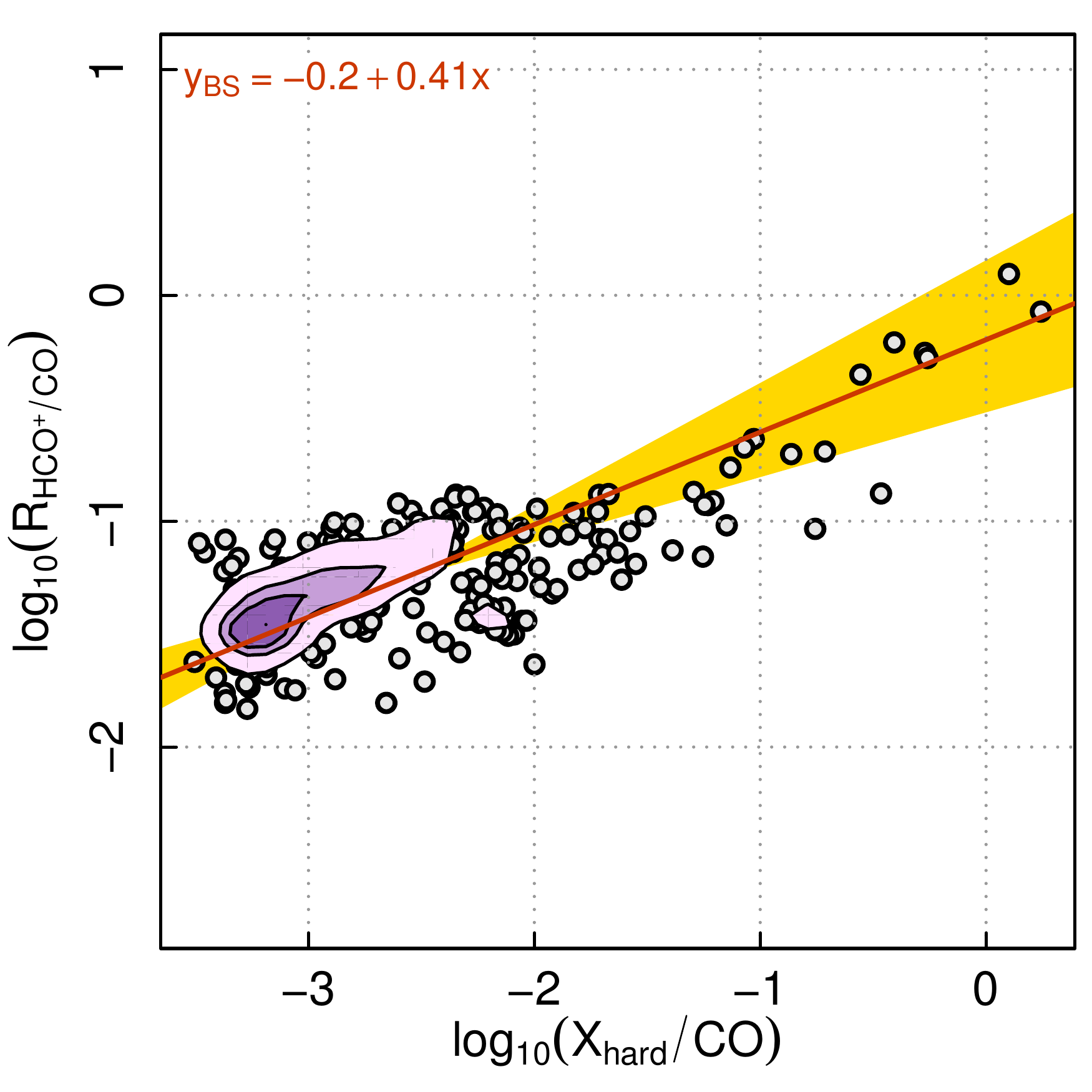} 
             \caption{{\bf a)}~({\it Left panel})~The HCN(4--3)/CO(3--2) velocity-integrated intensity ratio ($R_{\rm HCN/CO}$, in logarithmic scale) versus the ratio of X-ray flux in the 6--8~keV band (in counts~s$^{-1}$) to the CO(3-2) intensities (in K~km~s$^{-1}$) ($X_{\rm hard}$/CO, in logarithmic scale) measured at a common spatial resolution of 0.5$\arcsec$ over the CND (open circles). Isodensity contours illustrate the distribution of values in this parameter space for 10$\%$, 25$\%$, and 50$\%$ of the data points. The straight line represents the bisector linear fit to the data in logarithmic  scale [log($X_{\rm hard}$/CO), log($R_{\rm HCN/CO}$)].  The yellow-shaded area shows the range allowed by the ordinary least squares fits of $y$ as a function of $x$ and viceversa. The parameters for the bisector fit are indicated. {\bf b)}~({\it Right panel})~Same as  {\bf a)} but particularized for the HCO$^+$(4--3)/CO(3--2) ratio ($R_{\rm HCO^+/CO}$) versus $X_{\rm hard}$/CO. The linear correlation coefficients of the regressions are $r\sim0.7$}
              \label{scatter-plots}
\end{figure*}
    
%________________________________________________

 As discussed in Garc\'{\i}a-Burillo et al.~(\cite{Gar10}), the morphology of hard-X ray emission in the 6--8~keV band obtained by Chandra (Young et al.~\cite{You01}; Ogle et al.~\cite{Ogl03}) can be used to search for observational evidence of AGN feedback on the excitation/chemistry of molecular gas in the CND. Emission in the 6--8~keV band observed in NGC~1068 is dominated by reflection of X rays by cold neutral (presumably mostly molecular) gas (Iwasawa et al.~\cite{Iwa97}).  Figure~\ref{scatter-plots} shows the $R_{\rm HCN/CO}$ and $R_{\rm HCO^+/CO}$ ratios in the CND as a function of the irradiation of molecular gas by hard X-rays normalized by gas column density; the latter is derived from the ratio of X-ray flux in the 6--8~keV band to the CO(3-2) intensities ($X_{\rm hard}$/CO). All quantities have been measured at a common spatial resolution of 0.5$\arcsec$. With an admittedly large scatter, Fig.~\ref{scatter-plots} provides quantitative evidence of a correlation between the $R_{\rm HCN/CO}$ and $R_{\rm HCO^+/CO}$ ratios and  $X_{\rm hard}$/CO. This is  similar to the correlation found for the SiO/CO and CN/CO ratios as function of hard X-ray illumination of molecular gas in the CND, discussed by  Garc\'{\i}a-Burillo et al.~(\cite{Gar10}).     
  
  The existence of an AGN-driven outflow proves that the feedback of activity is shaping the gas kinematics in the inner $r\sim400$~pc of NGC~1068. The change in molecular line ratios in the CND described above can be taken as independent evidence that radiative and/or mechanical feedback from the AGN is at work. The feedback of activity on the excitation and chemistry of molecular gas can be radiative, in PDR/XDR environments, or mechanical, in MDRs. In paper II we use the line ratio maps derived in this work, complemented by those derived from interferometric observations of NGC~1068 done in other molecular tracers,  to model the excitation and chemistry of molecular gas in the different environments of the CND and the SB ring and find a link between the different sources of energy present in the disk.

 \section{Summary and conclusions}\label{summary} 

 We have used ALMA to map the emission of a set of dense molecular gas tracers (CO(3--2), CO(6--5), HCN(4--3), HCO$^+$(4--3), and CS(7--6)) and their underlying continuum emission in the central $r\sim2$~kpc of NGC~1068 with spatial resolutions $\sim0.3\arcsec-0.5\arcsec$ (20--35~pc). These observations have greatly improved the sensitivity and spatial resolution of any previous interferometric study of this Seyfert, allowing us to make a significant progress in our understanding of the fueling and the feedback of activity in NGC~1068.

We summarize the main results of our study as follows:

\begin{itemize}

\item

The CO(3--2) line and underlying continuum image the distribution of the dense molecular gas and dust emission from three main regions: the CND, the bar, and the SB ring. We also report the detection of CO(3--2) emission at different locations over the interarm region.

\item 

The CND, which is fully resolved  in the CO(6--5) map, is a highly-structured closed elliptical ring of 350~pc-size. Similar to the dust continuum ring, the CO ring is noticeably off-centered relative to the AGN.  

\item 

Outside the CND, the CO(3--2) emission is detected along two elongated offset lanes that run mostly parallel to the bar major axis. We also detect a gas component with anomalous velocities on the northeastern side of the disk at distances $\sim400$~pc from the AGN. This CO feature has a counterpart in dust continuum emission and is interpreted as the signature of a {\em bow-shock} in the molecular disk.

\item

As for dust emission, most of the CO(3--2) flux in the disk is detected in the SB ring: a two-arm spiral structure that starts from the ends of the stellar bar and unfolds in the disk over $\sim180^{\circ}$ in azimuth forming a pseudo-ring.

\item

In stark contrast with the CO(3--2) map, most of the emission in  HCO$^+$(4--3), HCN(4--3), and CS(7--6) stems from the CND.  This reflects the different line ratios measured in the CND and in the SB ring. The line ratio maps at the spatial resolution of ALMA show dramatic changes of up to an order of magnitude inside the CND. These changes are tightly correlated with the  UV/X-ray illumination by the AGN, which varies significantly on these spatial scales, hinting at the footprint of AGN feedback.

\item

We detect dust continuum and molecular line emission at the AGN.  The emission peak is nevertheless shifted to the northeast in a structure that connects the AGN with the CND. We used the ALMA fluxes in Bands 9 and 7 together with NIR/MIR data to constrain the properties of the putative torus using CLUMPY models. 
The fitted outer torus radius is $R_{\rm o}=20^{+6}_{-10}\,$pc.; the torus mass is $M_{\rm
  torus} = 2.1 (\pm 1.2) \times 10^{5}\,M_\odot$. The ALMA fluxes lie above the model predictions, an indication that a non-negligible fraction of cold dust emission not associated with the torus was included in the 20~pc--resolution data of ALMA. 

\item

The Fourier decomposition of the gas velocity field derived from the CO(3--2) data  indicates that rotation is perturbed by  an inward radial flow in the SB ring and the bar region, as expected if we are inside corotation of the perturbations in these regions.  However, the kinematics of the CND and the bow-shock arc are shaped by an outward radial pattern superposed to rotation. There is a tight spatial correlation between the AGN ionized nebulosity, the radio jet plasma and the molecular outflow signature identified in the CO velocity field from $r\sim50$~pc to $r\sim400$~pc, which suggests that the outflow is AGN driven.  
The molecular outflow is launched when the ionization cone of the NLR sweeps the disk in the CND and farther out north in the bow-shock arc region.

\item
 The estimated outflow rate in the CND, $\mathrm{d}M/\mathrm{d}t\sim63^{+21}_{-37}~M_{\odot}$~yr$^{-1}$, an order of magnitude higher than the SFR measured at these radii, confirms that the outflow is AGN driven. The molecular mass load rate implies a very short gas depletion timescale of $\leq1$~Myr in the CND. 

\end{itemize}

The AGN-driven molecular outflow in NGC~1068 could quench star formation in the inner $r\sim400$~pc of the galaxy on short timescales and at the same time regulate gas accretion in the CND. However, the molecular gas reservoir in the CND is likely replenished on longer timescales by efficient  gas inflow from the outer disk. The signature of gas inflow has been identified in the velocity field at larger radii and has been attributed to the combined action of the bar and the spiral arms. Self-regulation of star formation and gas accretion would be thus possible by a combination of density wave-driven inflows and AGN-driven outflows. A similar scenario of self-regulation has been invoked to be at work in gas-rich high-redshift disk galaxies, analyzed in the numerical simulations of Gabor \& Bournaud~(\cite{Gab14}). 

While the gas kinematics are dominated by the molecular outflow in the CND,  on theoretical grounds it is nevertheless  expected that the gas should be fueling the AGN at smaller radii. 
The NIR data of M{\"u}ller-S{\'a}nchez et al.~(\cite{Mue09}) showed elliptical streamers that bridge the CND and the central engine. They were interpreted as a tentative evidence of ongoing AGN fueling.
However, with the spatial resolution of the observations presented in this paper we cannot spatially resolve the kinematics of the dense molecular gas in the inner $r\sim50$~pc.  Obtaining higher-resolution data is important to studying the inflow/outflow signatures in the region that connects the CND with the AGN.   A significant improvement in spatial resolution would also allow us to resolve with quasi-thermal molecular lines the region where the jet is known to be interacting with the  gas, image the connection to the inner  $r\sim0.7$~pc H$_2$O megamaser disk (Gallimore et al.~\cite{Gal01}) and spatially resolve the emission from the putative molecular torus.

\begin{acknowledgements}
         We acknowledge the staff of ALMA in Chile and the ARC-people at IRAM-Grenoble in France 
for their invaluable help during the data reduction process. This paper
makes use of the following ALMA data: ADS/JAO.ALMA$\#$2011.0.00083.S.
ALMA is a partnership of ESO (representing its member states), NSF (USA), and NINS (Japan), together with NRC (Canada) and NSC and ASIAA (Taiwan),
in cooperation with the Republic of Chile. The Joint ALMA Observatory is
operated by ESO, AUI/NRAO, and NAOJ. %The National Radio Astronomy
%Observatory is a facility of the National Science Foundation operated under cooperative
%agreement by Associated Universities, Inc. 
We used observations made
with the NASA/ESA Hubble Space Telescope, and obtained from the Hubble
Legacy Archive.%, which is a collaboration between the Space Telescope Science
%Institute (STScI/NASA), the Space Telescope European Coordinating Facility
%(ST-ECF/ESA), and the Canadian Astronomy Data Centre (CADC/NRC/CSA).     
 SGB and IM acknowledge support from Spanish grants AYA2010-15169 and from the Junta de Andalucia through TIC-114 and the Excellence Project P08-TIC-03531. SGB, AL, and AF acknowledge support from MICIN within program CONSOLIDER INGENIO 2010, under grant 'Molecular Astrophysics: The Herschel and ALMA Era--ASTROMOL' (ref CSD2009-00038). 
 SGB, AU, LC, and PP acknowledge support from Spanish grant AYA2012-32295. 
 FC acknowledges the European Research Council for the Advanced Grant Program Num.~267399-Momentum.
 AAH acknowledges support from the Universidad de
Cantabria through the Augusto G. Linares programme and
from the Spanish Plan Nacional grants AYA2009-05705-E
and AYA2012-31447. CRA is supported by a Marie Curie Intra European Fellowship
within the 7th European Community Framework Programme (PIEF-GA-2012-327934).
CRA also ackowledges financial support from the Spanish Ministry of Science and Innovation (MICINN) through project 
PN AYA2010-21887-C04.04 (Estallidos).

\end{acknowledgements}

%-------------------------------------------------------------------

\appendix

%\begin{appendix} %First online appendix
\section{CLUMPY torus models}

Over the past few years, a number of studies have demonstrated clumpy
torus models (Nenkova et al.~\cite{Nen08a, Nen08b}; Schartmann et
al.~\cite{Sch08}; Hoenig \& Kishimoto~\cite{Hoe10a}) represent well the nuclear
infrared emission of local Seyfert galaxies, provided there is no
contamination from extended dust components not related to the torus 
(e.g., foreground absorbing dust, optically thin
emitting dust in ionization cones, etc). More specifically, these
models reproduce the non-stellar NIR and MIR SEDs and
MIR spectroscopy of local Seyfert galaxies  (Ramos Almeida et al.~\cite{Ram09,Ram11a}; Hoenig et
al.~\cite{Hoe10b}; Alonso-Herrero et al.~\cite{Alo11}; Lira et al.~\cite{Lir13}) and PG quasars (Mor et
al.~\cite{Mor09}). By modeling the nuclear infrared data of AGN we can
constrain the torus geometry, dust properties, and distribution, and the 
AGN bolometric luminosity. 
Moreover, adding far-infrared nuclear 
photometric points or even upper limits has proven to be useful to
constrain the torus sizes, especially if the dusty clumps have a
relatively uniform radial distribution along the torus 
(Ramos Almeida et al.~\cite{Ram11b}; Asensio Ramos \& Ramos Almeida~\cite{Ase13}; Ramos Almeida et al.~\cite{Ram14}). 

Alonso-Herrero et al.~(\cite{Alo11}) fitted the nuclear NIR and MIR
SED and  MIR spectroscopy of NGC~1068 using the so-called {\sc CLUMPY} torus models of Nenkova et al.~(\cite{Nen08a, Nen08b}). 
The  parameters in the {\sc CLUMPY} models are the torus size $Y$  defined as the ratio
between the outer radius $R_{\rm o}$ and the inner radius\footnote{The inner radius is set by the assumed dust 
 sublimation temperature of 1500~K and the AGN bolometric
 luminosity.} $R_{\rm
  sub}$, the torus angular size $\sigma$, the viewing
angle $i$, the number of clouds along the equatorial direction $N_0$,
the optical depth of the clouds $\tau_V$, and the index $q$ of the
radial distribution of the clouds $\propto r^{-q}$. Using a
Bayesian approach to fit the data with the {\tt BayesClumpy} tool (Asensio Ramos \& Ramos Almeida~\cite{Ase09}) the torus model
parameters of NGC~1068 were well constrained. In their fitting they restricted the torus size to small values based on the $12\,\mu$m interferometric sizes of a few parsecs inferred for nearby AGN (Burtscher et al.~\cite{Bur13}). 

We can use the ALMA Band 9 and Band 7 continuum thermal fluxes at
$435\,\mu$m  and $860\,\mu$m, respectively,  to investigate whether we
can set further constraints on the torus properties of NGC~1068. To do
so we also added the NGC~1068 SED 
and MIR spectroscopy presented in Alonso-Herrero et
al.~(\cite{Alo11}). 
%We added the ALMA Band 9 measurement at the full
%resolution of $0\farcs3$, which is similar to
%that of the MIR data. The ALMA Band 7
%continuum measurement  has a slightly worse resolution ($0\farcs5$)
%and a higher contribution from non-thermal emission as discussed in Sect.~\ref{cont-ratio},
%and therefore we used the estimated thermal component as an upper
%limit. For the ALMA Band 9 flux uncertainties we included both the
%error in the photometric calibration ($\sim 30\%$) and the
%uncertainties associated with the modeling of the non-thermal
%component at this wavelength estimated above ($\sim 10\%$). 
As in Alonso-Herrero et
al.~(\cite{Alo11}) we used the  detection of a maser in the nuclear region of
NGC~1068, which implies 
a close to edge-on view of the AGN accretion disk, to set the following prior
for the viewing angle $i=60-90^{\circ}$. Because we are including the ALMA
far-infrared photometry we allowed the full range for the torus size
$Y=5-100$. We used the new version of the BayesClumpy tool (Asensio
Ramos \& Ramos Almeida~\cite{Ase09}) which now interpolates linearly between
the {\sc CLUMPY} models.

Based on the fit, we can also estimate the gas mass in the torus. In the models of Nenkova et al.~(\cite{Nen08b}) the total mass in torus clouds can
be written as: 
\begin{equation}
M_{\rm torus}=4\pi m_{\rm H} (\sin \sigma)N_{\rm
  torus}^{\rm eq}R_{\rm sub}^2 Y I_q(Y)
 \label{Mtor}
\end{equation}  
  
\noindent where the function $I_q(Y)=1$ for $q=2$ and $N_{\rm torus}^{\rm eq}$
is the column density along the equatorial direction. The inner radius of 
the torus $R_{\rm sub}$ was computed from the AGN bolometric luminosity
inferred from the fit and we assumed a gas-to-$A_{\rm V}$ ratio of $N_{\rm H_2}/A_{\rm V} =
1\times10^{21}\,{\rm mol}\,{\rm cm}^{-2}\,{\rm mag}^{-1}$, taken from Bohlin et al~(\cite{Boh78}). The typical scatter  
reported by Bohlin et al.~(\cite{Boh78}) for this factor in H$_2$ dominated gas clouds is 30$\%$.

The results of the fit in NGC~1068 and their implications are discussed in Sect.~\ref{CLUMPY-models}.

%We obtained: $M_{\rm 
%torus}=2.1 (\pm 1.0)\times 10^5\, M_\odot$, where we considered that
%the main uncertainty comes from the relatively unconstrained torus
%size $Y$. This mass estimate is comfortingly similar to the estimated molecular gas mass 
%detected inside the central $r=20$~pc aperture derived from the CO(3--2) emission, as discussed in 
%Sect.~\ref{XCO}.

%\begin{figure*}
%\centering
%\includegraphics[width=16.4cm,clip]{1787f24.ps}
%\caption{Plotted above...}\label{appfig}
%\end{figure*}

%\end{appendix}
\end{document}